\documentclass[11pt,a4paper]{article}
\pdfoutput=1
\usepackage{jheppub}
\usepackage{lscape}
\usepackage{array}

\usepackage{amsfonts}
\usepackage{amssymb,amsmath}
\usepackage{mathtools}
\usepackage{bm}
\usepackage{bbm} 
\usepackage{graphicx}
\usepackage{caption}
\usepackage{subcaption}
\usepackage{amsmath}

\usepackage[normalem]{ulem}

\def\rme{{\rm e}}
\newcommand{\ii}{\mathrm{i}}

\newcommand{\CC}{\Omega}
\newcommand\be{\begin{equation}}
\newcommand\ee{\end{equation}}
\newcommand\bea{\begin{eqnarray}}
\newcommand\eea{\end{eqnarray}}

\newcommand{\nn}{\nonumber}
\newcommand{\dd}{\mathrm{d}}

\renewcommand{\=}{\;= \;}
\newcommand{\+}{\;+ \;}

\renewcommand{\a}{\alpha}

\newcommand{\s}{\sigma}
\renewcommand{\t}{\tau}
\newcommand{\g}{\gamma}

\renewcommand{\nn}{\nonumber}

\newcommand{\wt}{\widetilde}

\renewcommand{\Re}{\text{Re}}
\renewcommand{\Im}{\text{Im}}

\newcommand{\CN}{\mathcal{N}}
\newcommand{\CU}{\mathcal{U}}

\newcommand{\p}{\partial}

\newcommand{\CI}{\mathcal I}

\newcommand{\IZ}{\mathbb Z}
\newcommand{\IR}{\mathbb R}
\newcommand{\IC}{\mathbb C}
\newcommand{\IH}{\mathbb H}

\renewcommand{\v}{\varphi}

\renewcommand{\i}{{\rm i}}

\newcommand{\Ge}{\Gamma_\text{e}}
\newcommand{\e}{{\bf e}}

\newcommand{\half}{\frac12}
\newcommand{\z}{\zeta}
\newcommand{\n}{\nu}

\newcommand{\imt}{\tau_2}

\newcommand{\Seff}{S_\text{eff}}

\newcommand{\uu}{\underline{u}}

\newcommand{\taumn}{T}

\newcommand{\fli}{{\bf i}} 

\newcommand\llb{ab}
\newcommand\sllb{\sum_{a \to b}}
\newcommand\bi{\begin{itemize}}
\newcommand\ei{\end{itemize}}

\newcommand\Qabz[3]{Q_{#1,#2}(#3)}
\newcommand\bspl{\begin{split}}
\newcommand\espl{\end{split}}

\title{The large-$N$ limit of the 4d~$\mathcal{N}=1$ superconformal index}

\author[a]{Alejandro Cabo-Bizet,}
\emailAdd{alejandro.cabo\_bizet@kcl.ac.uk}
\author[b]{Davide Cassani,}
\emailAdd{davide.cassani@pd.infn.it}
\author[c,d]{Dario Martelli,}
\emailAdd{dario.martelli@unito.it}
\author[a]{Sameer Murthy}
\emailAdd{sameer.murthy@kcl.ac.uk}
\affiliation[a]{\textit Department of Mathematics, King's College London,\\
The Strand, London WC2R 2LS, U.K.}
\affiliation[b]{\textit INFN, Sezione di Padova, \\
Via Marzolo 8, 35131 Padova, Italy}
\affiliation[c]{\textit Dipartimento di Matematica ``Giuseppe Peano'', Universit\`a di Torino,\\
Via Carlo Alberto 10, 10123 Torino, Italy}
\affiliation[d]{\textit INFN, Sezione di Torino \& Arnold--Regge Center,\\ 
Via Pietro Giuria 1, 10125 Torino, Italy}

\abstract{
We systematically analyze the large-$N$ limit of the superconformal index of $\mathcal{N}=1$ superconformal theories having a quiver description. The index of these theories is known in terms of unitary matrix integrals, which we calculate using the recently-developed technique of elliptic extension. This technique allows us to easily evaluate the integral as a sum over saddle points of an effective action in the limit where the rank of the gauge group is infinite. For a generic quiver theory under consideration, we find a special family of saddles whose effective action takes a universal form controlled by the anomaly coefficients of the theory. This family includes the known supersymmetric black hole solution in the holographically dual~AdS$_5$ theories. We then analyze the index refined by turning on flavor chemical potentials. We show that, for a certain range of chemical potentials, the effective action again takes a universal cubic form that is controlled by the anomaly coefficients of the theory. Finally, we present a large class of solutions to the saddle-point equations which are labelled by group homomorphisms of finite abelian groups of order~$N$ into the torus. 
}

\begin{document}
 
\maketitle


\section{Introduction and summary \label{sec:Intro}}

The last couple of years have seen good progress in the study of the $\frac{1}{16}$-BPS superconformal 
index of four dimensional~$\CN=4$ super Yang-Mills theory (SYM) and, more generally, 
the $\frac{1}{4}$-BPS index in~$\CN=1$ superconformal field theories (SCFT). The index in question 
is a supersymmetric partition function which receives contributions from states that preserve 
two supercharges, which is the minimum amount of supersymmetry required to construct
such a quantity protected under supersymmetric deformations of the theory. 
Apart from its importance in capturing the protected spectrum of the field theory, this index also plays an important 
role in the gauge/gravity duality.
The holographic dual of a 4d~$\CN=1$ SCFT is a gravitational theory on AdS$_5$ 
which admits black hole solutions preserving two 
supercharges~\cite{Gutowski:2004ez,Gutowski:2004yv,Chong:2005hr,Chong:2005da,Kunduri:2006ek}. 
The AdS/CFT  correspondence predicts that the growth of states of the index in the large central charge limit 
should capture the Bekenstein-Hawking entropy of the black hole, and it is this aspect 
that has particularly motivated the recent progress.

These indices were first calculated in~\cite{Romelsberger:2005eg,Kinney:2005ej}
in the form of integrals over unitary matrices, and the recent progress involves a detailed 
study of these integrals. Independent studies in the last couple of years have reached the 
conclusion that the $\frac14$-BPS index in~$\CN=1$ theories (or its direct lifts like the~$\frac{1}{16}$-BPS 
in~$\CN=4$ SYM) indeed captures the entropy of the dual black hole at 
large~$N$~\cite{Hosseini:2017mds,Cabo-Bizet:2018ehj,Choi:2018hmj, Choi:2018vbz, Benini:2018ywd,
Honda:2019cio,ArabiArdehali:2019tdm,Zaffaroni:2019dhb,Kim:2019yrz,Cabo-Bizet:2019osg,Amariti:2019mgp,
Lezcano:2019pae,Lanir:2019abx,Cabo-Bizet:2019eaf,ArabiArdehali:2019orz}.
The basic idea of all the approaches is the same and can be paraphrased as follows: 
one calculates the index of the BPS states, and shows that it agrees with the ``entropy function" of the BPS black hole. 
The entropy function is a function of the chemical potentials dual to the charges 
whose Legendre transform yields the black hole entropy~\cite{Hosseini:2017mds}. 
More precisely, on the gravity side this function is a regularized on-shell action of the dual 
AdS$_5$ black hole geometry~\cite{Cabo-Bizet:2018ehj,Cassani:2019mms}. 
 
The different studies are essentially variants of three approaches, each of which have advantages and 
disadvantages. One approach is to study the index in a 
\emph{Cardy-like limit}~\cite{Choi:2018hmj,Honda:2019cio,ArabiArdehali:2019tdm,Kim:2019yrz,Cabo-Bizet:2019osg,Amariti:2019mgp,ArabiArdehali:2019orz}.
In this approach the rank~$N$ of the gauge group can be finite, but the disadvantage is that the method 
only applies in the infinite 
 charge limit, or equivalently, to infinitely large black holes. 
The advantage is that it applies to generic superconformal theories, and the answer only depends on 
universal quantities like the conformal anomaly coefficients.
Another advantage is that we can apply it to the  
index involving two independent angular momenta (presented in~\eqref{IndexHamDef}
below). The other two approaches, which we presently discuss, calculate the index involving  
only one combination of the angular momenta  (presented in~\eqref{Indexseqt} below), although 
this is a technical limitation which may be possible to overcome.

A second approach is the \emph{Bethe-ansatz}-like formalism which does not directly use the matrix integral 
formulation of~\cite{Romelsberger:2005eg,Kinney:2005ej}, but instead rewrites the index as a different 
contour integral which can be performed by a residue calculation. 
This approach, originally designed for the 3d topologically twisted index~\cite{Benini:2015noa,Benini:2016hjo} and the dual AdS$_4$ black holes~\cite{Benini:2015eyy,Benini:2016rke}, was  developed for 4d, $\mathcal{N}=1$ theories 
in~\cite{Closset:2017bse,Benini:2018mlo}, and applied to the  problem of black hole microstate counting 
in~\cite{Benini:2018ywd} (for~$\CN=4$ SYM) and~\cite{Lezcano:2019pae,Lanir:2019abx} 
(for more general toric quiver gauge theories). 
An advantage of this approach is its regime of applicability, which is that 
the rank~$N$ of the gauge group can be large, while the charge of the states can be finite in units of~$N^2$,
which is exactly the regime of parameters of the black hole solution in supergravity. 
A practical limitation is that it relies on finding solutions to the 
associated Bethe-ansatz-like equations, which have not been systematically studied so far.
It should be said  that some families of solutions for specific theories have been found 
in~\cite{Hosseini:2016cyf,Hong:2018viz,Benini:2018ywd,ArabiArdehali:2019orz}
and, importantly, this includes a solution corresponding to a black hole. 
 Although in this approach the large-$N$ index takes the form of a sum over solutions to the Bethe-ansatz 
 equations, an interpretation of the latter as saddle-points of the integral is not clear (see~\cite{Benini:2018ywd}).

The third approach, which we use here, is a \emph{direct saddle-point analysis} of the matrix integral that was developed for~$\CN=4$ SYM in~\cite{Cabo-Bizet:2019eaf}. The integral over unitary matrices in~\cite{Romelsberger:2005eg,Kinney:2005ej} reduces,  in a completely standard manner, to an integral over the corresponding eigenvalues which live on a circle. 
The essence of the approach of~\cite{Cabo-Bizet:2019eaf} is to extend the 
range of eigenvalues of the unitary matrix from a circle to a torus, one of whose cycles is the original circle. 
This prompts us to refer to this approach as that of \emph{elliptic extension}. 
As we review below, this approach allows us to find solutions of the saddle-point equations and, further, 
it allows us to calculate the effective action at each saddle point in a straightforward manner. 
In this paper we use this idea to lay down a simple 
and systematic approach to the calculation of the large-$N$ index of~$\CN=1$ quiver theories. 
We study the basic index, which may be defined for any ${\cal N}=1$ supersymmetric field theory 
with an R-symmetry, as well as the index refined by including chemical potentials for flavor (non-R) symmetries, 
and our focus will be to extract simple universal results for generic theories.

\vskip 0.4cm

In the rest of this introductory section, we present the context of the problem and our main results. 
We consider $\mathcal{N}=1$ superconformal theories on $S^1\times S^3$. The relevant conserved charges are the angular momenta~$J_1$, $J_2$ i.e.~the 
Cartan elements of~the $SO(4)$ isometry of $S^3$, the energy~$E$ generating translations around $S^1$,  
and the $U(1)$ R-charge~$Q$.
There is a choice of supercharge~$\mathcal{Q}$ that commutes with the 
bosonic charges $J_{1,2}+\frac{Q}{2}$, and for which 
\be
\{\mathcal{Q},\overline{\mathcal{Q}}\} \= E-J_1-J_2-\tfrac{3}{2}\,Q \,.
\ee
The \emph{superconformal index}, defined as the following trace,
\be\label{IndexHamDef}
 \mathcal{I}(\s,\t;n_0)\=\text{Tr}_{\mathcal{H}}\, (-1)^F \, \rme^{-\beta \{\mathcal{Q},\overline{\mathcal{Q}}\}} \, 
 \rme^{2\pi \i (\s-n_0) (J_1+\frac{Q}{2})+2\pi\i\, \tau\, (J_2+\frac{Q}{2})} 
\ee
is independent of $\beta$, as it only gets contributions from the cohomology of~$\mathcal{Q}$, namely 
states that obey the BPS condition $E-J_1-J_2-\frac{3}{2}Q=0$. (For this reason the 
factor~$\rme^{-\beta \{\mathcal{Q},\overline{\mathcal{Q}}\}}$ is sometimes suppressed.) 
The chemical potentials $\sigma, \tau$ are allowed to take complex values, with $\Im(\s), \Im(\t) > 0$. The integer parameter~$n_0$ was introduced in~\cite{Cabo-Bizet:2018ehj, Cabo-Bizet:2019osg} 
so as to facilitate the comparison with the gravitational results.
In particular, in the Cardy-like limit~$\s, \t \to 0$ studied in \cite{Kim:2019yrz,Cabo-Bizet:2019osg}, 
it is $n_0=\pm 1$, rather than $n_0=0$, that gives the $O(N^2)$ black hole entropy. 
Since~\eqref{IndexHamDef} is only a function of the two variables $(\sigma-n_0)$ and $\tau$, 
we can reabsorb $n_0$ by a shift of $\sigma$, that is
\be
 \mathcal{I}(\s,\t;n_0) \,\equiv\,  \mathcal{I}(\s-n_0,\t;0)\,,
\ee
as long as $\sigma$ and $\tau$ are independent variables.
In this paper we study the slice $\sigma=\tau$ in the space of variables, that is we study the index
\be
 \mathcal{I}(\t,\t;n_0) \; \equiv\;  \mathcal{I}(\t-n_0,\t;0)\,.
\ee
After this identification is made, the independent variables are~$\t$ and the discrete choice of~$n_0$.
In fact only $n_0=0$ and $n_0=1$ give inequivalent choices. This is seen by making the change 
of variable $\t = \t'+ \frac{n_0}{2}$, so that
\be
 \mathcal{I}(\t'+ \tfrac{n_0}{2},\t'+ \tfrac{n_0}{2};n_0)  \=\text{Tr}_{\mathcal{H}}\, (-1)^F \, 
 \rme^{-\beta \{\mathcal{Q},\overline{\mathcal{Q}}\}} \, 
 \rme^{2\pi \i \,\t'\, \left(2J_+ +Q\right) -2\pi\i\, n_0\, J_-}\,,
\ee
where $J_\pm=\frac{1}{2}(J_1\pm J_2)$ are the Cartan generators of the two $SU(2)$ factors in $SO(4)$. 
Since~$J_-$ takes half-integer values, the choice of $n_0$ is only relevant modulo 2.
For $n_0=0$, we have the usual expression for the index with the two chemical potentials~$\s$ and~$\t$ 
identified, that is $ \mathcal{I}(\t,\t;0)$. For $n_0=1$, we obtain from \eqref{IndexHamDef}
\be\label{thermal_index}
 \mathcal{I}(\t,\t;1) \;\equiv\;  \mathcal{I}(\t-1,\t;0) \=\text{Tr}_{\mathcal{H}} \; 
 \rme^{-\beta \{\mathcal{Q},\overline{\mathcal{Q}}\}} \, 
 \rme^{2\pi\i\, \tau\, (2J_+ +Q)}\,\rme^{-\pi \i\,Q}\,,
\ee
where we used that $\rme^{-2\pi \i\, J_1}=(-1)^F$. Written in this way, the index has the form 
of a thermal partition function where $\tau$ is a chemical potential for the charge $2J_++Q$ 
and $(-1)^F$ is replaced by an insertion of $\rme^{-\pi \i\,Q}$, which can be seen as a shift in 
the R-symmetry chemical potential. This interpretation matches the dual black hole asymptotics, 
where the supercharge naturally is anti-periodic while transported around the Euclidean time 
circle~\cite{Cabo-Bizet:2018ehj}.\footnote{The discussion above also makes it clear that the 
choices $n_0=-1$ and $n_0=+1$ are related as $\mathcal{I}(\t,\t;-1) \,=\,   \mathcal{I}(\t+1,\t+1;1)\,.$}
However, in our discussion we will find it convenient to keep $n_0$ generic, and we will denote the index under study by 
\be \label{Indexseqt}
\mathcal{I}(\tau)\,=\,\mathcal{I}(\tau,\tau;n_0) 
\=\text{Tr}_{\mathcal{H}}\, (-1)^F \, \rme^{-\beta \{\mathcal{Q},\overline{\mathcal{Q}}\}} \, 
 \rme^{-2\pi \i n_0 (J_1+\frac{Q}{2})+2\pi\i\, \tau\, (2J_++Q)} \,.
\ee

As mentioned above, the trace~\eqref{IndexHamDef} can be calculated in terms of an integral 
over unitary matrices.
Writing the eigenvalues of a unitary matrix as~$\rme^{2 \pi \i u_i}$, this can be 
expressed as an integral over the gauge variables~$u_i$ running over the interval~$[0,1]$, this will be 
the starting point of our analysis.
In this paper we consider~$\CN=1$ quiver theories with~$SU(N)$
gauge group at each node of the quiver.  The integral then runs over the gauge holonomies 
of all the gauge groups, this is presented in Equation~\eqref{IntegrandIndex}. 
The main idea of~\cite{Cabo-Bizet:2019eaf} is to deform the integrand of this integral, 
without changing its value on the real line, to a complex-valued function defined on the complex $u$-plane 
that is periodic under translations by the lattice~$\IZ \t + \IZ$. In other words, the 
integrand is now well-defined on the torus~$\IC/(\IZ\t+\IZ)$. 

In the large-$N$ approximation, we expect that the matrix integral can be written as a sum over 
solutions to the saddle point equations, 
\be \label{Itsim}
\CI(\t) \; \sim \; \sum_{\g \in \text{\{saddles\}}} \exp \bigl(-S_\text{eff} (\t; \g) \bigr) \,,
\ee
where the  saddles $\gamma$ contributing to the sum are those captured by a certain contour 
that is a deformation of the original contour along the real axis.
This leads to the following questions: What is the complete set of saddles?  
What is the effective action~$S_\text{eff}$ evaluated on a generic saddle? 
What is the final contour and what saddles does it pick? 
As explained in~\cite{Cabo-Bizet:2019eaf}, it is a straightforward consequence of the double periodicity 
mentioned above that the uniform distribution of eigenvalues wrapping the torus~$\IC/(\IZ\t+\IZ)$ 
along any cycle is a saddle-point configuration of the extended integral. Thus we obtain the infinite family 
of saddles labelled by integers~$(m,n)$ corresponding to the cycle wrapped by the string of eigenvalues. 
The extended integrand itself is governed by a certain special function---the Bloch-Wigner elliptic 
dilogarithm---which makes the calculation of the action of the saddle points also quite simple. 
Despite the fact that the extended integrand is not a meromorphic function, for each $(m,n)$ saddle 
one can show that the original contour can be deformed so as to pass through it, at leading order 
in the large-$N$ expansion.
These remarks answer the first two of the three questions raised above, and we leave the third question for future work. 
Relatedly, note that although the equation~\eqref{Itsim} gives the complete perturbative 
expansion around each saddle~$\gamma$ at large $N$, we do not yet have the exact non-perturbative 
answer---which would involve making sense of the infinite sum for every value of~$\tau$.

\vskip 0.4cm

With this background and context, we can now describe the main results of this paper.
\begin{enumerate}
\item \emph{Large-$N$ value of the index} 

The leading large-$N$ effective action of the~$(m,n)$ saddles described above has a simple expression in terms of 
the third Bernoulli polynomial 
\be
 \Seff(m,n)   \= \frac{\pi\i N^2}{3m(m\tau+n)^2} \sum_{\a \in \{\text{multiplets}\}} 
\!\! B_3 \bigl(z_{\a} + (m\t + n) (r_{\a}-1) \bigr)  \,,
\ee
up to a purely imaginary, $\tau$-independent term that we will discuss later. Here the sum over~$\a$ runs over all the~$\CN=1$ multiplets of the theory with~R-charge~$r_\a$, and 
\be
z_{\a}  \= \left\{ -(n_0 m+2n) \frac{r_{\a}}{2}  \right\}\,,
\ee
with~$\{x\}=x-\lfloor x\rfloor$ being the fractional part of the real number~$x$.
The precise expression is given in Equation~\eqref{Sintermediate1}.

\item\emph{Universal gravitational phases} 

Among all the~$(m,n)$ saddles, the saddles having~$n_0 m +2n = 0$ or $\pm 1$ 
are special in that their effective action 
is completely controlled by the R-symmetry anomaly coefficients of the SCFT. 
(See Equations~\eqref{Seff_family_with_0}, \eqref{SeffAnomalies} for the full expressions.) 
In particular, for $n_0=\pm1$ the action of the saddle $(m,n)=(1,0)$  corresponds precisely to the regularized on-shell action of the 
supersymmetric black hole in AdS$_5$~\cite{Cabo-Bizet:2018ehj}. 

The other solutions in this family also have action proportional to~$N^2$. 
Since, in addition, they only depend on the R-anomaly coefficients, we expect that
they should have a universal description as gravitational solutions of the five-dimensional gauged supergravity. 
Independent of its gravitational interpretation, we can use the expression~\eqref{Itsim} to calculate 
the phase diagram of the theory at large~$N$ within this class of solutions. 
At any given value of~$\t$, the dominant phase is the solution~$(m,n)$ which minimizes the real part of the action.
We discuss this in Section~\ref{sec:phases}.

\item \emph{Universal form of the action for flavored index} 

We then study the index refined by adding chemical potentials dual to arbitrary abelian flavor symmetries. This refined index is defined in Equation~\eqref{our_index_alternative}. The~$(m,n)$ 
configurations are also saddle-points for this index. The effective action as a function of the chemical potentials 
is cubic. We find that, remarkably, for a particular set of chambers in the space of chemical potentials, 
the effective action is, once again, completely controlled by the R-symmetry and flavor symmetry 
anomalies of the theory. 

In these chambers, the main term in the action is of order~$N^2$ and is controlled by the anomaly 
coefficient~$C_{IJK} = {\rm Tr} (Q_IQ_JQ_K)$, where~$Q_I$ are certain combinations of the flavor 
and R-symmetry charges defined in Section~\ref{sec:democratic_basis}. The precise expressions are 
given in Equations~\eqref{def_CIJK}--\eqref{actionCIJK_Deltaplus1}. 
The effective action of the flavored index has been calculated in many examples in the literature.
In particular, the papers~\cite{{Lezcano:2019pae},Lanir:2019abx} 
discuss various examples using the Bethe-ansatz method. 
In each case the general expression we present in this paper agrees with the corresponding 
expression in the literature.

\item \emph{General saddle-point configurations} 

The~$(m,n)$ saddles above describe a family of saddles that can be thought of as a ``string" of~$N$ eigenvalues 
winding around the torus. 
One can ask whether there are other possible shapes that the eigenvalues can take. 
We find a rich class of solutions to the saddle-point equations which can be 
described  as follows. Consider all possible finite abelian groups of order~$N$, 
the simplest such group is~$\IZ/N\IZ$ but there can be more general groups depending on the 
prime factors of~$N$, see Equation~\eqref{Gprimefacs}.
We find that every group homomorphism of a finite abelian group into the torus~$\IC/(\IZ\t + \IZ)$ (considered as an 
abelian group) leads to a solution of the saddle-point equations. 
The class of solutions that we find includes string-like
solutions carrying~$\IZ/N\IZ$ structure that have been discussed in the literature using the Bethe-ansatz 
method~\cite{Hosseini:2016cyf,Hong:2018viz,Benini:2018ywd,ArabiArdehali:2019orz}.
The details are presented in Section~\ref{sec:familysaddles}. 

\end{enumerate}

\vspace{0.2cm}

The plan of the rest of the paper is as follows.
In Section~\ref{sec:Saddles} we discuss the details of the elliptic extension approach and find 
the~$(m,n)$ string-like solutions of the  large $N$ saddle-point equations for a very general 
class of~$\CN=1$ superconformal quiver theories. 
In Section~\ref{sec:largeN_action_unflavoured} we calculate the action of these saddles 
and discuss universal solutions and the corresponding phase structure of the SCFT.
In Section~\ref{sec:saddles_flavoured_index} we introduce flavor chemical potentials and 
discuss the universal family of saddles with this refinement. 
In Section~\ref{sec:Action_Anomalies} we show that, in specific domains in the space of 
chemical potentials, the large-$N$ action of the $(m,n)$ saddles takes a universal form 
controlled by anomalies. 
In Section~\ref{sec:familysaddles} we find and discuss a large family of saddle-points which 
are classified by finite abelian groups. In Section~\ref{sec:outlook} we ouline some directions of future work.
In the appendices we present various technical details that are used at multiple 
points in the paper.

\section{Large-$N$ saddles of quiver theories \label{sec:Saddles}}

In this section we present the superconformal index for a very general class  
of four-dimensional quiver gauge theories, containing matter fields in bi-fundamental or adjoint representations.
In the first subsection we rewrite the index specialized to the case~$\s=\t$ in terms of a doubly-periodic 
non-holomorphic function associated to the torus~$\IC/(\IZ\t+\IZ)$. 
In Subsections~\ref{sec:saddlescont}, \ref{sec:saddlesdisc} we solve the saddle-point equations 
for the model in the continuum and the discrete formalisms, respectively, and show that 
a string of eigenvalues winding~$(m,n)$ times (with~$\text{gcd}(m,n)=1$) around the two cycles of the torus solves the 
saddle-point equations. In Subsection~\ref{sec:contour}, we show that the contour of the original
matrix integral can be deformed so as to pass through the~$(m,n)$ saddles, so that they contribute to the action.

\vspace{0.4cm}

We consider a $\mathcal{N}=1$ gauge theory defined by a quiver diagram with $\n$ nodes labelled by 
the index~$a=1,\ldots,\n$. Each node~$a$ is associated with the gauge group~$SU(N_a)$, so that 
the gauge group of the theory is~$G = \prod_{a=1}^\nu SU(N_a)$.
 The matter multiplets are described by arrows connecting pairs of nodes~$(a,b)$. 
Each arrow represents a chiral superfield, transforming in the bi-fundamental representation $({\bf N_a},{\bf \overline{N}_b})$ of~$SU(N_a) \times SU(N_b)$, and having R-charge $r_{ab}$. This notation includes chiral superfields transforming in the adjoint representation of a gauge group factor $SU(N_a)$, with R-charges $r_{aa}$.

For a quiver gauge theory of this type, the index~\eqref{IndexHamDef} can be represented 
as an integral over~$\n$ unitary matrices~\cite{Romelsberger:2005eg, Kinney:2005ej, Dolan:2008qi} which are interpreted as the holonomies of the gauge field factors around the~$S^1$. 
After integrating over angular variables, this matrix integral reduces to an integral over the eigenvalues of the matrices. 
Writing the eigenvalues of the unitary matrices as~$\rme^{2 \pi \i u_i^a}$ in terms of the \emph{gauge variables}~$u^a_i \in \IR/\IZ$,  the matrix integral can be written
as an integral over these gauge variables, one for each~$i$-th direction in the~$a$-th Cartan torus, 
with a certain measure factor. We use the notation~$\underline{u}$ to denote the set of all gauge variables $u^a_i$, $i=1,\dots, N_a$, $a=1,\dots, \n$. 
The precise form of the superconformal index~\eqref{Indexseqt} is as follows
(with~$q=\rme^{2 \pi \i \t}$, ${\rm Im}\,\tau >0$), 
\be\label{IntegrandIndex}
\begin{split}
& \mathcal{I} (\t;n_0)   \=  (q;q)^{2\sum_{a=1}^\nu N_a } \,
\int [D\underline{u}]\prod_{a=1}^{\nu} \prod_{i,j=1 \atop i\neq j}^{N_a}\Ge\bigl(u^a_{ij} + 2\t;\t,\t\bigr)\\ 
& \qquad \qquad \qquad \qquad \qquad \qquad\qquad \times \prod_{a\to b} \prod_{i=1}^{N_a} \, \prod_{j=1}^{N_b} 
\Ge\bigl(u^{a b}_{i j} + \frac{r_{\llb}}{2}(2\t-n_0);\t,\t\bigr) \,, \qquad  
\end{split}
\ee
with~$u^{a b}_{ij}=u^a_i-u^b_j$, $u^a_{ij}=u^a_i-u^a_j$. 
Here, the first line includes the vector multiplet contribution while the second line is the chiral multiplet contribution. The symbol $\prod_{a=1}^{\nu}$ denotes a product over the different gauge factors $SU(N_a)$, while $\prod_{a\to b}$ denotes the product over all chiral superfield contributions (namely, the contributions associated with arrows in the quiver diagram that start from any node $a$ and reach any node $b$). Again, allowing the head and tail of the arrows to be identified,
this notation incorporates the contributions of  chiral superfields transforming in the adjoint representation of a gauge group factor $SU(N_a)$, with $u^{a a}_{ij}=u^a_i-u^a_j$.
The Pochhammer symbol~$(w;q)$ and the elliptic gamma function~$\Ge$ are defined in~\eqref{defPoch}, \eqref{GammaeDef}, 
respectively.
The measure of integration is
\bea
[D\underline{u}]\=\prod_{a=1}^{\nu} \, \prod_{i=1}^{N_a} \, d u^a_i \, \frac{1}{N_a!} \, \delta\biggl( \sum_{i=1}^{N}u_i^a \biggr) \,,
\eea
and the contour of integration for each of the $u^a_i$'s is~$\IR/\IZ$ for which we can choose the representative~$[0,1)$. 
Note that effectively the vector multiplet associated with each gauge factor contributes as an adjoint chiral multiplet with R-charge 2. For the R-charge of the chiral multiplets we assume~$0< r_{ab}<2$,\footnote{This assumption ensures that there are no zeros or poles of the integrand when $u^{ab}_{ij}=0$.}
which is indeed satisfied for all the quiver theories that we consider, in particular for the superconformal quivers with a known supergravity dual.

\subsection{The elliptic form of the action for~$\CN=1$ quiver theories \label{sec:quivers}}

As mentioned in the introduction, in order to analyze the integral~\eqref{IntegrandIndex}, 
we deform the integrand so as to make it well-defined on the torus~$\IC/(\IZ\t+\IZ)$. 
The new integrand, which is doubly periodic in each eigenvalue~$u_i$, is no longer meromorphic in~$u_i$. 
Instead, the real and imaginary parts are real-analytic (except for at finite number of points in the 
fundamental domain).\footnote{We shall call such functions \emph{doubly periodic} or sometimes~\emph{elliptic}. 
This is an abuse of terminology as usually the notation elliptic is used for meromorphic functions.
Our terminology follows that of the elliptic dilogarithm~\cite{Bloch,ZagierOnBloch}---a non-meromorphic function---which 
is one of the main players in the analysis.}
In the large-$N$ approximation, one has to find configurations of eigenvalues on the torus 
which solve the variational problem.  
Due to the lack of meromorphy of the integrand, one has to study the variational problem 
in both the~$u_i^a$ and~$\overline{u}_i^a$ variables separately as the vanishing of one of these 
equations no longer guarantees the vanishing of the other.

In order to implement the deformation, we introduce two doubly-periodic functions. The first one~$P(z;\t)$,
defined in~\eqref{defP}, is closely related to the Jacobi theta function which should be reasonably familiar to most 
string theorists. This function has a long history starting from the 19th century (see \cite{Weil}), and its 
Fourier expansion along its two periods is well-known as the second Kronecker limit formula~\eqref{KroneckerLimit}. 
The second function~$Q(z;\t)$ \cite{DukeImamoglu, Pasol:2017pob} is relatively unknown in the physics literature, 
it is related to the so-called Bloch-Wigner elliptic dilogarithm~\cite{Bloch}.  
This function has been studied intensively by number theorists in the last few decades and, in particular, 
one knows the double Fourier expansion~\cite{ZagierOnBloch} which we present in~\eqref{QKron}. 

Using these building blocks, we construct the function
\be \label{defQcd}
Q_{c,d}(z) \= Q_{c,d}(z;\t) \= q^{\frac{c^3}{6}-\frac{c}{12}} \, \frac{Q(z+c\t+d)}{P(z+c\t+d)^c}\,,  \qquad c,d\in \IR \,,
\ee
This function is clearly elliptic as all its building blocks are, and it obeys the property  
\be \label{Qgamrel0}
Q_{c,d}(z) \=  \Ge(z+(c+1)\t+d;\t,\t)^{-1}  \quad \text{when~$z_2=0$} \,.
\ee 
In order to deform the integral expression for the index, one simply replaces  
each function $\Ge(z+(c+1)\t+d;\t,\t)^{-1}$ in the integrand of~\eqref{IntegrandIndex} by~$Q_{c,d}(z)$.

Following this procedure, we obtain the following expression for 
the integral~\eqref{IntegrandIndex},  
\be \label{IntSu}
\mathcal{I} (\t)  \=  \int [D\underline{u}] \, \exp\bigl( -S(\underline{u}) \bigr) \,,
\ee
where the deformed integrand, called the \emph{elliptic action} $S(\underline{u})$, is defined as
\be\label{EffAction}
S(\underline{u})  \= -2 \sum_{a=1}^{\nu}N_a  \log\bigl(q^{-\frac{1}{24}}\eta(\tau)\bigr) 
+ \sum_{a=1}^{\nu} \sum_{i,j=1 \atop i\neq j}^{N_a} V(u^a_{ij})  
+ \sllb \sum_{i=1}^{N_a}\sum_{j=1}^{N_b}  \, V_{ab}(u^{ab}_{ij}) \,,
\ee
where $\sum_{a\to b }$ denotes the sum over all chiral superfield contributions.
The ``potential'' functions describing the interaction between the gauge variables are given by
\begin{align}\label{DefVa}
V(z) &\= \log \Qabz{1}{0}{z}  \=  \frac{\pi\i \t}{6} + \log Q(z+ \t) - \log P(z+ \t)  \,,\nn\\[1mm]
V_{ab}(z) &\= \log \Qabz{c_{\llb}}{d_{\llb}}{z}\nn\\[1mm]
& \=   \frac{\pi\i \t}{6} \left(2c_{\llb}^3- c_{\llb}\right) +\log Q(z+c_{\llb}\, \t+d_{\llb})
- c_{\llb} \log P(z+c_{\llb}\, \t+d_{\llb})\,.
\end{align}
Here the doubly-periodic functions $Q_{(c,d)}$ are defined as above, and 
\be\label{cab}
c_{\llb} \= r_{\llb}-1 \,, \quad \quad \quad d_{\llb} \= - n_0\,\frac{r_{\llb}}{2} \,.
\ee
We note that the functions $P$ and $Q$ are invariant under the shift $d\mapsto d+1$, so that $Q_{c,d+1}(z)=Q_{c,d}(z)$.
Thus,~$\sum_{i,j}V(u^a_{ij})$ describes the contribution  of the  gauge superfield at node $a$ to the action~$S(\uu)$, 
while~$\sum_{i}\sum_jV_{ab}(u^{ab}_{ij})$ is the contribution of a chiral superfield associated to an arrow going 
from node~$a$ to node~$b$. For definiteness, we set~$V_{ab}=0$ if there is no arrow going from~$a$ 
to~$b$ in the quiver diagram. 

\subsection{The saddle-point equations and $(m,n)$ solutions in the continuum limit \label{sec:saddlescont}}

In this subsection we find stationary points for 
the class of~$\mathcal{N}=1$ superconformal quiver theories that we considered above.
We begin with the action~\eqref{EffAction} rewritten slightly:  
\begin{align}\label{LargeNActionLagrange}
S(\underline{u})\,&\=\,S_0 \+ \sum_{a=1}^{\nu} \sum_{i, j=1 \atop i\neq j}^{N_a} V \bigl(u^a_i-u^a_j \bigr)
\+ \sllb \sum_{i=1}^{N_a}\sum_{j=1}^{N_b} V_{ab} \bigl(u^a_i-u^b_j \bigr)\nn\\
\,&\, \qquad - \sum_{a=1}^{\nu} N_a \, \biggl(\lambda^a \,\sum_{i=1}^{N_a} u^a_i 
\+ \wt \lambda^a \,\sum_{i=1}^{N_a} \overline{ u^a_i }\biggr) \,.
\end{align}
Here the functions $V(z)$ and $V_{ab}(z)$ are doubly periodic complex-valued functions as 
discussed above, and~$S_0$ is independent of~$\uu, \bar \uu$.\footnote{We use the 
notation $\overline{z}=z^*$ for the complex conjugate.} The function $V$ encodes the contribution 
from the vector multiplets, while $V_{ab}$ describes the contribution of the chiral multiplets going 
from node $a$ to node $b$, and having R-charge $r_{\llb}$.
Since the action is not meromorphic\footnote{This is sometimes denoted by having the complex conjugate of the argument
as an additional variable of the function~$V(z,\bar z)$, here use the notation~$V(z)$ and think of it 
as a non-holomorphic function of~$z$.}, 
we have to solve the saddle point equations for~$u^a_i$ 
and~$\overline{u^a_i}$ separately. 
The Lagrange multipliers $\lambda^a$, $\wt \lambda^a$, implement the $SU(N_a)$ constraints 
on the full complexified gauge holonomies, i.e.,
\be \label{sunconstraint}
\sum_{i=1}^{N_a} \,u^a_i\=0\,, \qquad \sum_{i=1}^{N_a}\, \overline{u^a_i}\=0\,.
\ee 
Note that $\lambda^a$, $\wt \lambda^a$ are a priori independent variables as we are allowing for 
complex saddles and the extended action is not meromorphic. 
Here we have defined the Lagrange multipliers with a factor of~$N_a$ in anticipation of the fact, 
that we will see below, that the value of~$\lambda_a$ is~$O(1)$.
In principle we could define a large-$N$ limit by keeping the different values of~$N_a$ distinct and taking all of them 
large in some specified way. However for simplicity we will assume~$N_a=N$ for all~$a$, and then take~$N$ large. 
We note that there is no obstruction to carrying this analysis in the general case. 
The superconformal quivers that we are mainly interested in do satisfy this condition.  

\vspace{0.2cm}

In the large-$N$ limit, it is convenient to pass to the continuum formulation by using the following identifications 
at each node
\bea
\frac{i}{N} \mapsto x \,, \qquad   \frac{1}{N}\mapsto dx \,,\qquad 
u^{a}_i\mapsto u^a(x) \,, \qquad \sum_{i=1}^{N} \mapsto N \int_0^1 dx\,,
\eea
where $x \in [0,1)$.
In this way the action \eqref{LargeNActionLagrange} becomes the functional
\begin{align}\label{continuumaction}
\frac{1}{N^2} \, S[u] &\= \sum_{a=1}^{\nu} \,\int_0^1 dx \int_0^1 dy\, V \bigl(u^a(x)-u^a(y) \bigr)
+ \sllb\,\int_0^1  dx \int_0^1 dy\, V_{ab} \bigl(u^a(x)- u^b(y) \bigr)\nn\\ 
& \qquad -  \sum_{a=1}^\nu \biggl( \lambda^a \int_0^1 dx \, u^a(x) +  \wt \lambda^a \int_0^1 dx \, \overline{ u^a(x)} \biggr) \,.
\end{align}
Notice that we have dropped the term $S_0$, as it is subleading at large~$N$. 
On the other hand, we cannot drop the Lagrange multiplier term.

Let us discuss the extremization equations. Varying with respect to $u^a(x)$ gives
\begin{align}\label{extr_eqs_continuum}
& \int_{0}^1 d y \, \left[ \partial V(u^a(x)-u^a(y))  - \partial V(u^a(y)-u^a(x))  \right]\nn\\
& + \!\!\sum_{\mathrm{fixed}\,a\to b}\,  \int_{0}^1 d y  \, \partial V_{ab}(u^a(x)-u^b(y)) 
 - \sum_{\mathrm{fixed}\,a\leftarrow b} \, \int_{0}^1 d y \, 
\partial V_{ba}(u^b(y)- u^a(x)) 
- \lambda^a \= 0  \,,
\end{align}
where~$\partial$ denotes the holomorphic derivative with respect to the argument of the function, and the 
sums are over all chiral fields that go from the fixed node $a$ to any node $b$ (``fixed$\,a\to b$''), or that reach the same node $a$ starting from any node $b$ (``fixed$\,a\leftarrow b$'').
The equations arising from varying~$\bar u^a(x)$ have the same form as~\eqref{extr_eqs_continuum}
with the replacement~$\p V \to \bar \p V$ and similarly with~$\p V_{ab}$. 
Note that~$\bar \p V(u) \neq \overline{\p V (u)}$, and so these equations are genuinely independent equations.
Moving on, varying with respect to the Lagrange multipliers $\lambda^a$, $\wt \lambda^a$ yields the constraints
\be\label{normalization_u(x)}
\int_0^1 d x\, u^a(x) \= 0\,, \qquad  \int_0^1 d x\, \overline{ u^a(x)} \= 0\,,
\ee
meaning that the unimodularity constraint is imposed on both the real and the imaginary part of the gauge variables $u^a(x)$.

We now show that the gauge variable configuration
\be\label{saddles}
u^a(x) \= x\,T - \frac{T}{2}
\ee
is a saddle of the large-$N$ action for any period $T$ of the action~\eqref{LargeNActionLagrange}. 
These periods correspond to the points of the lattice which are labelled by two integers~$(m,n)$.
Equivalently, the uniform distribution~\eqref{saddles} from~0 to~$m\t +n$ can be 
thought of as the uniform distribution wrapping~$(m,n)$ times around the two cycles 
of the torus~$\IC/(\IZ\t+\IZ)$. 
In order to count independent configurations in the large-$N$ limit, we should consider lattice points
with the addition condition~$\text{gcd}(m,n)=1$.

Now, obviously \eqref{saddles} solves the constraint \eqref{normalization_u(x)}. We now show that it 
also satisfies the equation~\eqref{extr_eqs_continuum}. 
Plugging \eqref{saddles} in  \eqref{extr_eqs_continuum}, we obtain
\be
\int_{0}^1 d y  \Big[ \partial V(T(x-y))  - \partial V(T(y-x))\ \ +\!\! \sum_{\mathrm{fixed}\,a\to b}\!\!  \partial V_{ab}( T(x-y))\ \ 
-\!\! \sum_{\mathrm{fixed}\,a\leftarrow b}  \!\!  \partial V_{ba}(T(y-x)) \Big] \= \lambda^a \,.
\ee
This equation is of the form
\be
 \int_0^1 dy \, f_a(y-x) \= \lambda^a\,,
\ee
where the integrand~$f_a$ is periodic under the shift of the real variable~$y\to y+1$. 
Since we are integrating over the full period, the result of the integral is simply
\be
 \int_0^1 dy \, f_a(y) \= \lambda^a\,,
\ee
which does not depend on $x$.
Thus we obtain the value of the Lagrange multiplier $\lambda^a$ to be
\begin{align}\label{expression_Lagrange_mult}
\lambda^a &\= \int_{0}^1 d y \, \Big[ \partial V(-Ty)  - \partial V(Ty)  \  +\!\! \sum_{\mathrm{fixed}\,a\to b} \!\! \partial V_{ab}( -Ty)\ -\!\! \sum_{\mathrm{fixed}\,a\leftarrow b}\!\! \partial V_{ba}(Ty)   \Big] \nn\\
&\= \int_{0}^1 d y \, \Big[\, \sum_{\mathrm{fixed}\,a\to b}  \partial V_{ab}( Ty)\ - 
\sum_{\mathrm{fixed}\,a\leftarrow b} \partial V_{ba}(Ty) \,  \Big] \,.
\end{align}
Here, to reach the second line we use the fact that the integral of $\partial V(-Ty)$ 
equals the integral of $\partial V(Ty)$ because of periodicity. (For a periodic function~$f$ with period~1,
we have~$\int_0^1 f(y) dy = \int_0^1 f(y-1) dy$, which is equal to~$\int_0^1 f(-y') dy'$ by the change of 
variable~$y'=1-y$.) 
Similarly the integral of $\partial V_{ab}( -Ty)$ equals the integral of $\partial V_{ab}( Ty)$ for the same reason.
Note that $\lambda^a=O(1)$, as anticipated. 
The equations arising from varying with respect to $\bar u^a$ and $\wt \lambda^a$ are solved in exactly 
the same way, with $\wt \lambda$ being determined as
\be\label{expression_Lagrange_mult_tilde}
\wt\lambda^a \= \int_{0}^1 d y \, \Big[\, \sum_{\mathrm{fixed}\,a\to b} \!\bar\partial V_{ab}( Ty)\ - 
\sum_{\mathrm{fixed}\,a\leftarrow b}\!\bar\partial V_{ba}(Ty) \,  \Big]\,,
\ee
which in general is not the complex conjugate of \eqref{expression_Lagrange_mult}.

In some special cases, one may find that the expressions \eqref{expression_Lagrange_mult}, \eqref{expression_Lagrange_mult_tilde} vanish, hence $\lambda^a=\wt \lambda^a=0$ at the extremum; this means that the extremization equations are also solved for quivers with $U(N)$ gauge groups, and not just $SU(N)$.\footnote{One should recall, however, that $U(N)$ quivers have more severe restrictions from anomaly cancellation.} For instance, this happens for non-chiral quivers, where for every arrow going from node $a$ to node $b$ leading to the potential $V_{ab}$, there is an arrow going from node $b$ to node $a$, with identical potential $V_{ba}=V_{ab}$; this implies that two terms in the last line of \eqref{expression_Lagrange_mult} cancel against each other. 
 One finds $\lambda^a=\wt \lambda^a=0$ also for chiral quivers where the R-charges of bifundamental chiral multiplets are all the same ($r_{a,b}=r$ for all $a\neq b$). In this case, $V_{ab}=V_r$ for all $a\neq b$, hence the expression for the Lagrange multiplier becomes 
\be\label{FactorGen}
\lambda^a \=  \left(    n_{{\rm out},a} - n_{{\rm in},a}  \right)\int_{0}^1 d y\, \partial V_r( Ty)\,,
\ee
where $n_{{\rm out},a}$ is the number of arrows going out of node $a$, while $n_{{\rm in},a}$ 
is the number of arrows pointing towards node $a$.
Now, cancellation of the gauge anomaly implies that at each node of the  quiver the number of 
outgoing arrows equals the number of ingoing arrows, that is
 $n_{{\rm out},a} - n_{{\rm in},a} = 0$, $a =1, \dots, \nu$,
thus showing that the saddle-point equations are solved with $\lambda^a=0$. 
The same argument leads to $\wt \lambda^a=0$. 
Examples of chiral quivers where the R-charges are all equal are provided by the~$Y^{p,p}$ 
and~$Y^{p,0}$ infinite families \cite{Benvenuti:2004dy}, the former being~$\mathbb{Z}_{2p}$ orbifolds of~$\mathcal{N}=4$ SYM, 
and the latter being~$\mathbb{Z}_{p}$ orbifolds of the conifold theory~\cite{Klebanov:1998hh}.

\subsection{The discrete case \label{sec:saddlesdisc}}

We can also offer a finite-$N$, discrete version of the continuum discussion given above, which is useful later. 
The main steps are the same, so we will be more brief.
We show that the gauge variable configuration
\be \label{discansatz1}
u^a_i \= T \Bigl( \frac{i}{N} - \frac{N+1}{2N} \Bigr) \,, \qquad i \= 1, \dots, N \,, \qquad a \= 1, \dots , \nu \,,
\ee
extremizes the finite-$N$ action~\eqref{LargeNActionLagrange}.
Clearly~\eqref{discansatz1} obeys the constraint
\be \label{suNconstraint}
\sum_{i=1}^{N} u^a_i \= 0 \,,
\ee
arising from the variation of~\eqref{LargeNActionLagrange} with respect to the Lagrange multiplier $\lambda^a$,
as consistent with the~$SU(N)$ gauge group. 
Varying with respect to $u^a_i$, we obtain the following saddle-point equations 
\be \label{saddleptdisc}
\begin{split} 
&\sum_{j=1}^{N} \Bigl(  \partial V_a(u^a_i-u^a_j) - \partial V_a(u^a_j-u^a_i) \\
&\hspace{1.4cm} +  \sum_{\mathrm{fixed}\,a\to b}\!\!  \p V_{ab} \bigl(u^a_i-u^b_j \bigr) \;
- \sum_{\mathrm{fixed}\,a\leftarrow b}\!\! \p V_{ba} \bigl( u^b_j-u^a_i \bigr)   \Bigr) \= \lambda^a\,,
\end{split}
\ee
and then plugging \eqref{discansatz1} in, yields
\begin{align} 
&\sum_{j=1}^{N} \Bigl(  \partial V_a\left(\tfrac{T}{N}(i-j)\right) - \partial V_a\left(\tfrac{T}{N}(j-i) \right)\nn\\
&\hspace{1.4cm} +  \sum_{\mathrm{fixed}\,a\to b}\!\!  \p V_{ab} \left(\tfrac{T}{N}(i-j)\right) \; 
- \sum_{\mathrm{fixed}\,a\leftarrow b}\!\! \p V_{ba} \left(\tfrac{T}{N}(j-i)\right)   \Bigr) \= \lambda^a\,.
\end{align}
We can now exploit the fact that the functions are periodic under $i\to i+N$ 
(because this sends $u^a_i \to u^a_i+T$, and all terms are $T$-periodic 
when seen as functions of $u$) together with the fact that we are summing 
over all $j=1,\ldots,N$, to argue that the left hand side does not depend on the 
value of $i$, and that we can change $-j$ into $+j$ in the first and third term. We thus arrive at 
\begin{align} 
&\sum_{j=1}^{N} \Bigl( \;  \sum_{\mathrm{fixed}\,a\to b}\!\! \p V_{ab} \left(\tfrac{T}{N}j\right) 
-  \sum_{\mathrm{fixed}\,a\leftarrow b}\!\! \p V_{ba} \left(\tfrac{T}{N}j \right) \Bigr) \= \lambda^a\,,
\end{align}
which just fixes the value of the Lagrange multiplier~$\lambda^a$. 
Again, the equations for~$\bar\lambda^a$ and~$\bar u^a_i$ are solved in an analogous manner.

Although we do not take a large-$N$ limit in solving the saddle-point equations in the discrete method, 
we note that the validity of the saddle-point approximation to the original matrix integral needs a large-$N$ limit; this gives the same result as the continuum limit described above.
Instead of using Lagrange multipliers, we can equivalently satisfy the~$SU(N_a)$ constraint by explicitly 
solving~\eqref{sunconstraint} for, say, $u^a_{N_a}$ in terms of the other eigenvalues
from the very beginning, 
and then extremizing with respect to the remaining variables. This leads us to the same final 
result as the procedure above.

\subsection{The contour deformation  \label{sec:contour}}

We have shown above that the uniform distribution of the gauge variables between~$0$ and the 
lattice point~$m\t+n$, $m,n \in \IZ$ solves the saddle-point equations of the matrix integral~\eqref{IntSu}. 
In order to show that these configurations contribute to the integral, we also need to show that the 
contour of integration passes through the saddle-point. A contour deformation argument is 
not a priori obvious because the integrand of~\eqref{IntSu} is not meromorphic. 
The discussion below is an adaption of the procedure used in~\cite{Cabo-Bizet:2019eaf}
 for~$\CN=4$ SYM to the class of~$\CN=1$ theories that we discuss in this paper.

The main point is to use the interplay between the two 
representations of the superconformal index: \eqref{IntegrandIndex} whose integrand is meromorphic,
and~\eqref{IntSu} whose integrand is doubly periodic. 
Both these integrals are defined using  the same contour in which the variables~$u^a_i$ 
go from~$0$ to~$1$ on the real axis.
Since the integrand of~\eqref{IntegrandIndex} is meromorphic, 
we can deform its contour without changing the value of the integral as long as we do not cross any poles of the 
integrand.\footnote{The residues picked up from crossing of these poles could lead to important physical phenomena. We do not pursue this interesting direction here.} 
Following this idea, we deform the contour of the meromorphic integrand to a new contour which passes 
through a given saddle, and then show that on this new contour we can replace the meromorphic integrand by the 
doubly-periodic integrand without changing the value of the integral at large~$N$.

As explained in~\cite{Cabo-Bizet:2019eaf}, the new contour~$\mathcal{C}$ 
consists of three pieces in each variable~$u^i_a$, which we denote 
as~$\mathcal{C}_\text{hor} + \mathcal{C}_\text{vert} + \mathcal{C}_\text{saddle}$. 
The piece~$\mathcal{C}_\text{hor}$ runs over a subset of the real axis, here 
Equation~\eqref{Qgamrel0} shows that the two integrands agree. The piece~$\mathcal{C}_\text{vert}$
consists of two closely placed oppositely oriented vertical lines, and the integral along this piece of 
either of the two integrands vanishes (and therefore the replacement is valid).
The third piece~$\mathcal{C}_\text{saddle}$, which is the non-trivial piece, is an infinitesimal 
horizontal strip passing through the saddle-point value of~$u^a_i$. It was shown in~\cite{Cabo-Bizet:2019eaf}
that the value of the two integrands for~$\CN=4$ SYM agree at the saddle-point value, and that consequently
one can make the replacement in an infinitesimal small neighborhood of the saddle-point to good approximation. 
One then uses the saddle-point approximation on the new contour so that the value of the integral 
is the value of the integrand at the saddle-point in the leading large-$N$ approximation.

The part of the argument that depends in a non-trivial manner on the theory under consideration is 
the agreement of the meromorphic and the doubly-periodic action when evaluated on the saddle-point. 
As we now show, this holds generically for the~$\CN=1$ theories discussed here.  
We recall, from the discussion in the previous subsections (in particular, see Equation~\eqref{discansatz1}), 
that the~$(m,n)$ saddle point of the action~\eqref{IntSu}  is described by the following gauge variable 
configuration
\be \label{discansatz2}
u^a_i \= (m\t+n) \frac{i}{N} + u_0 \; \equiv \; u_i \,,
\qquad  \qquad i \= 1, \dots, N \,, \qquad a \= 1, \dots , \nu \,,
\ee
with the value of the constant~$u_0$ chosen so as to obey the~$SU(N)$ constraint.

We start by recalling a relation,  that involves the doubly periodic functions~$P$ and~$Q$, 
and the elliptic Gamma function~$\Ge$~\cite{DukeImamoglu, Pasol:2017pob}\cite{Cabo-Bizet:2019eaf},
\be \label{defQtl}
 Q(z;\t) \=  e^{2 \pi \i\alpha_{Q}(z_1,z_2)} q^{\frac13 B_3(z_2) - \frac12 z_2 B_2(z_2)} \, \frac{P(z;\t)^{z_2} }{\Ge(z+\t;\t,\t)} \,,
\ee
where the function $\alpha_{Q}$ is a real function of $z_1$ and $z_2$ which is not doubly periodic.
The function~$\a_Q$ can be written as a sum of an explicit non-periodic function and a 
doubly-periodic function\footnote{We recall that the Fourier expansions of the doubly periodic functions~$P$ and~$Q$ 
defined in Equations~\eqref{logpi} and~\eqref{QKron}, have implicit ambiguities that we parameterize 
by two real and doubly periodic functions~$\wt \Psi_P$ and~$\wt \Psi_Q$, respectively. The function~$\Psi_Q$
is determined by these two functions. To be concrete, for the purpose of this discussion 
we fix~$\wt \Psi_P=0$. In this case~$\Psi_Q$ is determined by~$\wt \Psi_Q$.}~$\wt \Psi_{Q}$ 
(to be determined below), as follows~\cite{Cabo-Bizet:2019eaf},
\be\label{avalue}
\a_{Q} \= -\frac{1}{4} (1\,+\,2\, \{z_1\}) \lfloor z_2\rfloor \,(1\,+\,\lfloor z_2\rfloor)\,+\, \frac{1}{2} \Psi_Q(z_1,z_2) \,.
\ee
The function~$\a_Q-\frac{1}{2} \Psi_Q$ is piecewise continuous and it vanishes 
in the region~$-1\leq z_2<1$. 
Upon substitution of the function $Q$ as given in Equation \eqref{defQtl}, in the definition of 
the function $Q_{c,d}$ in terms of $Q$ and $P$, as given in Equation \eqref{defQcd}, it follows that 
\be \label{QabPgamrel}
Q_{c,d}(z) \=  e^{2\pi \i\alpha_Q(z_1+d,\,z_2+c)} \, q^{-A_{c}(z_2)} \,
\frac{P(z+(c+1)\t+d;\t)^{z_2}}{\Ge(z+(c+1)\t+d;\t,\t)}  \,,
\ee
the cubic polynomial~$A_c$ is
\be \label{cubicA}
A_{c}(x) \=  \tfrac16  \, x^3  \, +\, \tfrac12 c\, x^2\,+\,  \tfrac{1}{2} c^2 \, x - \tfrac{1}{12} \, x\,.
\ee

The doubly-periodic action~\eqref{EffAction},~\eqref{DefVa} is a linear combination of the functions~$Q_{c,d}$,
evaluated on the gauge variables. Each one of the summands in~\eqref{EffAction} corresponds to a 
specific multiplet.   
We show below that after summing  over all the gauge variables in the ansatz~\eqref{discansatz2} and over all the matter multiplets,  the contributions coming from the polynomial~$A_c(z_2)$, and the 
function~$z_2 \log P(z+(c+1)\t+d;\t)$ vanish. 
Thus we reach the conclusion that the absolute value of the integrands of~\eqref{IntegrandIndex}
and~\eqref{IntSu} are equal on the $(m,n)$ saddle point configurations.
Next we choose the phase~$\Psi_{Q}$  such that the phases of the doubly periodic and meromorphic integrands are also 
equal when evaluated on the $(m,n)$ saddles.\footnote{ Here a question arises as to whether this
prescription for~$\Psi_{Q}$ is well-defined. 
In particular, it could happen that a certain point~$z$ on the torus lies on the string of eigenvalues 
for two different saddles~$(m,n)$ and~$(m',n')$. The point~$z$ would correspondingly lift to 
two different points in the complex plane which differ by a lattice translation. 
The question then is whether the value of the phase of~$\Ge$ and in particular the value of~$\a_{Q}$ 
agrees at these two points. 
This is a subtle question whose complete analysis will be posted elsewhere. For our 
purposes here, we restrict our analysis to a set of saddles with an upper cutoff on~$m$.
In this situation if we take the first term in the right-hand side of~\eqref{avalue}, the difference of evaluating this between two points
differing by a lattice translation, is a rational number with a bounded
denominator. We can then lift our discussion to a larger torus (which is still finite) 
on which~$\Psi_Q$ is well-defined. We note that all the calculations of the action are done by considering 
configurations of gauge variables that are extended on the complex plane (not just restricted to the fundamental domain), so that they are not affected by this cutoff.
}

First we analyze the contribution that comes from the cubic polynomial~$A_c$ given in~\eqref{cubicA}.
The integrand in question involves a product over all supermultiplets in the theory, that here we label by an index $\alpha$ (this includes the vector multiplet). Each factor contributes with a corresponding polynomial~$A_{c_\alpha}$. Let~$\rho^{(a)}_\alpha$ 
denote the weights of the representation~R$^{(a)}_\alpha$  that the supermultiplet~$\alpha$ carries under the gauge group at the~$a$-th node of the quiver.  After summing over all the  
weights~$\rho^{(a)}_\alpha$ and then over all the supermultiplets, 
the contributions coming from the four terms on the right-hand side of~\eqref{cubicA} can be organized in 
linear combinations of the following four expressions,
\be\label{AnomalyCanc}\begin{split}
\Bigl(\sum_{\alpha}\sum_{\rho}\rho_\alpha^{(a)i}\rho_\alpha^{(b)j} \rho_\alpha^{(c)k}\Bigr) \,\,u^a_{2i}\, u^b_{2j}\, u^c_{2k}\,,\\ 
\Bigl(\sum_{\alpha}\sum_{\rho}(r_\alpha-1)\,\rho_\alpha^{(a)i} \rho_\alpha^{(b)j}\Bigr)\,  \,u^a_{2i}\,u^b_{2j}\,, \\
\Bigl(\sum_{\alpha}\sum_{\rho} (r_\alpha-1)^2\,\rho_\alpha^{(a)i}\Bigr) \,\, u^a_{2i}\,,\\
\Bigl(\sum_{\alpha}\sum_{\rho} \rho_\alpha^{(a)i} \Bigr)\,\, u^a_{2i}\,.
\end{split}
\ee
 Here the indices~$i$,~$j$ and~$k$ are summed over all possible values, 
while the indices~$a$,~$b$~and~$c$~labeling the nodes of the quiver are kept fixed. 
 Finally,  $(r_\alpha-1)$ is the R-charge of the fermion field in the multiplet $\alpha$ (we formally assign $r_\alpha=2$ to the vector multiplet, so that the gaugino has the correct R-charge 1). 
The sum over $\rho$ means that one needs to sum over all the weights $\rho_\alpha^{(a)}$ that belong to the 
representation R$^{(a)}_\alpha$. 
The $u$-independent terms in \eqref{AnomalyCanc} are the Gauge-Gauge-Gauge, R-Gauge-Gauge, R-R-Gauge and mixed Gauge-gravitational anomaly coefficients, respectively, for the Cartan generators of the gauge group. These vanish in anomaly-free theories that have an R-symmetry  conserved at the quantum level, as we assume here.
 Thus we conclude that the contribution of the cubic polynomial~$A_{c}$ to the integrand 
vanishes.

Then we move to the function~$z_2 \log P$. 
The contribution to the action of the function~$P$ associated to a given multiplet can be 
written as the exponential of
\be \label{SSdiff}
\begin{split}
\sum_{i,j=1}^N &\, (u_{ij})_2 \,  \Bigl(\log P \bigl(u_{ij}+ (c+1) \t+ b \bigr) \Bigr)\,.
\end{split}
\ee
We can evaluate this expression on the saddle point~$u_i=\frac{i}{N} (m\t+n)+u_0$ 
using the double Fourier expansion~\eqref{logpi} for the function~$\log P$.
In this manner we obtain a sum over the integers~$\wt n, \wt m$ of terms that are proportional to
\be \label{ijsumzero}
 \sum_{i,j=1}^N \, (i-j) \, \e \bigl(\tfrac{i-j}{N}(\wt n m - \wt m n) \bigr) \,,
\ee
where we are using the notation $\e(x) = \rme^{2\pi \i x}$.
These terms can be proven to vanish as follows. Let us define~$k=\wt n m - \wt m n$ then
\be \label{Identity1}
\begin{split}
 \sum_{i,j=1}^N \, (i-j) \, \e \bigl(\tfrac{i-j}{N} k \bigr)  
& \= \sum_{i=1}^N \, i \, \e \bigl(\tfrac{i}{N} k \bigr)\sum_{j=1}^N  \, \e \bigl(\tfrac{-j}{N} k \bigr) \; - \;  
 \sum_{j=1}^N \, j \, \e \bigl(\tfrac{-j}{N} k \bigr)\sum_{i=1}^N  \, \e \bigl(\tfrac{i}{N} k \bigr)  \\
&  \= \sum_{i=1}^N \, i \, \e \bigl(\tfrac{i}{N} k \bigr) \, \delta_{k,0} \; - \;  
 \sum_{j=1}^N \, j \, \e \bigl(\tfrac{-j}{N} k \bigr) \, \delta_{k,0} \\
& \=  \delta_{k,0} \Bigl( \, \sum_{i=1}^N \, i \; - \; \sum_{j=1}^N \, j  \Bigr) \= 0 \,.
\end{split}
\ee

Let us recapitulate the procedure that we followed. 
We begin with the meromorphic integral~\eqref{IntegrandIndex} whose contour can be 
deformed freely up to potential residues.
Then we argue that there exists a contour which passes through the $(m,n)$ configuration 
such that the value of the meromorphic integral~\eqref{IntegrandIndex} equals the value 
of the doubly-periodic integral~\eqref{IntSu} along the contour. 
Since we have already checked that the~$(m,n)$ configurations solve the saddle-point equations of 
the doubly-periodic action separately for the real and imaginary parts,
we use the doubly-periodic action to implement the saddle-point approximation. 
This leads to the conclusion that the integral on that contour is dominated by the 
value of the integrand in the vicinity of the saddle.
We stress that a rigorous global analysis remains to be done.\footnote{For meromorphic integrands,
this can be done using the formalism of Picard-Lefschetz theory \cite{Witten:2010cx,Aniceto:2018bis}.}
Such an analysis is outside the scope of this paper. In Section~\ref{sec:phases} we perform a naive analysis of relative dominance of the saddles.

\section{The effective action of the $(m,n)$ saddle \label{sec:largeN_action_unflavoured}}

In this section we compute the action of the large-$N$ saddles \eqref{saddles} with period~$T=m\tau+n$. 
The action that has the least real part will dominate and thus provide our estimate for the index \eqref{IntSu} 
in the grand-canonical ensemble, wherein the angular chemical potential $\tau$ is the independent variable.

\subsection{Evaluation of the action}

Upon evaluating the continuum action \eqref{continuumaction} on the configurations \eqref{saddles}, one obtains the large-$N$ effective action
\be\label{effAction}
\begin{split}
S_\text{eff}({m},{n};\t) & \= \nu N^2\! \int^1_0 \!dx \int^1_0 \!dy \, V\bigl(T (x - y) \bigr)  + N^2\! \sllb \, \int^1_0 \!dx  \int^1_0 \!dy \,V_{ab}\bigl(T(x -y)\bigr) \,,
\end{split}
\ee
which depends on the  complex parameter $\tau$ as well as on the integers $m,n$ that  appear in $T=m\tau+n$.
We can reduce each double integral to a single integral as follows,
\be
\begin{split}
 \int^1_0 dy \int^1_0 dx \, V\bigl(T (x - y) \bigr) \= \int^1_0 dy \int^1_0 dx \, V\bigl(T x  \bigr)
 \=  \int^1_0 dx \, \bigl(T x  \bigr) \,,
\end{split}
\ee
where we have used periodicity of the potential in establishing the first equality. Recalling the definitions \eqref{DefVa}, we obtain
\be\label{SeffGeneralForm}
 S_\text{eff}({m},{n};\t) = \nu N^2\! \int^1_0 \!dx  \log Q_{1,0} ((m\tau+n) x )  \,+\, 
N^2 \sllb  \int^1_0 \!dx   \log \Qabz{c_{\llb}}{d_{\llb}}{(m\tau+n) x } \,,
\ee
where each function $Q_{c,d}$ denotes the contribution of a chiral multiplet, 
and $Q_{1,0}$ is the contribution of the $SU(N)$ vector multiplet. Evaluating these  integrals using formulae provided in Appendix~\ref{App:defs_and_identities} we reach our final expression for the large-$N$ action, to be presented below. 
One can see that the result does not depend on any common divisor of $m$ and $n$. Also, notice from \eqref{effAction} that $S_\text{eff}(-{m},-{n};\t)=S_\text{eff}({m},{n};\t)$, since a change of sign $T \to -T$ just amounts to swapping the integration variables. Hence without loss of generality from now on we assume that {\it $m$ and $n$ are relatively prime, with $m\ge 0$}.

\paragraph{The saddle $\bm{m=0, n \neq 0}$.} We first discuss the special case $m=0$, $n\neq0$, where the gauge variables take {\it real} values $u^a(x)= n\left(x-\frac{1}{2}\right)$. In the large-$N$ limit, the eigenvalue distribution of 
all these saddles on the torus~$\IC/(\IZ\t+\IZ)$ are equivalent. 
Recalling that we assume that the R-charges of all chiral 
multiplets satisfy $0< r_{\llb}< 2$, the identity  
\eqref{identities_m=0_final} implies that the real part of the action vanishes at order $O(N^2)$. 
This saddle in the form~$(m,n)=(0,1)$ corresponds to the saddle discussed in~\cite{Kinney:2005ej}. 
Indeed, in the saddle of~\cite{Kinney:2005ej} 
the gauge variables---which are assumed to be real---take the uniform density~$\rho(u)\equiv \frac{\dd x}{\dd u} =1$
which corresponds to $u(x)= x + {\rm constant}$, and the corresponding action is independent of $N$ at leading order. 

\medskip

From now on we take $m>0$. Evaluating the doubly-periodic potentials \eqref{DefVa} at the saddles and using Identities \eqref{Id11}, \eqref{Id12}, we find that
the effective action \eqref{effAction} 
can be expressed in terms of Bernoulli polynomials 
\begin{align}\label{Bernoulli_main_text}
 B_2(z) &\= z\left(z-1\right)+\tfrac{1}{6}\,,\nn\\
 B_3(z) &\= z \left(z-\tfrac{1}{2}\right)\left(z-1\right)\,,
\end{align}
 depending on the variable
\be\label{zab_variable}
z_{\llb} \= \{m d_{\llb}-n c_{\llb} \} \= \left\{ -(n_0 m+2n) \frac{r_{\llb}}{2}  \right\}\,,
\ee
where in the second equality we used \eqref{cab}, and for any real $x$ we define the {\it fractional part} $\{x\}= x-\lfloor x\rfloor $.\footnote{We note that $B_2(\{x\} ) -\frac{1}{6}= -\{x\}(1-\{x\})\equiv -\vartheta(x)$ and \hbox{$B_3(\{x\}) = \frac{1}{2}\{x\} \left(1-\{x\}\right)\left(1-2\{x\}\right)\equiv\frac{1}{2}\kappa(x)$,} where $\vartheta(x)$ and $\kappa(x)$ are the functions used {\it e.g.}~in \cite{Ardehali:2015bla,Honda:2019cio,ArabiArdehali:2019tdm,Cabo-Bizet:2019osg}. 
 Some more details on Bernoulli polynomials are given in Appendix~\ref{App:defs_and_identities}.}
The action then reads
\begin{align}\label{Seffmn}
\Seff(m,n;\t)  
&\= \frac{\pi\i\tau}{6}\left(2{\rm Tr}R^3 - {\rm Tr}R\right)+  \frac{\pi\i}{ m (m\t+n)}\bigg[  \frac{{\rm Tr}R}{6}+ N^2 \sllb (r_{\llb}-1) 
\big(B_2 (z_{\llb}) -\tfrac{1}{6}\big) \bigg] \nn\\ 
&\qquad  +  \frac{\pi\i N^2}{3 m (m \tau+n)^2}  \sllb \, B_3 (z_{\llb})  \, +\,  \pi\i N^2 \Phi  \,,
\end{align}
where~$\Phi$ is a real~$\tau$-independent function that we discuss below. Before that we note that 
the expression~\eqref{Seffmn} can be rewritten in a compact form by using the 
following identity involving Bernoulli polynomials,
\be
\label{B3expanded}
B_3(x+y)  \=  B_3 (x) + 3 B_2 (x) y + 3 B_1(x) y^2 + y^3 \,, \qquad x, y \in \IC \,.
\ee
Applying this to the right-hand side of~\eqref{Seffmn} we obtain
\be\label{Sintermediate1}
\begin{split}
\Seff(m,n;\tau)  & \=  \frac{\pi\i N^2}{3m(m\tau+n)^2} \Big[\nu \,  B_3(m\tau +n) + \sllb   
 B_3 \big(z_{\llb} + (m\t + n) (r_{\llb}-1) \big) \Big]  \\
 &\  -\,\frac{\pi\i\tau}{6}\mathrm{Tr}R + \pi\i N^2(\CC+\Phi) \,,
  \end{split}
\ee
where 
\be
\label{theconstant1}
\begin{split}
N^2 \CC 
& \ \equiv -\frac{n}{3m}{\rm Tr}R^3 + \frac{N^2}{2m}\Big[ \nu + \sllb  \,(r_{ab}-1)^2\left(1    - 2    \left\{ -(n_0 m+2n) \frac{r_{\llb}}{2}  \right\} \right) \Big]\,
\end{split}
\ee 
is  $\tau$-independent and purely real. This rewriting will be useful in Section~\ref{sec:saddles_flavoured_index}.

In Eq.~\eqref{Seffmn}, the term linear in $\tau$ is the result of resumming the corresponding terms in~\eqref{DefVa} into the R-symmetry anomaly coefficients \eqref{tHooftA} 
\be\label{intermediate_comp}
\nu N^2   +  N^2 \sllb   \left(2(r_{ab}-1)^3- (r_{ab}-1)\right) \= 
2{\rm Tr}R^3 - {\rm Tr}R \= \frac{16}{9}\left(\textbf{a}\,+\,3\, \textbf{c}\right)\,.
\ee
The last equality in  \eqref{intermediate_comp} shows the combination of~$\textbf{a}$ and~$\textbf{c}$ Weyl anomaly coefficients that is obtained using the relations~\eqref{centralCharges} for superconformal theories. We remark that this term is proportional to the supersymmetric Casimir energy on a round $S^3\times S^1$ \cite{Assel:2014paa,Assel:2015nca}. 
For the $SU(N)$ quivers we are considering, cancellation of the R-Gauge-Gauge ABJ anomaly 
implies~${\rm Tr}R=0$ at leading $O(N^2)$ order, see Appendix \ref{App:anomalies}  for details.
However we temporarily keep the ${\rm Tr}R$ term in the result with the purpose of showing a remarkable agreement 
with the Cardy-like limit of the index at finite $N$, to be discussed momentarily.

 \subsection{The $\tau$-independent part of the action}\label{sec:discussion_phase}
  
The term~$\Phi$ appearing in \eqref{Seffmn} is a real, $\tau$-independent function of $m,n$, as well as of the number of nodes $\nu$ and the R-charges $r_{\llb}$, that remains not determined 
by our technology as it has been developed so far. It arises from the Fourier modes of the function $\Psi_Q(z)$ discussed in Section~\ref{sec:contour}, see 
Appendix~\ref{App:defs_and_identities} for its definition from the integrals  in~\eqref{SeffGeneralForm}. Following the discussion in Section~\ref{sec:contour},  the constant $\Phi$ should be determined by demanding that $\Seff(m,n;\t)$ matches the meromorphic extension of the integrand of \eqref{IntegrandIndex}, evaluated on the gauge variable configurations \eqref{saddles}.  In the rewriting \eqref{Sintermediate1} the terms $\Phi$ and $\CC$ are naturally combined, as they are both real and $\tau$-independent, and  in Section~\ref{sec:saddles_flavoured_index} we will see that a comparison with other results in the literature indeed  relates  $\Phi$  to $\CC$.

Before continuing, we discuss to what extent the value of $\Phi$ affects the results of our analysis, in particular in relation to the comparison with the gravity side.
 Since it yields a purely imaginary,  $\tau$-independent  contribution to the action, $\Phi$ is not relevant for the phase structure  of the index in the grand-canonical ensemble, in the case where only one saddle dominates. Indeed in the grand-canonical ensemble, for each value of the chemical potential $\tau$ the dominating large-$N$ saddle is the one with least real part of the action $S_{\rm eff}$, and the corresponding value of the partition function is $\log Z_{\rm grand}=-S_{\rm eff}$.\footnote{If there are multiple saddles that have the same minimum value of ${\Re}(\Seff)$, then knowing the phase of their exponential contributions $\rme^{-\Seff}$ to the index becomes crucial to determine how they are resummed. In this case $\Phi$ plays an important role. However in order to determine this phase we would need to know the subleading corrections to the large-$N$ limit, which is  out of the scopes of the present work. See \cite{Benini:2018ywd} for a discussion of this phenomenon in the present context.} If instead we discuss the microcanonical ensemble, where the large-$N$ partition function is given by the Legendre transform $\log Z_{\rm micro} = (\tau\, \partial_\tau -1)S_{\rm eff} $, things are more subtle. Being independent of $\tau$, $\pi \ii N^2 \Phi$ appears in the Legendre transform  $(1-\tau\, \partial_\tau)S_{\rm eff}$ precisely in the same way as it appears in $S_{\rm eff} $,  and thus just contributes to the imaginary part of $\log Z_{\rm micro}$. While a priori one could imagine discarding the imaginary part and regarding the entropy as the real part of $\log Z_{\rm micro}$, it has been shown \cite{Hosseini:2017mds,Cabo-Bizet:2018ehj,Choi:2018hmj} that the correct procedure is more delicate.
  In fact one should impose the vanishing of the imaginary part of $\log Z_{\rm micro}$ in order to reproduce the $O(N^2)$ entropy of  known supersymmetric AdS$_5$ black hole solutions.
 This means that  $\Phi$ would play a relevant role, as it appears in ${\rm Im}(\log Z_{\rm micro})=0$.
As illustrated in \cite{Cabo-Bizet:2018ehj}, the latter condition corresponds to a constraint on the $J$ and $Q$ variables in the supersymmetric microcanonical ensemble. Relatedly, after imposing the constraint the expectation values for $J$ and $Q$ in the grand-canonical ensemble depend on $\Phi$.  It appears that only a specific choice of $\Phi$ gives the correct charges that match the dual gravitational solution. One way to fix $\Phi$ that is in agreement with the gravitational results is to regard the action $\Seff$ as a holomorphic function of the chemical potentials $\tau$ {\it and} $\varphi$ that are conjugate to the angular momentum $2J_+ $ and to the R-charge $Q$, respectively, before imposing the relation $\varphi = \tau-\frac{n_0}{2}$ that leads to the index \eqref{Indexseqt}.\footnote{This would lead us slightly off the supersymmetric sector that is captured by the index. Related to this idea, interesting progress has been reported on near-extremal limits of black hole solutions in~\cite{Larsen:2019oll,Nian:2020qsk,Goldstein:2019gpz}.} We will discuss a concrete example in the next subsection.

The upshot is that despite the fact that in our treatment we have not determined the form of $\Phi$, we have argued that in principle it should be determined by  demanding that $\Seff(m,n;\t)$ matches the meromorphic extension of the integrand of \eqref{IntegrandIndex}. On the other hand, for saddles that can be compared with gravity solutions, there is a distinguished value of $\Phi$ that leads to a complete matching of the results on the two sides.  
In the following sections we will see that these two choices are not entirely in agreement.

\subsection{Special families of saddles}\label{sec:special_families}

We further analyze the structure of the saddle-point action \eqref{Seffmn} by identifying some notable cases. In general this depends 
on the details of the quiver considered.  A convenient way to classify the $(m,n)$ saddles in view of evaluating the action is to consider the families 
defined by the different values of the integer 
\be\label{def_ell}
\ell \,\equiv\, -n_0m-2n
\ee
which appears in the argument
\be
z_{\llb}= \Big\{ \ell\, \frac{r_{\llb}}{2}  \Big\}\,
\ee 
of the Bernoulli polynomials in~\eqref{Seffmn}.
Here we always assume $m>0$, and we recall that $m,n$ are coprime.
The evaluation of the action is straightforward if 
\be\label{condition_ell}
-1\,< \, \ell\, \frac{r_{\llb}}{2} \,< \,1 \qquad \text{for all $r_{\llb}$}\,,
\ee
as in this case we can trivially take care of the fractional part for all chiral superfields and evaluate the 
Bernoulli polynomials in  \eqref{Seffmn}. This condition is obviously satisfied for $\ell =0$ and, since we assume that all R-charges lie in the 
range $0 < r_{\llb} < 2$, by $\ell = \pm 1$. 
We will see that the families of saddles characterized by $\ell=0,\pm1$ lead to an action 
which is universal, in the sense that it depends on the field theory data only 
through R-symmetry anomaly coefficients. 

If all chiral multiplets have R-charge $r=2/3$, as for $\mathcal{N}=4$ SYM and its orbifolds, 
then $z = \{\frac{\ell}{3}\}$ is determined by $\ell \ ({\rm mod}\ 3)$ and can only take the values $z=0,\frac{1}{3},\frac{2}{3}$. 
In this case  
 evaluation of the action is straightforward for any~choice of $\ell$, see \cite{Cabo-Bizet:2019eaf} for a thorough analysis of this case. 
For more general theories this is not true. If the R-charge $r$ of a given chiral multiplet is rational, then 
there will be finitely many possible values of the corresponding variable $z$,\footnote{Say $r=p/q$, with $p,q$ 
relatively prime. Then $z$ is determined by $\ell \, ({\rm mod}\, q)$ if $p$ is even, and 
by $\ell \, ({\rm mod}\, 2q)$ if $p$ is odd.} while for the generic case where the R-charge is  irrational there 
are no equivalent choices of $\ell$ and one has infinitely many possible values of $z$, which makes a detailed 
study of the action complicated. 

If all R-charges satisfy $0 < r_{\llb} < 1$, which is true for many theories, it is also straightforward to 
evaluate the action of saddles such that $\ell=2$, however as we will see its expression is not entirely captured by anomaly coefficients. For $|\ell| >2$, the condition \eqref{condition_ell} is not satisfied in a generic theory, so the analysis becomes case-dependent and we will not discuss it further.

After illustrating these consideration in more detail, below we study the case of the conifold theory 
as a simple example where the exact superconformal R-charges are rational and the complete 
phase structure of the $(m,n)$ saddles can be worked out.

\paragraph{The family $\bm{\ell =0}$.}  When $n_0$ is odd, the coprime integers solving the condition $\ell=0$
are $(m = 2$, $n=- n_0)$, while if $n_0$ is even we need to take $(m= 1$, $n=- n_0/2)$.
In both cases, the action \eqref{Seffmn} evaluates to
\begin{align}\label{Seff_family_with_0}
\Seff(\ell=0;\t) &\= \frac{\pi\i\t}{6}\left(2{\rm Tr}R^3 - {\rm Tr}R\right)
 \,+\, \frac{\pi\i\, {\rm Tr}R}{6m (m\t+n)}  \, +\,  \pi\i N^2 \Phi \,,
\end{align}
where we used the definition of ${\rm Tr}R$ in \eqref{tHooftA} at large-$N$. 
 Since this saddle-point action is guaranteed to be correct only at $O(N^2)$ order, we should set ${\rm Tr}R=0$ as a consequence of the R-Gauge-Gauge anomaly cancellation for the quivers of interest in this paper. This leaves us with
\begin{align}\label{Seff_family_with_0_TrR=0}
\Seff(\ell=0;\t) &\=  \frac{\pi \i\t}{3}{\rm Tr}R^3 \, +\, \pi \i  N^2 \Phi \,.
\end{align}
Notice that the Legendre transform of this $S_{\rm eff}$ is purely imaginary, $(1-\tau\,\partial_\tau)S_{\rm eff} = \pi \ii N^2 \Phi$. 
We conclude that, independently of the value of $\Phi$, these saddles carry no $O(N^2)$ entropy.

We can also compare the expression above with the Cardy-like limit of the index. Take $n_0=0$, namely consider the standard index with no shift of the angular chemical potential, {\it cf.}~\eqref{IndexHamDef}. Then we have $(m=1, n=0)$, and upon taking the small-$\tau$ limit the leading order ${\rm Tr}R/\tau$ term in \eqref{Seff_family_with_0} remarkably agrees with the Cardy-like formula of \cite{DiPietro:2014bca}, which is derived at finite-$N$. This indicates that \eqref{Seff_family_with_0} correctly captures at least a part of the finite-$N$ action of the $(m,n)$ saddles considered here. We leave the analysis of the subleading corrections to the $O(N^2)$ result for future work.

\paragraph{The family $\bm{\ell =\pm1}$.}  In this case both $n_0$ and $m$ are odd. In particular, this family of saddles exists for the index \eqref{thermal_index} but not for the $n_0=0$ index. Also recall that
we assume that all R-charges satisfy $0 <r_{\llb}<2$. Then the Bernoulli polynomials evaluate to
\begin{align}\label{BernoulliPoly_n=pm1_class}
B_2(z_{\llb}) & \= \frac{1}{4}r_{ab}(r_{\llb}-2)+\frac16\,,\nn\\
B_3(z_{\llb})  &\=  \pm \frac{1}{8}\,r_{\llb}(r_{\llb}-1)(r_{\llb}-2)\,,
\end{align}
where the sign choice is correlated with $\ell=\pm1$.
Plugging this in the effective action~\eqref{Seffmn} and using
\be 
\sllb N^2 \,r_{\llb}(r_{\llb}-1)(r_{\llb}-2) \= {\rm Tr}R^3 - {\rm Tr}R\,,
\ee
 we arrive at
\be
\begin{split}\label{SeffAnomalies}
\Seff(\ell=\pm1;\t) 
 &\= \frac{\pi\i\t}{6}\left(2{\rm Tr}R^3 - {\rm Tr}R\right) \,+\, \frac{\pi\i}{12 m (m\t+n)}\left( 3 {\rm Tr}R^3 - {\rm Tr}R    \right)\\ 
&\qquad  \pm  \frac{\pi\i }{24 m (m \tau+n)^2} \left({\rm Tr}R^3 - {\rm Tr}R\right)  
  \, +\, \pi\i N^2 \Phi  \,.
\end{split}
\ee

Before omitting the ${\rm Tr}R$ terms, let us consider the sub-case $n_0=\mp1$, $m=1,n=0$, in which case the action \eqref{SeffAnomalies} reads
\be
\begin{split}\label{n0m=1_n=0_action}
 \frac{1}{\pi\i}\,\Seff(m=1, n=0,n_0=\mp1;\t) 
 &\=  \frac{8\t^3+6\tau -n_0}{24 \t^2} \, {\rm Tr}R^3 - \frac{2\tau -n_0 }{24 \t^2} \,{\rm Tr}R  - \frac{\t}{6} {\rm Tr}R 
  \, +\,  N^2 \Phi  \,.
\end{split}
\ee
Upon taking the Cardy-like limit $\tau\to 0$, we can compare the $O(\tau^{-2})$ and $O(\tau^{-1})$ terms 
in this expression with  the results of \cite{Cabo-Bizet:2019osg,Kim:2019yrz}, finding perfect agreement. 
The asymptotic methods used in \cite{Cabo-Bizet:2019osg} hold at finite-$N$ but only provide the divergent 
terms in the small-$\tau$ regime. Here we 
find complementary results which hold at leading order in the large-$N$ expansion but are valid at finite $\tau$. 
We emphasize again that for the quivers considered, the terms proportional to ${\rm Tr}R$ are subleading 
in the large-$N$ limit and thus in principle we have no control on them, nonetheless it is remarkable that all 
the $O(\tau^{-2})$ and $O(\tau^{-1})$ terms in \eqref{n0m=1_n=0_action} agree with the finite-$N$ results 
obtained from the Cardy limit of the index. As for the $\ell=0$ saddle, it would be 
interesting to clarify to what extent the action~\eqref{n0m=1_n=0_action} captures the finite-$N$ contribution 
to the index from these saddles.

Let us now restrict ourselves to the leading $O(N^2)$ order and hence take ${\rm Tr}R=0$ and discuss the comparison with the gravity side. 
We observe that we can rewrite (\ref{SeffAnomalies}) in the suggestive form
\be
 \Seff(\ell=\pm1;\t) \,=\,\frac{\pi\i }{24}\, \frac{\left(  2m\tau +2n\pm 1\right)^3}{m (m\tau+n)^2}\, {\rm Tr}R^3        -\frac{\pi \i  }{6m} ( 2n \pm  3)  {\rm Tr}R^3 + \pi \i N^2    \Phi  \, ,
 \ee
 and that the choice 
 \be\label{choicePhi}
N^2 \Phi_{\ell=\pm1} \, = \,   \frac{2n\pm 3}{6m}\,{\rm Tr}R^3  
\ee
 gives the action the form of a ``perfect cube'', namely 
 \be
 \label{perfectcube}
 \Seff(\ell=\pm1;\t) \,=\,\frac{\pi\i }{24}\, \frac{\left(  2m\tau +2n+\ell \right)^3}{m (m\tau+n)^2}\, {\rm Tr}R^3 
 \= \frac{\pi\i }{24}\, \frac{\left(  2\tau -n_0\right)^3}{ (\tau+n/m)^2}\, {\rm Tr}R^3 \,.
\ee
For superconformal theories we can convert the R-symmetry anomaly coefficient into Weyl anomaly coefficients 
setting $\frac{9}{32}{\rm Tr}R^3 = {\bf a}= {\bf c}$, which yields
\be\label{action_simple}
 \Seff(\ell=\pm1;\t) \=  \frac{4\pi\i\,{\bf a}}{27}\, \frac{\left( 2\tau -n_0\right)^3}{(\tau+n/m)^2} \,.
\ee
For $\mathcal{N}=4$ SYM, this agrees with the  result found in \cite{Cabo-Bizet:2019eaf}.\footnote{This is 
given in Equation~(4.13) of \cite{Cabo-Bizet:2019eaf}. To compare, one should recall that for $\mathcal{N}=4$ SYM 
at large-$N$ ${\bf a}={\bf c}=\frac{N^2}{4}$, and that the Dirichlet character $\chi_1(-n_0m + n)$ appearing there evaluates to $\chi_1(-n_0m + n) = +1$ for $-n_0m + n = 1 $ (mod 3) and $\chi_1(-n_0m + n) = -1$ for $-n_0m+ n = -1 $ (mod 3).}

The expression \eqref{action_simple} provides a  prediction for the on-shell action $I$ of putative dual supergravity 
solutions, which would compete in the semiclassical approximation to the gravitational path integral in the same way 
as the saddles compete in the large-$N$ expression for the superconformal index. For definiteness, let us assume 
the theory is dual to type IIB supergravity on Sasaki-Einstein manifolds. Using the 
dictionary ${\bf a}={\bf c}=\frac{\pi L^3}{8G_5}$, where $L$ is the AdS$_5$ radius and $G_5$ the five-dimensional 
Newton constant, we obtain
\begin{align}\label{better_on_shell_action}
 I(\ell=\pm1;\t)  
 \= \frac{ \i\pi^2 L^3}{54G_5} \frac{\left(  2\tau -n_0\right)^3}{(\tau+n/m)^2} \,.
\end{align}
For $m=1$, $n=0$,  and $\ell=-n_0=\pm 1$ this  matches the supergravity on-shell action of supersymmetric AdS$_5$ black holes computed using the prescription of~\cite{Cabo-Bizet:2018ehj}. For more general values of $(m,n)$, we expect it to be the action of solutions that are yet to be 
described. Due to the universality of the expression, these may be solutions to minimal gauged 
supergravity in five dimensions, that it would be very  interesting to construct.

Let us now come back to the choice of constant phase $\Phi$.  As we explained in Section \ref{sec:contour}, on general grounds 
this phase should be determined by demanding agreement with the meromorphic action evaluated at the saddles. 
Anticipating the results of the next section, which will discuss the more general  index refined by flavor fugacities, in the present family of saddle with $\ell=\pm 1$, 
comparing with the results in the literature \cite{Benini:2018ywd,Lezcano:2019pae,Lanir:2019abx} implies that we should set $\CC+\Phi=0$ in (\ref{Sintermediate1}), at least for $m=1$.    
Specifically,  this yields
\be
\label{theconstant1}
\begin{split}
N^2 \Phi_{\ell=\pm 1} 
&=  \frac{2n\pm 3}{6} {\rm Tr}R^3  + \left\{ \begin{array}{cl}
- N^2 \nu & \qquad \mathrm{for}~\ell=+1\\
0 &\qquad \mathrm{for}~\ell=-1\\
\end{array} \right. \, ,
\end{split}
\ee
and it agrees with the choice (\ref{choicePhi}) in the case $\ell=-1$, while it differs from this  by the constant term $-N^2 \nu$ in the case $\ell=+1$. Therefore, 
 this choice leads to the  effective action (for $m=1$)
\be 
 \Seff(\ell=\pm1;\t) \=  \frac{4\pi\i\,{\bf a}}{27}\, \frac{\left( 2\tau -n_0\right)^3}{(\tau+n)^2}  + \left\{ \begin{array}{cl}- \pi \i N^2 \nu & \qquad \mathrm{for}~\ell=+1\\
0 &\qquad \mathrm{for}~\ell=-1\\
\end{array} \right. \, .
\ee
In this case,  while the first term is proportional to the Weyl anomaly ${\bf a}$ and therefore it can be translated to the gravity side using the holographic dictionary, the second term, 
being proportional to $N^2\nu$, that is the dimension of the gauge group of the quiver theory, cannot be matched to any computation involving the gravitational action.

\vskip 5mm

\paragraph{The family $\bm{\ell=\pm2}$.} We discuss this case as an example of a non-universal class of 
saddles, where the action depends on more field theory data than just the R-symmetry anomaly coefficient. For 
simplicity, we only consider theories where the R-charges of all chiral multiplets satisfy $0 < r_{\llb}< 1$. 
Then $z_{\llb}=r_{\llb}$ for the upper sign choice and $z_{\llb}=1-r_{\llb}$ for the lower sign choice, which gives
\begin{align}\label{BernoulliPoly_n=pm2_class}
B_2(z_{\llb}) &\= r_{\llb}(r_{\llb}-1) + \frac16\,,\nn\\[1mm]
B_3(z_{\llb}) &\= \pm \,r_{\llb}\Big(r_{\llb}-\frac{1}{2}\Big)(r_{\llb}-1)\,.
\end{align}
Plugging these expressions in the general form of the action \eqref{Seffmn}, we find
\begin{align}\label{Seff_mp2_family}
\Seff(\ell=\pm2;\t) & \= \frac{\pi\i\tau}{3}{\rm Tr}R^3 \,+  \frac{\pi\i N^2}{ m (m\t+n)}  \,
\sllb  r_{ab}(r_{ab}-1)^2  \nn\\ 
&\ \pm  \frac{\pi\i N^2}{3 m (m \tau+n)^2} \, \sllb  r_{\llb} 
\left(r_{\llb}-\tfrac{1}{2}\right)(r_{\llb}-1)  \, +\,  \pi\i N^2 \Phi  \,,
\end{align}
where we have omitted the terms proportional to ${\rm Tr}R$, as they vanish at order $O(N^2)$. 
Generically the sums over the chiral fields appearing in this expression can be expressed in terms 
of ${\rm dim}G = \nu N^2$, ${\rm Tr}R^2$ and ${\rm Tr}R^3$, hence they are not fully determined 
by anomalies. For each specific theory in the class of quivers we are considering, these terms 
are proportional to $N^2$ and thus to ${\rm Tr}R^3$, but the proportionality coefficient is theory-dependent.
In other words, in this family of saddles the action does not take the form of one universal function 
of $(m,n;\tau)$, multiplied by ${\rm Tr}R^3 $. Let us illustrate this non-universality further by considering two simple examples. 

For $\mathcal{N}=4$ SYM and its orbifolds, this family is analogous to the family $ n_0m+2n=\mp1$ 
discussed above, in agreement with the observation made earlier that in order to determine $z$ we only 
need to know $\ell$ (mod 3).\footnote{Since the chiral fields have R-charge $r=2/3$, they satisfy the 
condition $r=1-\frac{r}{2}$, so that the Bernoulli polynomials \eqref{BernoulliPoly_n=pm2_class} can be 
rewritten as in \eqref{BernoulliPoly_n=pm1_class}. Then the action takes the same form \eqref{SeffAnomalies} 
(where one should set ${\rm Tr}R=0$), with the only difference that now  $m,n$ are constrained by the 
condition $n_0m+2n=\mp2$, instead of $n_0m+2n=\mp1$.} Hence all choices of $m,n$ lead to either the 
action~\eqref{Seff_family_with_0_TrR=0}, or to the action \eqref{SeffAnomalies}, where one should set ${\rm Tr}R=0$ as this is an exact relation in $\mathcal{N}=4$ SYM.

For the conifold theory \cite{Klebanov:1998hh}, the quiver is made of two gauge nodes and four bifundamental fields, all 
with R-charge $r=1/2$. Then the action \eqref{Seff_mp2_family} takes the form
\begin{align}
\Seff(\ell=\pm2;\t) 
  &\= \frac{\pi\i}{3}\Big(\tau \,+  \frac{1}{ m (m\t+n)}+ 2\Phi\Big) {\rm Tr}R^3  \,,
\end{align}
where  we used ${\rm Tr}R^3 = \frac{3}{2}N^2$ at order $O(N^2)$. We see that this class of saddles 
gives an action where ${\rm Tr}R^3$ is multiplied by a function of $(m,n,\tau)$ different
from \eqref{Seff_family_with_0_TrR=0} and  \eqref{SeffAnomalies}. For the conifold theory the cases we 
have discussed exhaust the possible forms of the action. This follows from the fact that for $r=1/2$, the 
possible values of the variable \eqref{zab_variable} are determined by $\ell$ (mod 4) and 
are $z =0, \frac{1}{4}, \frac{1}{2},\frac{3}{4}$. 
For more general theories, even if the R-charges are all between 0 and 1, there will be further families of 
saddles, where the action takes a yet different form.

\subsection{Phase structure in the grand-canonical ensemble \label{sec:phases}}
The phase structure of the grand-canonical ensemble is determined by evaluating the real part of the large-$N$ 
action for the different $(m,n)$ saddles, and picking the one with the least value while the chemical potential $\tau$ 
is varied. After replacing the expressions of the Bernoulli polynomials 
and setting ${\rm Tr}R=0$, the real part of the action \eqref{Seffmn} is
\begin{align}\label{eq:Re_Seff}
{\rm Re}\,S_{\rm eff}(m,n;\tau) &\= - \frac{\pi \tau_2}{3}\, {\rm Tr}R^3 + \frac{\pi\tau_2 }{|m\tau+n|^2} 
\,N^2\sllb  \,(r_{\llb}-1)\,\big\{ \ell\, \frac{r_{\llb}}{2} \big\} \left(\big\{ \ell\, \frac{r_{\llb}}{2}  \big\}-1 \right)\nn\\[1mm]
&\ + \frac{2\pi (m\tau_1+n)\tau_2 }{3|m\tau+n|^4} \,N^2\sllb \, \big\{ \ell\, \frac{r_{\llb}}{2}  \big\} 
\Big(\big\{ \ell\, \frac{r_{\llb}}{2}  \big\}-\frac{1}{2}\Big)\left(\big\{ \ell\, \frac{r_{\llb}}{2}  \big\}-1 \right)\,,
\end{align}
where $\tau=\tau_1+\i \tau_2$, with $\tau_1 \in\mathbb{R}$, $\tau_2 \in \mathbb{R}_+$\ and we recall that the 
integer $\ell$ is defined in \eqref{def_ell}. The first term in this expression, that can be seen as a vacuum energy, is independent of $(m,n)$ and is thus 
irrelevant for comparing the action of the different saddles. The second term in the first line is positive-definite 
as long as the R-charges $r_{\llb}$ are all smaller than 1. On the other hand the term in the second line does not 
have a definite sign, and it dominates ${\rm Re}\,S_{\rm eff}$ when it becomes large and negative.

Since for the $\ell=0$ saddles the real part of the action is (recall \eqref{Seff_family_with_0_TrR=0})
\be\label{ReS_ell_0}
{\rm Re}\,S_{\rm eff}(\ell=0;\tau) \= - \frac{\pi \tau_2}{3}\, {\rm Tr}R^3\,,
\ee
where one should recall that $\tau_2>0$,
any other saddle dominates over this one if and only if the sum of the second and third term in~\eqref{eq:Re_Seff} is negative.
It is also interesting to notice that for $(m=0, n\neq0)$ the real part of the action vanishes at order $O(N^2)$, hence this saddle never dominates over the $\ell=0$ saddle.

\paragraph{The example of the conifold.} For $\mathcal{N}=4$ SYM the phase structure was studied 
in~\cite{Cabo-Bizet:2019eaf}. Another example where we can easily analyze the complete phase structure 
of the $(m,n)$ saddles is the one of the conifold theory.
It is convenient to study the real part of the action according to the different  values of $\ell$, which in this case are just $\ell=\{0,\pm1,\pm2\}$. 
For $\ell=0$ the real part of the action is given in \eqref{ReS_ell_0}, for $\ell=\pm1$ we have
\be
\frac{1}{{\rm Tr}R^3}\,{\rm Re}\,S_{\rm eff}(\ell=\pm1;\t) \=  - \frac{\pi \tau_2}{3} +\frac{\pi \tau_2}{4 |m\tau+n|^2} \pm \frac{\pi (m\tau_1+n)\tau_2 }{12|m\tau+n|^4} \,,
\ee
while for $\ell=\pm2$ we have 
\be\label{ell=2_conifold}
\frac{1}{{\rm Tr}R^3}\,{\rm Re}\,S_{\rm eff}(\ell=\pm2;\t)
 \= -\frac{\pi \tau_2}{3} + \frac{\pi\tau_2 }{3|m\tau+n|^2} \,.
\ee
Since the second term in \eqref{ell=2_conifold} is larger than 0, the $\ell=\pm2$ saddles are always subdominant compared to the $\ell=0$ ones.
If $n_0$ is even the $\ell=\pm1$ saddles do not exist and the $\ell = 0$ saddle dominates the grand-canonical 
ensemble. For odd $n_0$ the $\ell=\pm1$ saddles exist and dominate over the $\ell=0$ one in the domain given by 
\be
-\frac{1}{3}<\pm\,(m\tau_1+n)< 0\,,\qquad \tau_2^2< (m\tau_1+n)\Big(\,\mp\frac{1}{3}-m\tau_1-n\Big) \, \qquad\text{for}\ \ell=\pm1\,.
\ee
 Identifying the dominating $(m,n)$ saddle while $\tau$ spans this domain requires a more refined study.
 However the result is the same as for $\mathcal{N}=4$ SYM and it has been discussed in~\cite{Cabo-Bizet:2019eaf}.

\section{Large-$N$ limit of the refined index}\label{sec:saddles_flavoured_index}

\subsection{Including flavor chemical potentials}

The supersymmetric theories we are considering in general admit global symmetries that commute with the $\mathcal{N}=1$ supercharges and are thus non-R-symmetries. Depending on whether these are manifest in the Lagrangian or not, it is common to distinguish between ``flavor'' and ``baryonic'' symmetries. However, in our analysis this distinction will not play a role and we will refer  to as flavor symmetries all such non-R-symmetries.
We denote by $d$ the total number of Abelian factors in the global symmetry group, and denote by $\widetilde{Q}_\fli$, $\fli=1,\ldots,d-1$, the flavor charges and by $\widetilde{Q}_d$ the R-charge (not necessarily the exact superconformal one).\footnote{In this section we reserve the symbol $Q$ for a different basis of charges to be defined later.} These satisfy the commutation relations
\be\label{commrel_charges}
[\widetilde{Q}_{\fli},\mathcal{Q}]=0\,, \qquad [\widetilde{Q}_d,\mathcal{Q}]=-\mathcal{Q}\,,
\ee
where $\mathcal{Q}$ is the supercharge that is used to define the index.
Then we refine the index \eqref{IndexHamDef} by turning on chemical potentials for the flavor symmetries, and consider the following trace over the Hilbert space of the theory,
\be\label{our_index_alternative}
\mathcal{I}\=   {\rm Tr}_{\mathcal{H}}\,  (-1)^F \, \rme^{-\beta \{\mathcal{Q},\overline{\mathcal{Q}}\}+
2\pi \i\, (\sigma - n_0)  (J_1 + \frac{1}{2}\widetilde{Q}_d) + 2\pi \i
\tau\,  (J_2 + \frac{1}{2}\widetilde{Q}_d) + 2\pi \i\,\varphi^{\fli} \widetilde{Q}_{\fli}} \,,
\ee
where $\varphi^{\fli}$, $\fli=1,\ldots,d-1$, are the flavor chemical potentials. This is a well-defined expression, as all the charges $(J_1 + \frac{1}{2}\widetilde{Q}_d)$, $(J_2 + \frac{1}{2}\widetilde{Q}_d)$, $\widetilde{Q}_{\fli}$, commute with the supercharge $\mathcal{Q}$.
Using  $\rme^{-2\pi \i n_0 J_1}=\rme^{\pi \i n_0 F}$, Eq.~\eqref{our_index_alternative} may be rewritten as
\be\label{our_flavored_index}
\mathcal{I}\=   {\rm Tr}_{\mathcal{H}}\,  \rme^{\pi \i (n_0+1) F} \, \rme^{-\beta \{\mathcal{Q},\overline{\mathcal{Q}}\}+
2\pi \i \,\sigma  J_1 + 2\pi \i\, \tau  J_2 + 2\pi\i\,(\varphi^{\fli} \widetilde{Q}_{\fli} + \varphi^d  \widetilde{Q}_d)} \,,
\ee
where as before the R-symmetry chemical potential is fixed to
\be\label{expr_varphid_us}
\varphi^d \,=\, \frac{\sigma+ \tau -n_0}{2}\,.
\ee 
In Appendix~\ref{app:rewriting_index} we compare the expression above with other formulations that have appeared in the recent literature, spelling out the precise dictionary between the different variables that have been used and thus showing their mutual consistency.

We now consider the integral representation of \eqref{our_index_alternative}, setting $\s=\t$ as before. 
Given a chiral superfield going from node $a$ to node $b$ of the quiver, we introduce the variables
\be
\varphi_{\llb} = \varphi^{\fli}\, (\widetilde{Q}_{\fli})_{\llb}\,,
\ee
where $(\widetilde{Q}_{\fli})_{ab}$ is the value that the flavor charge operator $\widetilde{Q}_{\fli}$ takes on the chiral superfield under consideration. 
 The contribution of the flavor chemical potentials to the integrand of the index is obtained from~\eqref{IntegrandIndex}
 by modifying the contribution of each chiral multiplet as
\be\label{shift_Gamma_fct}
\Ge\bigl(u^{a b}_{i j}+  \frac{2\t -n_0}{2}\, r_{\llb};\,\t,\,\t\bigr)\quad \mapsto\quad \Ge\bigl(u^{a b}_{i j}+ \frac{2\t -n_0}{2}\,r_{\llb}+ \varphi_{\llb};\,\t,\,\t\bigr)\,.
\ee
It is then convenient to collect the R-charges $r_{\llb}$ and the variables $\varphi_{\llb}$   into the new variable
\be\label{eq:defDeltaab}
\Delta_{\llb}\equiv    \frac{2\t-n_0}{2}\,r_{\llb} + \varphi_{\llb}\, = \,   \varphi^{\fli} (\widetilde{Q}_{\fli})_{\llb}+  \varphi^d \,r_{\llb} \, = \, \varphi^I (\widetilde{Q}_I)_{\llb}\, ,
\ee 
with $I=1,\ldots,d$.
This corresponds to the charge of a chiral superfield going from node $a$ to node $b$ under the linear combination $ \varphi^I \widetilde{Q}_I$, which involves both the flavor symmetries and the R-symmetry. The reason for introducing this new set of variables is that it
 is invariant under a change of basis for the charges, in particular under a transformation that mixes the R-symmetry with the flavor symmetries (that we will implement in Section~\ref{sec:Action_Anomalies}).
 We allow all chemical potentials, and thus the variables $\Delta_{ab}$, to take complex values. Again, we will use the decomposition 
\be\Delta_{\llb}=(\Delta_{\llb})_1 +\tau\, (\Delta_{\llb})_2\,,
\ee
with ${(\Delta_{\llb})}_1$ and ${(\Delta_{\llb})}_2\in \mathbb{R}$.

 Using the formalism of Section~\ref{sec:quivers}, the integrand of the index can be expressed in terms of the action \eqref{IntSu}, \eqref{EffAction},
the only difference being that the arguments $c_{\llb}$ and $d_{\llb}$ of the potential functions $V$ and $V_{ab}$ defined in \eqref{DefVa} should now be identified with
\be\label{some_definitions}
c_{\llb} \= r_{\llb}+{(\varphi_{\llb})}_2-1  \,=\, (\Delta_{\llb})_2 - 1\,, \quad \quad \quad d_{\llb} \= - \frac{n_0}{2} r_{\llb} +{(\varphi_{\llb})}_1 \,=\, (\Delta_{\llb})_1 \,.
\ee
The configurations \eqref{saddles} remain large-$N$ saddles of this more general action, since the proof given in Section~\ref{sec:Saddles} still applies. Evaluating the action on these saddles as before, we arrive at the large-$N$ action
\begin{align}\label{Seffmnflavoured}
\Seff(m,n;\t,\varphi^{\fli})  
&= \frac{\pi\i\tau}{6}\big(2{\rm Tr}\hat{R}^3 - {\rm Tr}\hat{R}\big)\nn\\
&\quad +  \frac{\pi\i}{ m (m\t+n)}\bigg[ \frac{{\rm Tr}\hat{R}}{6}+ N^2 \sllb ( (\Delta_{\llb})_2-1) 
\big(B_2 (z_{\llb}) -\tfrac{1}{6}\big) \bigg] \nn\\ 
&\quad  +  \frac{\pi\i N^2}{3 m (m \tau+n)^2}  \sllb \, B_3 (z_{\llb})  \, +\,  \pi\i N^2 \Phi  \,.
\end{align}
This has the same form as (\ref{Seffmn}), however it contains non-trivial contributions from the flavor chemical potentials that appear in  the variable
\be\label{zabflavoured}
z_{\llb} \equiv  \{m d_{\llb} -n c_{\llb}  \} = \left\{ -(mn_0+2n) \frac{r_{\llb}}{2}     + m{(\varphi_{\llb})}_1  - n {(\varphi_{\llb})}_2 \right\} = \{m (\Delta_{\llb})_1 -n  (\Delta_{\llb})_2 \} \, , 
\ee
as well as in the traces
\be
\mathrm{Tr}\hat R  =  \nu + \sllb \left( (\Delta_{\llb})_2 - 1 \right)\,,\qquad 
\mathrm{Tr}\hat R^3  =  \nu + \sllb \left( (\Delta_{\llb})_2 - 1 \right)^3\,.
\ee
These are anomaly coefficients for the trial R-charge operator
\be
\hat{R}  \,=\, \varphi_2^I\, \widetilde{Q}_I\,,
\ee
corresponding to the part along $\tau$ of $\varphi^I \widetilde{Q}_I$,
 under which the chiral superfields have charge $(\Delta_{\llb})_2$. Note that the flavor chemical potentials $\varphi_2^{\fli}$ correspond to mixing parameters for this trial R-charge.

As in the case with no flavor chemical potentials, we can use identity \eqref{B3expanded} to recast the action in the compact form
\be\label{Sintermediate}
\begin{split}
\Seff(m,n)  & =  \frac{\pi\i N^2}{{3m(m\tau+n)^2}} \Big[\nu\,B_3(m\tau +n)+ \sllb    B_3 \big(z_{\llb} + (m\t + n)( (\Delta_{\llb})_2 - 1) \big)  \Big] \\
 & -\frac{\pi\i\tau}{6}\,\mathrm{Tr}\hat R + \pi\i N^2(\CC+\Phi) \,,
  \end{split}
\ee
where 
\be
\label{theconstant}
\begin{split}
N^2 \CC  
& \ \equiv -\frac{n}{3m}\,{\rm Tr}\hat{R}^3 + \frac{N^2}{2m}\,\Big[ \nu + \sllb  ((\Delta_{\llb})_2 - 1)^2\left(1    - 2    \{m (\Delta_{\llb})_1 -n  (\Delta_{\llb})_2 \}  \right) \Big]\,
\end{split}
\ee 
is again real and independent of $\tau$.

We now elaborate further on this expression, recasting it in a form that makes it clear that it generalizes to the $(m,n)$ saddles some results that have appeared in the literature for the case $m=1$.
The argument of the second $B_3$ in (\ref{Sintermediate})  can be written as
\be
\begin{split}
z_{\llb} + (m\t + n) ((\Delta_{\llb})_2 - 1) & =  [\Delta_{\llb}]^m_T + 1 - T\, ,
\end{split}
\ee
where recall that  $\taumn =  m\tau + n$  and $[\Delta]^m_T$ is a new function defined as follows
\be
\label{defmybracket}
  [\Delta]^m_T   \,\equiv\,  \{m \Delta_1 -n  \Delta_2 \} + T \,  \Delta_2  -1 
\ee
for any $\Delta = \Delta_1+\tau \Delta_2$. When $m=1$ this reduces to the function  $[\Delta]_\tau$ defined in \cite{Benini:2018ywd},\footnote{More precisely, the definitions agree for $\Delta_1 \notin\mathbb{Z}$. Indeed for $\Delta_1\in \mathbb{Z}$ the function $[\Delta]_\tau$ of \cite{Benini:2018ywd} is not defined, while we have $\{\Delta_1\}=0$.} but 
notice that (\ref{defmybracket}) is not equivalent  to the original  $[\Delta]_\tau$,   replacing $\tau$ with $T$ in the latter.  We appended the superscript  $m$ to emphasize this
fact.\footnote{Using the alternative decomposition $\Delta=\tilde{\Delta}_1+ \taumn \tilde{\Delta}_2$, with $\tilde{\Delta}_1,\tilde{\Delta}_2\in\mathbb{R}$, we can express the definition \eqref{defmybracket} as $[\Delta]^m_T= \{m\tilde\Delta_1\}+mT\tilde\Delta_2-1$. This means that the function $[\Delta]^m_T$ takes $m\Delta$ and shifts it by an integer in such a way that the result falls inside the strip in the complex plane bounded on the left by the line passing through $-1$ and $-1+\taumn$, and on the right by the line passing through $0$ and $\taumn$. 
}
 It will be useful to record the following properties satisfied by the function $[\Delta]^m_\taumn$:
\bea
[\Delta+1]^m_{\taumn}  & = &  [\Delta]^m_{\taumn}\, , \label{prope1}\\
~
 [\Delta+\tau]^m_\taumn & = &  [\Delta]^m_{\taumn} +\taumn\, ,\label{prope2}\\
 ~  [-\Delta]^m_\taumn  &= & -[\Delta]^m_\taumn -1\, .
  \label{prope3}
\eea
Notice that the properties (\ref{prope1}) and  (\ref{prope3}) are exactly the same as those obeyed by  the function $[\Delta]_\tau$, while  (\ref{prope2}) is a generalization to arbitrary $m$. 

Using this bracket and expanding out the $B_3$ Bernoulli polynomials, the action \eqref{Sintermediate} can be written more explicitly as
\be\label{mostgeneralniceS}
\begin{split}
\Seff(m,n)   &   =       \frac{\pi \i N^2 }{3m\taumn^2}\,\nu\, \taumn (\taumn-\tfrac{1}{2}) (\taumn-1)\\[1mm]
& +\frac{ \pi \i N^2}{3m\taumn^2} \sllb \, \big( [\Delta_{\llb} ]^m_{\taumn}  -\taumn \big)  \big([\Delta_{\llb} ]^m_{\taumn} -\taumn + \tfrac{1}{2}\big)   \big([\Delta_{\llb} ]^m_{\taumn} -\taumn  + 1 \big)  \\
& 
 +  \pi \i  N^2 (\CC+\Phi )\,,
\end{split}
\ee
where we have now set ${\rm Tr}\hat{R}=0$ as we are working in the strict large-$N$ limit and this trace vanishes at  $O(N^2)$ order due to ABJ anomaly cancellation. 
This form of the action can be compared immediately with previous results in the literature. In particular, for $(m,n)=(1,r)$, the first two lines in (\ref{mostgeneralniceS}) coincide precisely with 
the expression presented in \cite{Lanir:2019abx}, \emph{c.f.}~Eq.~(2.38) therein and subsequent discussion.\footnote{A comparison with \cite{Benini:2018ywd,Lanir:2019abx} shows that their gauge variable configurations $u^{a}_{i} = (\tau+r)\left( \frac{N+1}{2N} - \frac{i}{N}  \right)$ are recovered from \eqref{discansatz1} by taking $m=-1$, $n=-r$, however the symmetry under $(m,n)\to (-m,-n)$ of our large-$N$ action ensures that the same result is obtained choosing $m=1$, $n=r$.}  
This result has been obtained in \cite{Lanir:2019abx}  through the  Bethe Ansatz approach originally proposed in \cite{Benini:2018ywd} for ${\cal N}=4$ SYM. As we already discussed, this method is different from the one that we have used, and it is therefore remarkable that both routes yield consistent results. 
 Comparing (\ref{mostgeneralniceS}) with the result presented in    \cite{Benini:2018ywd,Lezcano:2019pae,Lanir:2019abx} and demanding that the terms independent of 
  $\tau$ 
 coincide  fixes the undetermined phase as $\Phi = -\CC$ in the $(m=1,n)$ case. In particular, plugging $m=1$ into \eqref{theconstant}, we obtain\footnote{Invariance under $(m,n)\to (-m,-n)$ implies that this is also the expression of $\Phi(m=-1,-n)$.} 
\be
\begin{split}\label{eq:Phi_m=1}
\Phi(m=1,n) 
& = \frac{n}{3}\frac{{\rm Tr}\hat{R}^3}{N^2} - \frac{1}{2}\Big( \nu + \sllb  ((\Delta_{\llb})_2 - 1)^2\left(1    - 2   \{ (\Delta_{\llb})_1 -n  (\Delta_{\llb})_2 \}\right) \Big)\,.
\end{split}
\ee

It should be emphasized that the  $(m,n)$   solutions are more sophisticated than the $(1,n)$ ones. In particular, while the effective action for the $(1,n)$ saddles can be obtained simply replacing  $\tau\to \tau +n$ in the result for the basic $(1,0)$ solution, this is not the case for the $(m,n)$ family.
  Indeed, although these  saddles were also identified as solutions to the Bethe Ansatz Equations in the approach of~\cite{Benini:2018ywd}, their effective action had not been computed so far for generic theories.\footnote{For $\mathcal{N}=4$ SYM with no flavors, the effective action of these saddles was computed in~\cite{Cabo-Bizet:2019eaf}.}  
The term $\Phi$ however remains undetermined for general $m$, as discussed in Section~\ref{sec:largeN_action_unflavoured}. This is in principle fixed by demanding agreement with the value of the original, meromorphic integrand of \eqref{IntegrandIndex}, evaluated on the gauge variable configurations \eqref{discansatz1}, in the large-$N$ limit.

The phase structure of the large-$N$ index in the grand-canonical ensemble---where the chemical potentials $\tau,\varphi^{\fli}$ are the independent variables---is obtained by minimizing  ${\rm Re}(\Seff)$ over the different saddles. As we discussed previously, recall that the purely imaginary term $\pi \i N^2(\CC+\Phi)$ does not play a role in this extremization as long as there are no competing saddles with equal real part of the action, although it plays a role in the reality condition for the entropy. It would be interesting to study the details of  this phase structure.

\subsection{Special families of saddles in the flavored setup}\label{special_saddles_flavoured}

It is straightforward to generalize the $\ell=0,\pm1$ special cases discussed in Section~\ref{sec:special_families} to the present flavored setup. The generalization we discuss here requires the chemical potentials, seen as complex variables, to be aligned in a specific direction (but allows for completely generic values of the charges), while in the next section we will discuss a generalization allowing the chemical potentials to take values in two-dimensional domains of the complex plane (but we will need to assign values to the charges in order to determine the value of the action in the different domains).
 
For any given saddle with $m\neq0$, let us consider the situation where all chemical potentials, seen as complex variables, are proportional to $T=m\tau+n$. Decomposing as usual $\varphi^{\fli} = \varphi^{\fli}_1+\tau\varphi^{\fli}_2$, we can write this condition as
\be\label{eq:nice_flavor_chempot}
\varphi^I =   \varphi_2^I\, \frac{ m\tau + n}{m}\,, \qquad I=1,\ldots,d\,,
\ee
which can be seen as an equation that fixes $\varphi_1^I$ in terms of $\varphi_2^I$. 
Taking $I=\fli$, for $\fli=1,\ldots,d-1$, the condition above gives $m(\varphi^{\fli})_1 - n (\varphi^{\fli})_2 =0$, while taking $I=d$, and recalling that the R-symmetry chemical potential is always fixed to $\varphi^d = \tau-\frac{n_0}{2}$ (that is, $\varphi_2^d=1,\varphi_1^d=-\frac{n_0}{2}$), we obtain $\ell\equiv -mn_0-2n=0$. Hence we obtain a generalization of the $\ell=0$ case of Section~\ref{sec:special_families} to a setup where the flavor chemical potentials are switched on.
 Using these constraints, the argument of the fractional part in~\eqref{zabflavoured} vanishes, and 
 the action \eqref{Seffmnflavoured} simplifies to
\begin{align}
\Seff  &\=
 \frac{\pi\i\tau}{6} \,\big(2 {\rm Tr}\hat{R}^3 - {\rm Tr}\hat{R}\big)  
  +  \frac{\pi\i}{6 m (m\t+n)}\,{\rm Tr}\hat{R}  \, +\, \pi\i N^2 \Phi  \,.
\end{align}
The term linear in $\tau$ is the supersymmetric Casimir energy in the presence of flavor chemical potentials \cite{Bobev:2015kza}.
Notice that \eqref{zabflavoured} has a discontinuity when the argument of the fractional part vanishes, since $\{\epsilon\}$ evaluates to  $\{\epsilon\}=0$ for $\epsilon\to 0^+$ and $\{\epsilon\}=1$ for $\epsilon \to 0^-$. While the Bernoulli polynomials $B_2(\{\epsilon\})\to \frac{1}{6}$ and $B_3(\{\epsilon\})\to 0$ are continuous there, the same may not be true for the purely imaginary  contribution $(\i\,\Phi)$. In this case, the jump in the action should be studied by taking $\ell=0$, $\,m\varphi^{\fli}_1 - n \varphi^{\fli}_2 =\epsilon$, and evaluating $\lim_{\epsilon\to 0^\pm}\Phi$. In particular, for $m=1$ the expression of $\Phi$ is given in \eqref{eq:Phi_m=1}, and we find
\begin{align}
\lim_{\epsilon\to 0^+}\Seff  - \lim_{\epsilon\to 0^-}\Seff &= \pi\i N^2  \sllb  ((\Delta_{\llb})_2 - 1)^2\left(  \{ 0^+ \} - \{ 0^- \}\right)\nn\\ 
&= -\pi\i N^2  \sllb  ((\Delta_{\llb})_2 - 1)^2\,.
\end{align}
For $\mathcal{N}=4$ SYM, this is in agreement with the findings of~\cite{Benini:2018ywd}.

The generalization of the $\ell=\pm 1$ saddles is obtained by requiring 
\be\label{generalizing_pm1}
\varphi^I =   \varphi_2^I\, \frac{ m\tau + n \pm 1}{m}\,.
\ee
Taking $I=d$ the condition indeed implies $\ell\equiv-mn_0-2n=\pm1$.
Assuming that all trial R-charges satisfy $0< (\Delta_{ab})_2 <2$,
the same computation leading to \eqref{SeffAnomalies} yields
\begin{align}
\Seff
 &\= \frac{\pi\i\t}{6}\,\big(2{\rm Tr}\hat{R}^3 - {\rm Tr}\hat{R}\big) \,+\, \frac{\pi\i}{12 m (m\t+n)}\,\big( 3 {\rm Tr}\hat{R}^3 - {\rm Tr}\hat{R}    \big)\nn\\[1mm]
&\qquad  \pm  \frac{\pi\i }{24 m (m \tau+n)^2}\, \big({\rm Tr}\hat{R}^3 - {\rm Tr}\hat{R}\big)  
  \, +\, \pi\i N^2 \Phi  \,.
\end{align}
As in the setup where there are no flavor chemical potentials, we should set ${\rm Tr}\hat{R}=0$ as we are working in the strict large-$N$ limit and this trace vanishes at  $O(N^2)$ order due to ABJ anomaly cancellation. However, as in that case we observe that if we fix $(m,n)=(1,0)$ and take $\tau\to 0$, both the ${\rm Tr}\hat{R}^3$ and the ${\rm Tr}\hat{R}$ terms agree with the Cardy-like limit expression derived in \cite[Eq.~(2.34)]{Kim:2019yrz}, which is valid at finite $N$.

Note that the choices \eqref{eq:nice_flavor_chempot}, \eqref{generalizing_pm1} of chemical potentials are rather special, as they are aligned along $T$, or along $T\pm1$. 
  In the next section, we will see how this limitation can be removed while maintaining the nice property that the action is controlled by anomalies.

\section{The effective action written in terms of Tr\,$Q_I Q_J Q_K$}\label{sec:Action_Anomalies}

We show that in specific domains of the complexified chemical potentials, the large-$N$ action of the $(m,n)$ saddles takes a universal form controlled by anomalies. This generalizes previous results appeared in the literature in two ways: first, our proof does not rely on specific classes of examples, and second, it applies to all $(m,n)$ saddles; by contrast,  the gauge variable configurations studied in \cite{Benini:2018ywd,Lezcano:2019pae,Lanir:2019abx} are specified by $(m=1,n)$.

\subsection{A democratic basis of charges}\label{sec:democratic_basis}

So far we have distinguished between the R-symmetry $\widetilde{Q}_d$ and the flavor symmetries $\widetilde{Q}_{\fli}$. We now make a change of basis towards new charges $Q_I$, $I=1,\ldots,d$, that share the same commutation relation with the supercharge. Namely, we make a linear transformation
\be
\widetilde{Q}_J \= Q_I\, a^I{}_J\,,
\ee
with  $\widetilde{Q}_J= (\widetilde{Q}_{\bf j},\widetilde{Q}_d)$ and   $a^I{}_J$ is a real matrix, and require that the new charges $Q_I$  satisfy the commutation relation  
\be\label{commrelQ}
[Q_I,\mathcal{Q}] = -\frac{1}{2}\mathcal{Q}\,,\qquad I=1,\ldots,d\,,
\ee
 where $\mathcal{Q}$ is the supercharge used to define the index. This means that $2Q_I$ are all R-charges.
Recalling the commutation relations \eqref{commrel_charges}, the condition above is satisfied if
\be\label{constraints_a}
\sum_{I=1}^d a^I{}_{\bf j} = 0\,,\qquad\qquad  \sum_{I=1}^d a^I{}_d = 2\,.
\ee
To fix the ideas, an explicit transformation satisfying these conditions is \cite{Amariti:2019mgp}
\be\label{explicit_basis_charges}
\widetilde{Q}_{\fli} = Q_{\fli} - Q_d\,,\qquad \widetilde{Q}_d = \frac{2}{d}\sum_{I=1}^d Q_I\,,
\ee
but we do not need to stick to it.
This transformation of the charges will allow us to make some simple, integer charge assignements later on, when we will discuss toric quiver gauge theories, while the charges of the chiral superfields under the original R-symmetry $\widetilde{Q}_d$ may take rational or even irrational values, as it often happens for the exact superconformal R-charge. 

We also introduce new chemical potentials, $\Delta^I = a^I{}_J \varphi^J$, so that
\be
\Delta^I Q_I = \varphi^J \widetilde{Q}_J \,.
\ee
Requiring that the linear combinations $\Delta^IQ_I$ and $\varphi^J\widetilde{Q}_J$ satisfy the same commutation relation with the supercharge, and recalling that we chose the original R-symmetry chemical potential $\varphi^d$ as in \eqref{expr_varphid_us}, we obtain that the new chemical potentials satisfy the constraint~\cite{Hosseini:2017mds}
\be\label{eq:constraintDelta}
\sum_{I=1}^d \Delta^I = \tau+\sigma -n_0\,.
\ee
The refined index \eqref{our_flavored_index} now reads
\be\label{our_index_with_Delta}
\mathcal{I} = {\rm Tr}\,  \rme^{\pi \i (n_0+1) F}\,\rme^{-\beta \{\mathcal{Q},\overline{\mathcal{Q}}\}} \rme^{2\pi\i\,\left(\sigma J_1 + \tau J_2 +\Delta^I Q_I\right)}\,,
\ee
which is a more symmetric expression,
where  we do not distinguish between the R-symmetry and the flavor charges.
In Appendix~\ref{app:rewriting_index} we compare this reformulation with a closely related expression given in \cite{Amariti:2019mgp}. 

Crucially, the variables $\Delta_{ab}$ introduced in Eq.~\eqref{eq:defDeltaab} remain unchanged in the new basis, namely we can write
\be
\Delta_{\llb} = \varphi^I (\widetilde{Q}_I)_{\llb} = \Delta^I (Q_I)_{\llb}\,,
\ee
where by $(\widetilde{Q}_I)_{\llb}$ and $(Q_I)_{\llb}$ we denote the value of the old and new charges, respectively, on a chiral superfield going from node $a$ to node $b$ of the quiver.
This means that the result \eqref{mostgeneralniceS} for the large-$N$ saddle point action, which is expressed in terms of the variables $\Delta_{\llb}$, can immediately be interpreted in terms of the chemical potentials $\Delta^I$ and the charges $Q_I$. This will be exploited in the analysis below.

\subsection{Action controlled by anomalies}

We are now ready to show that in some specific domain of the chemical potentials, the action of the large-$N$ saddles is essentially controlled by anomalies. Namely, we will show that imposing suitable conditions on the chemical potentials, the action can be expressed in terms of cubic anomaly coefficients.

We start from the action \eqref{mostgeneralniceS}, omitting to write the purely imaginary, $\tau$-independent term  $\pi \i N^2(\CC+\Phi)$ for simplicity; this can be reinstated at the end of the computation. 
  The rest of the action can be expressed as
\begin{align}\label{eq:towards_anomaly}
\Seff(m,n)   &   =  \frac{\pi \i N^2}{24m\taumn^2} \Big[  \big(2\taumn-1\big)^3\, \nu +   \sllb  \big( 2[\Delta_{\llb} ]^m_{\taumn}  -2\taumn+1  \big)^3  \nn\\
&\qquad \qquad- \big(2\taumn-1 \big) \,\nu - \sllb \big( 2[\Delta_{\llb}]^m_\taumn -2\taumn+1 \big) \Big] \,,
\end{align}
where we omitted the last line following the discussion above. 
Note that the terms in the first and second line of (\ref{eq:towards_anomaly}) formally take the form of cubic and linear traces, respectively. In what follows we shall take advantage of
this observation, showing how it can be made very precise in some universal domains of the chemical potentials, thus extending the discussion of the universal 
 families in Section  \ref{sec:quivers} to the setting with flavor fugacities.  The key point will be to determine under which conditions $2[\Delta_{\llb}]^m_\taumn $ can understood as 
 charges of the chiral fields under an auxiliary R-symmetry; this, loosely speaking, corresponds to certain linearization conditions of the bracket function, that we will discuss below.

To proceed, recall that the chemical potentials satisfy the constraint \eqref{eq:constraintDelta}, that we can write in the form (after setting $\sigma=\tau$)  
\be\label{def_Deltad}
\Delta^d = 2\tau-n_0 - \sum_{\fli=1}^{d-1}\Delta^{\fli}  \,.
\ee
Applying $[~\cdot~]_{\taumn}^m$ to both sides of this relation and using (\ref{prope1}), (\ref{prope2}) and (\ref{prope3}) we obtain
\be
\big[\Delta^d \big]^m_{\taumn}  = -  \big[ \sum_{\fli=1}^{d-1}\Delta^{\fli} \big]^m_{\taumn} + 2\taumn - 1\, . 
\ee
From this relation, that is true in general, it is apparent that if we impose the condition
\be\label{sum_d-1_Deltas}
\big[  \sum_{\fli=1}^{d-1}\Delta^{\fli}\big]^m_\taumn =  \sum_{\fli=1}^{d-1}\big[\Delta^{\fli}\big]^m_\taumn +k(d-2)  
\ee
for some integer  $k(d-2)$ with $-1< k(d-2)< d-1$, we find that the bracketed chemical potentials $[\Delta^I]^m_T$ will obey a constraint analogous to (\ref{def_Deltad}), namely 
\be\label{sum_d_Deltas}
\sum_{I=1}^d  [\Delta^I]^m_\taumn  = 2\taumn-1 + k(2-d)\,.
\ee

 Assuming  $d\geq3$, {\it i.e.}~that we have at least three global symmetries, it follows that the values of $k$ allowed by \eqref{sum_d-1_Deltas}  are
\be\label{def_our_integer}
k=\frac{h}{d-2}\,,\qquad\text{with}\ h=0,1,\ldots,d-2\,.
\ee
 However, not all such values of $h$ will be permitted, as in addition to \eqref{sum_d_Deltas} we will also impose further  conditions (see  \eqref{linearization_condition} below) which, depending on the theory considered and the charge assignements, in general entail further constraints. While for now we just assume that these conditions are satisfied and study their consequences, in Section~\ref{sec:toric_quivers}  we will discuss how this is true for toric quivers.

Now we make the further assumption that 
\be\label{linearization_condition}
\big[\sum_{I=1}^d \Delta^IQ_I\big]^m_T = \sum_{I=1}^d \big[\Delta^I\big]^m_T\, Q_I + k \sum_{I=1}^d Q_I - k\qquad \text{on all chiral superfields}\,,
\ee
where $k$ is the  parameter introduced above. This is the linearization condition that we mentioned above.\footnote{This  reproduces  the condition 
(\ref{sum_d-1_Deltas})   by formally postulating the existence of a chiral field with charges $Q^I=(1,\dots,1,0)$.
 In the toric setting, that we will discuss below, this corresponds to viewing the $d$-th chiral field as a composite 
 of the first $d-1$ basic fields.  Namely setting $I=1$ and $J=d$ in the corresponding condition (\ref{eq:our_subsums}).}
It is a non-trivial condition that depends on the charges of the chiral fields and constrains the chemical potentials $\Delta^I$.

Equipped with these assumptions, we consider the linear combination
\be
R_{\rm trial} = \gamma^I Q_I\,,
\ee
with coefficients
\be\label{defgammacoeff}
\gamma^I \equiv  \frac{2}{2\taumn -1 + 2k}\left( [\Delta^I]^m_\taumn + k \right)\,,
\ee
which satisfy $\sum_{I=1}^d\gamma^I =2$ because of \eqref{sum_d_Deltas}. Recalling that the charges $Q_I$ satisfy the commutation relation \eqref{commrelQ}, we see that $R_{\rm trial}$ satisfies $[R_{\rm trial},\mathcal{Q}] = - \mathcal{Q}$
and can thus be seen as a trial R-charge. More precisely, since the parameters $\gamma^I$ that control the mixing of the charge generators are allowed to take complex values, this is a complexified R-charge.\footnote{Here we are elaborating on a point of view suggested in \cite{Hosseini:2018dob}, Appendix~A.}
 Denoting by $r^{\rm trial}_{\llb} $ the charges of each chiral superfield under $R_{\rm trial}$, the condition \eqref{linearization_condition} implies that
\be
[\Delta_{\llb}]^m_\taumn = \frac{ 2\taumn -1+2k }{2}\,r^{\rm trial}_{\llb} -k\qquad \text{on all chiral superfields}\,.
\ee
Plugging this in \eqref{eq:towards_anomaly} we obtain
\begin{align}
\Seff(m,n)  
&\,=\, \frac{\pi \i N^2}{24m \taumn^2} \bigg[  \big(2\taumn-1 \big)^3\, \nu +   \big(2\taumn-1+2k\big)^3\sllb \big(  r^{\rm trial}_{\llb}   - 1\big)^3  \nn\\
&\qquad\ \qquad-\big(2\taumn-1\big)\, \nu - \big(2\taumn- 1 + 2k\big) \sllb \big( r^{\rm trial}_{\llb}   - 1\big) \bigg] \nn\\
&\,=\,\frac{\pi \i }{24m\taumn^2} \Big[  \big(2\taumn-1+2k\big)^3\, {\rm Tr}R^3_{\rm trial} - \big(2\taumn-1+2k\big)\,{\rm Tr}R_{\rm trial} \Big] + \pi \i N^2 \nu f_k(\taumn)\,,
\end{align}
where 
\be
{\rm Tr}\,R_{\rm trial}^3 = \nu + \sllb\big(  r^{\rm trial}_{\llb}   - 1\big)^3\,,\qquad {\rm Tr}\,R_{\rm trial} = \nu + \sllb\big(  r^{\rm trial}_{\llb}   - 1\big)\,,
\ee
and we introduced the function
\begin{align}
f_k(\taumn) &=  \frac{B_3 (\taumn)  -B_3(\taumn+k)}{3m\taumn^2} \nn\\
& = -\frac{k}{6m\taumn^2} \left[ 2k^2 + 3k(2\taumn-1) + 6\taumn^2 - 6\taumn  +1 \right]\,.
\end{align}
We can now recall the large-$N$ property
\be
{\rm Tr}R_{\rm trial}=0\,,
\ee
which follows from the fact that ${\rm Tr}\,Q_I = 0$ at order $O(N^2)$ as a consequence of $U(1)_I$--gauge--gauge anomaly cancellation.
We also express the cubic trace as
\be
{\rm Tr}R_{\rm trial}^3 =  \gamma^I \gamma^J \gamma^K  {\rm Tr} \left(Q_IQ_JQ_K\right) = C_{IJK}\gamma^I \gamma^J \gamma^K \,,
\ee
where we defined the anomaly coefficients 
\be\label{def_CIJK}
C_{IJK} = {\rm Tr} (Q_IQ_JQ_K)\,,
\ee  
which generically  are of order $O(N^2)$ and are completely symmetric in their indices, $C_{IJK}=C_{(IJK)}$. 
 Plugging the definition \eqref{defgammacoeff} in,  we arrive at the result
\be\label{result_action_CIJK_k}
\Seff(m,n)  = \frac{\pi \i}{3m\taumn^2} C_{IJK}  \left( [\Delta^I]^m_\taumn + k \right) \left( [\Delta^J]^m_\taumn + k \right) \left( [\Delta^K]^m_\taumn + k \right) + \pi \i N^2\nu f_k(\taumn)  \,,
\ee
that is an expression for the action essentially controlled by anomaly coefficients.
As it will be clear in Section~\ref{sec:toric_quivers}, the most relevant values of $k$ are $k=0$ and $k=1$. For $k=0$ the action reads
\be\label{actionCIJK_Delta}
\Seff(m,n)  = \frac{\pi \i }{3m\taumn^2} C_{IJK}   [\Delta^I]^m_\taumn    [\Delta^J]^m_\taumn  [\Delta^K]^m_\taumn  \,,
\ee
while for $k=1$ it reads
\be\label{actionCIJK_Deltaplus1}
\Seff(m,n)  = \frac{\pi \i}{3m\taumn^2} C_{IJK}  \left( [\Delta^I]^m_\taumn + 1 \right) \left( [\Delta^J]^m_\taumn + 1 \right) \left( [\Delta^K]^m_\taumn + 1 \right) - \frac{\pi \i N^2 \nu}{m}  \,.
\ee
We see that the action \eqref{actionCIJK_Deltaplus1} takes the same form as \eqref{actionCIJK_Delta}, modulo the shift $[\Delta^I]^m_\taumn \to [\Delta^I]^m_\taumn + 1$, and apart for the constant term. 
 It should  be noted that the constant shift by $-\pi \i N^2\nu/m$ appearing in \eqref{actionCIJK_Deltaplus1} is purely imaginary and independent of the chemical potentials, so it does not affect the phase structure in the grand-canonical ensemble, which is controlled by ${\rm Re}(S_{\rm eff})$.
In addition to this,  we have to add the  constant phase $\pi \i N^2(\CC+\Phi)$ that we omitted at the beginning of this section, to both  (\ref{actionCIJK_Delta}) and (\ref{actionCIJK_Deltaplus1}).
Recall that while $\CC$ is a specific term defined  in (\ref{theconstant}), we have not determined $\Phi$.

For $\mathcal{N}=4$ SYM, and for the gauge variable configurations corresponding to our $(m=1,n)$ saddles, the expressions \eqref{actionCIJK_Delta}, \eqref{actionCIJK_Deltaplus1} were first obtained in \cite{Benini:2018ywd} by using the Bethe Ansatz approach. In particular, for $m=1$ the constant terms in the expressions in Eq. (4.33) of \cite{Benini:2018ywd}   coincide
 with those  in  (\ref{actionCIJK_Delta}) and (\ref{actionCIJK_Deltaplus1}), setting to zero the term  $\CC+\Phi$ that here we have not included explicitly. Related results appeared in \cite{Lezcano:2019pae,Lanir:2019abx} for several examples of toric quiver gauge theories.\footnote{See also \cite{Kim:2019yrz,Amariti:2019mgp} for an analysis for theories with different non-R-symmetries, focusing on the Cardy-like limit of the index rather than on the large-$N$ limit.} We turn to this class of theories next.

\subsection{Toric quiver gauge theories}\label{sec:toric_quivers}
 
We now consider a generic toric quiver and show that there always exists a domain of the chemical potentials where the conditions \eqref{linearization_condition} are satisfied with either $k=0$ or $k=1$, hence the action of the $(m,n)$ saddles takes either the form \eqref{actionCIJK_Delta}, or the form  \eqref{actionCIJK_Deltaplus1}.
We will see that our general expressions reproduce the results obtainted in \cite{Lanir:2019abx,Lezcano:2019pae} in some specific examples of toric quiver gauge theories, when $m=1$. 

Toric quiver gauge theories arise as the theories on a stack of $N$ D3-branes probing the tip of a toric Calabi-Yau cone, where ``toric'' essentially means that the cone admits a $U(1)^3$ symmetry. These theories admit $d\geq 3$ global $U(1)$ symmetries, of which three (including the R-symmetry) come from the toric action, and $d-3$ are baryonic symmetries.
Their properties are nicely encoded in the toric diagrams of the cones, see \cite{Benvenuti:2004dy,Benvenuti:2005ja,Franco:2005sm,Butti:2005vn,Benvenuti:2006xg}  for some of the original references that will be relevant for our discussion. For instance, the number of global $U(1)$ symmetries is the same as the number of external points in the toric diagram.
We will refer to the ``minimal toric phases'' of toric quivers, where the following statements can be made \cite{Butti:2005vn}. The gauge group is a product of  $SU(N)$ factors, with the same number of colours in each factor. There are $d(d-1)/2$ types of bi-fundamental chiral superfields, such that all superfields of a given type carry the same charges. Of these, $d$ types are associated with a basis of toric divisors in the Calabi-Yau cone and may be referred to as ``basic'', while the remaining types are associated with unions of toric divisors and may be seen as ``composite''.\footnote{It should be noted that this is true when the Sasaki-Einstein base of the Calabi-Yau cone is smooth. When the Sasaki-Einstein base is singular, there are more global $U(1)$ symmetries than basic fields \cite{Butti:2005vn}. This is because in such cases there are integer points that are not vertices but lie on the edges of the toric diagram; while both a vertex and an integer point along the edges give rise to a $U(1)$ symmetry, only the vertices are associated with toric divisors in the Calabi-Yau cone and give rise to basic fields. Examples with toric diagrams of this type are the Suspended Pinch Point (SPP) and the Pseudo del Pezzo (P)dP$_4$ quivers. Anyway, these details will not be important for us: the argument we are going to present applies regardless of whether there are integer points lying on the edges of the toric diagram or not. In our discussion, the variable $d$ should always be understood as the number of global $U(1)$ symmetries, namely $d =$ (number of vertices) + (number of integer points along the edges). We refer to \cite{Butti:2005vn} for more details.} 
In a basis of charges given by $Q_I$, $I=1,\ldots,d$, where $2Q_I$ are all R-symmetries, the $I$-th type of basic fields can be assigned charge 1 under  $Q_I$, and 0 under the others, so that its charge under the linear combination $\Delta^J Q_J$ is just $\Delta^I$.  The composite fields carry the same charges as if they were made by specific products of the basic fields, so the charge of the composite fields under $\Delta^IQ_I$ is of the form
  \be\label{charge_composite}
 \sum_{K=I}^{J-1} \Delta^K = \Delta^{I} + \Delta^{I+1} + \ldots + \Delta^{J-1}\,,
  \ee
where $I < J-1<d$ and the integers $I,J$ are determined from the toric diagram in an algorithmic way explained in \cite{Butti:2005vn}, that is not important for the present discussion. The specific charge assignement we have chosen can be found in~\cite{Benvenuti:2006xg}.
  
We now study our conditions \eqref{linearization_condition}, which as shown above lead to an expression for the large-$N$ action controlled by anomalies. On the basic fields the condition is automatically satisfied given the charge assignement above. So if there are no composite fields, we are free to pick $k=\frac{h}{d-2}$, for any choice of $h=0,\ldots,d-2$, and the corresponding saddle point action is given by \eqref{result_action_CIJK_k}. An example with no composite fields is $\mathcal{N}=4$ SYM, where $d=3$ and the only two options are $k=0,1$, yielding the expressions \eqref{actionCIJK_Delta}, \eqref{actionCIJK_Deltaplus1}, respectively, where one should replace $\nu=1$ and $C_{123}=\frac{N^2}{2}$.
 For the $m=1$ saddles this reproduces the results of \cite{Benini:2018ywd}. Another example with no composite  fields is the conifold theory, where $d=4$; in this case in addition to $k=0,1$ we can also take $k=\frac{1}{2}$. Recalling that there are two gauge groups, $\nu=2$, and using $C_{123}=C_{124}=C_{134}=C_{234}=\frac{N^2}{2}$ for the anomaly coefficients,
we conclude that the large-$N$ action of the $(m,n)$ saddles for the conifold theory is
\be
\frac{\Seff(m,n)}{\pi\i N^2}  = 
\left\{ \begin{array}{ll}
F\big([\Delta^I]^m_\taumn;\taumn\big)  
\,\,\,\,\,\,\,\,\,\,\,\,\,\,&{\rm for}\quad [  \sum_{\fli=1}^{3}\Delta^{\fli}]^m_\taumn =  \sum_{\fli=1}^{3}[\Delta^{\fli}]^m_\taumn  
\\[2mm]
F\big([\Delta^I]^m_\taumn + \tfrac{1}{2};\taumn\big)  - \tfrac{1}{m} + \tfrac{1}{2m\taumn}    \,\,\,\,\,\,\,\,\,\,\,\,\,\,&{\rm for}\quad [  \sum_{\fli=1}^{3}\Delta^{\fli}]^m_\taumn =  \sum_{\fli=1}^{3}[\Delta^{\fli}]^m_\taumn + 1  \,,
\\[2mm]
F\big([\Delta^I]^m_\taumn + 1;\taumn\big) - \frac{2}{m} \,\,\,\,\,\,\,\,\,\,\,\,\,\,&{\rm for}\quad [  \sum_{\fli=1}^{3}\Delta^{\fli}]^m_\taumn =  \sum_{\fli=1}^{3}[\Delta^{\fli}]^m_\taumn + 2 
\end{array}\right.
\ee
where we introduced the function 
\be
F\big(x^I;\taumn\big) = \frac{1}{m\taumn^{2}}\left( x^1 x^2x^3 + x^1x^2x^4 + x^1x^3x^4+ x^2x^3x^4\right).
\ee
For $m=1$ this is the result presented in \cite{Lanir:2019abx,Lezcano:2019pae}, that we generalized to  $(m,n)$ gauge variable configurations. We emphasize that it  covers the full domain of chemical potentials for this theory.

For toric quivers with composite fields, Eq.~\eqref{linearization_condition} gives the additional conditions
  \be\label{eq:our_subsums}
\big[\sum_{K=I}^{J-1} \Delta^K \big]^m_\taumn = \sum_{K=I}^{J-1} \big[\Delta^K\big]^m_\taumn + k(J-I-1)\,.
\ee
 These equations can be solved if and only if $k(J-I-1)\equiv \frac{h}{d-2}(J-I-1)$ is an integer, for all allowed choices of $I$ and $J$ and for some choice of $h=0,\ldots,(d-2)$. 
Clearly, the choices $h=0$ (that is, $k=0$) and $h=d-2$ (that is, $k=1$) are always permitted. This completes our proof that for any toric quiver there is a regime of chemical potentials where the saddle point action takes either the form \eqref{actionCIJK_Delta}, or the form \eqref{actionCIJK_Deltaplus1} and is thus fully controlled by the anomaly coefficients \eqref{def_CIJK}. It was shown in~\cite{Benvenuti:2006xg} that these  coefficients can also be extracted from the toric diagram as the area of the triangles defined by three vertices of the diagram, and agree with the Chern-Simons terms $C_{IJK}A^I\wedge F^J\wedge F^K$ in a dual five-dimensional supergravity theory. The expressions \eqref{actionCIJK_Delta}
 or  \eqref{actionCIJK_Deltaplus1} should thus match a dual supergravity on-shell action, defined along the lines of \cite{Cabo-Bizet:2018ehj,Cassani:2019mms}. It was shown in \cite[App.$\:$C]{Cassani:2019mms} that the Legendre transform of the function \eqref{actionCIJK_Delta} with $(m=1,n=0)$ indeed leads to the entropy of the asymptotically AdS$_5$ BPS black holes of \cite{Gutowski:2004yv,Kunduri:2006ek}.

 Intermediate values of the integer $h$, such that $0< h<d-2$, are generically not allowed. Indeed generically there is a composite field with two basic components, namely such that $J-I=2$. Hence for this field we need to require that $\frac{h}{d-2}$ is integer, which is only true for $h=0$ and $h=d-2$.

We now briefly discuss how the results above compare with the analysis of \cite{Lanir:2019abx} (see also \cite{Lezcano:2019pae}), where several examples of toric quivers are considered and it is found that the contribution to the large-$N$ index from each of the gauge variable configurations $(m=1,n)$ takes different functional forms, depending on the regime of the chemical potentials $\Delta^I$.  In order to make the comparison, we will set $m=1,n=0$.
In \cite{Lanir:2019abx}, the different regimes of the chemical potentials are distinguished by inequalities of the type 
\be\label{defCasen}
{\rm Im}\left(-\frac{\mathfrak{n}}{\tau}\right) > {\rm Im}\left(\frac{y}{\tau}\right) > {\rm Im}\left(-\frac{\mathfrak{n}-1}{\tau}\right) \,,
\ee
where  $\mathfrak{n}$ is an integer, 
 while  $y$ can be the sum of the first $d-1$ chemical potentials, $y=\sum_{\fli=1}^{d-1} [\Delta^{\fli} ]_\tau$, or a ``sub-sum'' involving only a subset of the $\Delta^{\fli}$'s, that we denote as $y =\sum_{i\,\in\,{\rm subset}} \Delta^{\fli}$. These sub-sums are determined by the values of $\Delta^{\fli}\widetilde{Q}_{\fli}$ on the composite fields in the quiver, where $\widetilde{Q}_{\fli}$, $\fli=1,\ldots,d-1$, are flavor charges. The change of basis relating our charges $Q_I$ used above to the flavor charges $\widetilde{Q}_{\fli}$ and the R-symmetry $\widetilde{Q}_d$ used in  \cite{Lanir:2019abx} is given precisely by Eq.~\eqref{explicit_basis_charges}.  
 In Table~\ref{Labc_table} we illustrate these slightly different charge assignements using the $L^{a,b,c}$ family of quivers as an example.

\begin{table}
\begin{center} 
$$\begin{array}{|c|c|c|c|c|}  \hline&&&&\\[-1.2em] 
\ \mathrm{Field} \ &\ \mathrm{divisor}\ \,&\ \Delta^IQ_I \ &\ \Delta^{\fli}\widetilde{Q}_{\fli}\ &  \ \widetilde{Q}_4\  \\ \hline\hline&&&&\\[-1.3em] 
Y &\ D_1\ \,&\ \Delta^1\ &\ \Delta^1\ &\ 1/2\ \, \\ \hline &&&&\\[-1.3em] 
U_1   &\ D_2\ \,&\ \Delta^2\ &\ \Delta^2 &\ 1/2\ \,\\ \hline&&&&\\[-1.3em]  
Z &\ D_3\ \,&\ \Delta^3\ & \ \Delta^3 &\ 1/2\  \,\\ \hline &&&&\\[-1.3em] 
U_2   &\ D_4\ \,&\ \Delta^4\ &\ -\Delta^1-\Delta^2-\Delta^3\ \,&\ 1/2 \ \,\\ \hline &&&&\\[-1.3em] 
V_1 & \ D_3\cup D_4\ \,& \ \Delta^3+\Delta^4\  \,&\ -\Delta^1-\Delta^2\ \ \,&\ 1 \ \,\\ \hline &&&&\\[-1.3em] 
V_2 & \ D_2\cup D_3\ \,& \ \Delta^2+\Delta^3\  \,&\ \Delta^2+\Delta^3\ \,&\ 1\  \,\\ \hline 
\end{array}$$ 
\caption{Charge assignments for the $L^{a,b,c}$ family of toric quivers. These have $d=4$ global symmetries, and include the $Y^{p,q}$ family (given by $Y^{p,q}=L^{p+q,p-q,p}$), which in turn includes $ \mathbb{Z}_{2p}$ orbifolds of $\mathcal{N}=4$ SYM (given by $Y^{p,p}$), as well as the conifold theory and its $\mathbb{Z}_{p}$ orbifolds (given by $Y^{p,0}$). The table displays the six different types of fields, where we are using the notation of~\cite{Franco:2005sm}; the divisors in the Calabi-Yau cone that determine the charges of the composite fields in terms of the first four, basic types; the values of the charges $Q_I$, $I=1,\ldots,4$, through the combination $\Delta^IQ_I$; the flavor charges $\widetilde{Q}_{\fli}$, $\fli=1,\ldots,3$, through the combination $\Delta^{\fli}\widetilde{Q}_{\fli}$; the  R-charge $\widetilde{Q}_4$. The charges $\widetilde{Q}_I$ are obtained from the $Q_I$ chosen here via the transformation~\eqref{explicit_basis_charges}, and coincide with those appearing in~\cite{Lanir:2019abx}.} 
\label{Labc_table}
\end{center} 
\end{table}

For $y=\sum_{\fli=1}^{d-1} [\Delta^{\fli} ]_\tau$, Eq.~\eqref{defCasen} can equivalently be expressed as 
\be\label{linearization_bracket_toric}
[\sum_{\fli=1}^{d-1} \Delta^{\fli} ]_\tau  = \sum_{\fli=1}^{d-1} [\Delta^{\fli} ]_\tau + \mathfrak{n}-1  \,,
\ee
and the possible values of $\mathfrak{n}$ are $\mathfrak{n} =1,2,\ldots , d-1$.
Indeed, given the definition of $[\ldots]_\tau$, the two sums in this equation can only differ by an integer, and the value of the latter is specified by the inequality \eqref{defCasen}.\footnote{In these manipulations, the following observations are useful. Decomposing any complex number as $y = y_1 + \tau y_2$, and taking $\tau_2>0$, the inequality \eqref{defCasen} can be re-expressed as
$
-\mathfrak{n} < y_1 < -\mathfrak{n} + 1\,.
$
We can  also convert the square bracket $[\ldots]_\tau$  into the bracket $\{\ldots\}$ via the relation $[y]_\tau \equiv \{ y_1\} -1 + \tau y_2  \, .$
Therefore in this notation the inequality for $y=\sum_{\fli=1}^{d-1} [\Delta^{\fli} ]_\tau$ reads
$$
-\mathfrak{n} < \sum_{\fli=1}^{d-1} \{ \Delta^{\fli}_1\} -d+1  < -\mathfrak{n}+1\,.
$$
Since $0\leq  \{ \Delta^{\fli}_1\}<1$, this expression makes it clear that the possible values of $\mathfrak{n}$ are $\mathfrak{n}=1,\ldots,d-1$. It is also immediate to show that this inequality is the same as~\eqref{linearization_bracket_toric}. 
}
Analogously, when the sum involves only a subset of the $\Delta^{\fli}$'s, the corresponding inequality can be re-expressed as
\be\label{condition_subsum}
[\sum_{\fli\,\in\,{\rm subset}} \Delta^{\fli} ]_\tau  = \sum_{\fli\,\in\,{\rm subset}} [\Delta^{\fli} ]_\tau + \mathfrak{n}_{\rm subset}-1  \,,
\ee
where there is a different integer $\mathfrak{n}_{\rm subset}$ for each subset. Since in \cite{Lanir:2019abx}  these sub-sums are determined by the value of $\Delta^{\fli}\widetilde{Q}_{\fli}$ on the fields, they are not precisely the same as the charges \eqref{charge_composite} of the composite fields under $\Delta^I Q_I$, which also involve $\Delta^d$.  
 Nevertheless, the conditions \eqref{condition_subsum} are equivalent to those obtained by using the sub-sums \eqref{charge_composite} and eliminating $\Delta^d$ from the latter by means of \eqref{def_Deltad}. 
The relations \eqref{condition_subsum} can thus be written as
\be\label{condition_subsum_toric}
[\sum_{K=I}^{J-1} \Delta^K ]_\tau  = \sum_{K=I}^{J-1}  [\Delta^K ]_\tau + \mathfrak{n}_{IJ}-1  \,,
\ee
where $\mathfrak{n}_{IJ}$ is an integer whose possible values are $\mathfrak{n}_{IJ} =1,2,\ldots , J-I$, and the integers $I,J$ are again those characterizing the composite fields, determined by the toric diagram.

Our discussion should make it clear that for any toric quiver, the different possible domains of chemical potentials are reached by varying the integers  $\mathfrak{n}$, $\mathfrak{n}_{IJ}$ defined above. In each of these chambers, the saddle point action takes a different expression, obtained by plugging the values of $[\Delta_{ab}]_\tau = [\Delta^I Q_I]_\tau$ for each chiral field in \eqref{mostgeneralniceS}. 
 The explicit expressions can be found in \cite{Lanir:2019abx} for a number of examples. 
Comparing \eqref{linearization_bracket_toric}, \eqref{condition_subsum_toric} with \eqref{sum_d-1_Deltas}, \eqref{eq:our_subsums}, we see that our conditions yielding the universal expressions for the saddle point action correspond to the choices $\mathfrak{n} = h+1$, and $\mathfrak{n}_{IJ} = \frac{h}{d-2}(J-I-1) 
+1$. Hence $h=0$ corresponds to the ``minimal chamber'' $\mathfrak{n}=\mathfrak{n}_{IJ}=1$, while $h=d-2$ corresponds to the ``maximal chamber'' $\mathfrak{n}= d-1$, $\mathfrak{n}_{IJ}=J-I$.  For the special case of the conifold, the additional possibility $h=1$ corresponds to $\mathfrak{n}=2$ (while there are no integers $\mathfrak{n}_{IJ}$ to consider). While for the ``minimal chamber'' agreement with the form of the action   \eqref{actionCIJK_Delta}  for several examples was  pointed out in \cite{Lanir:2019abx}, for the ``maximal chamber'' the form of the action \eqref{actionCIJK_Deltaplus1} was not noticed.

\subsection{Recovering the unflavored action}\label{sec:recovering_unflavoured}

We now consider the limit of the results above in which all flavor chemical potentials are switched off, so that the only charge appearing in the definition of the index is the R-symmetry (which may be the exact superconformal R-symmetry or a different one).  We will show that the expressions \eqref{actionCIJK_Delta} and \eqref{actionCIJK_Deltaplus1} agree with the action  \eqref{SeffAnomalies} derived the unflavored setup for the $\ell=\mp1$ saddles.

Switching off the flavor chemical potentials corresponds to setting
\be\label{switching_off_flavours}
\varphi^{\fli} = 0\,,\qquad \varphi^d = \frac{2\tau-n_0}{2}\,.
\ee
Recalling that $\Delta^I = a^I{}_J \varphi^J$, this is equivalent to
\be\label{value_Delta_unflavoured}
\Delta^I \,=\, a^I \, \frac{2\tau-n_0}{2}\,,
\ee
where we have denoted $a^I \equiv a^I{}_d$ so as to simplify the notation in the next formulae.
Recalling the definition \eqref{defmybracket} of the brackets,
 we have 
\be\label{bracket_Delta_unflavoured}
[\Delta^I]^m_\taumn  \,=\, \Big\{\ell\,\frac{a^I}{2} \Big\} + a^I \taumn -1   \,,
\ee
with $\ell = -mn_0-2n$.
Using that the $a^I$ add up to 2 (this is the second relation in \eqref{constraints_a}), the constraint \eqref{sum_d_Deltas} takes the form
\be
\sum_{I=1}^d \Big\{ \ell\,\frac{a^I}{2} \Big\} \,=\, d-1+k (2-d).
\ee
We will assume $a^I>0$, which implies $a^I<2$. Then it is easy to check that the condition above is solved by $(\ell=-1,k=0)$, as well as by $(\ell=1, k=1)$.\footnote{While these are the generic solutions, other solutions may be possible when the $a^I$ take some special values. For instance for $a^I = 2/d$, corresponding to the choice \eqref{explicit_basis_charges} for the R-charge  $\widetilde{Q}_d$, one can take $\ell=-1\ ({\rm mod}\ d)$ when $k=0$, and $\ell=1\ ({\rm mod}\ d)$ when $k=1$. }
 The conditions \eqref{linearization_condition}, to be imposed on all chiral superfields, now read
\be
\Big[\frac{2\tau-n_0}{2}\,\widetilde{Q}_d\Big]^m_\taumn \,=\,  \Big[  a^I \, \frac{2\tau-n_0}{2} \Big]^m_\taumn \, \widetilde{Q}_J (a^{-1})^J{}_I + k \sum_{I=1}^d \widetilde{Q}_J (a^{-1})^J{}_I - k\,,
\ee
which after some manipulations using \eqref{defmybracket} can be rewritten as
\be
\Big\{\frac{\ell}{2}\,\widetilde{Q}_d\Big\} \,=\,  (a^{-1})^J{}_I \Big\{  a^I \, \frac{\ell}{2} \Big\}  \,\widetilde{Q}_J + (1-k) \Big(1- \sum_{I=1}^d \widetilde{Q}_J (a^{-1})^J{}_I \Big)\,.
\ee
As in the unflavored case, we make the assumption that the value of the R-charges $\widetilde{Q}_d$ on the superfields all lie between 0 and 2.\footnote{
For toric quivers, this follows from the assumption $a^I>0$, $I=1,\ldots,d$, and from the fact that $\sum_{I=1}^d a^I=2$. Indeed for toric quivers $a^I$ are the R-charges of the $I$-th type of basic fields in the basis defined by $\widetilde{Q}_J$ (this parameterization was proposed in \cite{Butti:2005vn}). The R-charges of the composite fields are of the form $a^I+\ldots +a^{J-1}$, and all these quantities lie between 0 and 2.}
Then it is easy to check that both choices $(\ell=-1,k=0)$ and $(\ell=1, k=1)$ solve the conditions.

We can now write down the saddle-point action for these two choices. As a preliminary step, we note that
\be
C_{IJK}a^Ia^Ja^K \equiv {\rm Tr }\big( (a^IQ_I) (a^JQ_J) (a^KQ_K) \big) = {\rm Tr }\big( \widetilde{Q}_d\, \widetilde{Q}_d\, \widetilde{Q}_d \big) \equiv {\rm Tr}R^3\,,
\ee
where ${\rm Tr}R^3$ is the cubic anomaly of the R-symmetry $\widetilde{Q}_d$.
Then, replacing \eqref{bracket_Delta_unflavoured} with the choice $\ell=-1$ into the $k=0$ action \eqref{actionCIJK_Delta}, and with the choice $\ell=1$ into the $k=1$ action \eqref{actionCIJK_Deltaplus1}, 
we obtain
\be\label{eq:recover_unflavoured_action}
\Seff (\ell=\mp  1; \tau ) =   \pi \i\,\frac{(2\taumn \mp 1)^3}{24 m \taumn^2}\, \mathrm{Tr}R^3 + \left\{ \begin{array}{cl}
0 &\qquad \mathrm{for}~\ell=-1\\
-\pi \i\,\frac{N^2 \nu}{m}& \qquad \mathrm{for}~\ell=1
\end{array} \right.\, .
\ee
To compare with the Section~\ref{sec:special_families}, we should take into account the term $\pi\i N^2(\CC+\Phi)$, that we omitted in this section. 
Evaluating $\CC$ for $\ell = \mp1$, we obtain 
\be
\pi\i N^2(\CC_{\ell=\mp 1}+\Phi) = \pi\i \left(-\frac{n}{3m} \pm \frac{1}{2m}\right){\rm Tr}R^3 + \pi\i N^2\Phi + \left\{ \begin{array}{cl}
0 &\qquad \mathrm{for}~\ell=-1\\
\pi \i\,\frac{N^2 \nu}{m}& \qquad \mathrm{for}~\ell=1
\end{array} \right. \, .
\ee
Adding this to the action \eqref{eq:recover_unflavoured_action} reproduces  precisely the expression \eqref{SeffAnomalies}, where we used $\mathrm{Tr}R=0$.
 Besides providing a consistency check, this computation shows the direct relation between the universal action controlled by anomalies in Section \ref{sec:special_families} and the one in the present section.
 
Notice that choosing $\Phi$ such that $\CC+\Phi=0$,  the first term in \eqref{eq:recover_unflavoured_action} can be translated to  gravity  using the holographic dictionary, while the term proportional to the dimension of the gauge group $N^2 \nu$ cannot be matched to a computation done within five-dimensional supergravity.
The choice of $\Phi$ that leads to the gravity-friendly expression (\ref{perfectcube}) corresponds to 
\be
\pi\i N^2(\CC_{\ell=\mp 1}+\Phi) =  \left\{ \begin{array}{cl}
0 &\qquad \mathrm{for}~\ell=-1\\
\pi \i\,\frac{N^2 \nu}{m}& \qquad \mathrm{for}~\ell=1
\end{array} \right. \, .
\ee
This makes clear that the two  possibilities differ by a  slight, albeit non-universal, term and it  strongly suggests that in the
 expressions  (\ref{actionCIJK_Delta}),  (\ref{actionCIJK_Deltaplus1}) the choice of $\Phi$ 
  consistent with gravity  expectations \cite{Cassani:2019mms} should be
 \be
\pi\i N^2(\CC_{k=0,k=1}+\Phi) =  \left\{ \begin{array}{cl}
0 &\qquad \mathrm{for}~k=0\\
\pi \i\,\frac{N^2 \nu}{m}& \qquad \mathrm{for}~k=1
\end{array} \right. \, .
\ee

\section{More general solutions to the saddle-point equations \label{sec:familysaddles}}

In this section we discuss saddle-point configurations that are more general than the string-like 
saddles that we have discussed so far. In the first subsection we discuss how these general saddles 
are classified by finite abelian groups. Other solutions discussed in the literature in the context of 
the Bethe-ansatz are included in this classification for a particular choice of the abelian group.
We then construct examples of saddles that are defined by two periods~$T_1$ and~$T_2$. 
We show that such configurations are finite-$N$ saddles of our doubly periodic effective action,
and proceed to compute their action in two different large-$N$ limits.

\subsection{Group-theoretic description of the general saddles \label{sec:groupsaddles}}

Consider the saddle-point equations in the discrete variables, presented in~\eqref{saddleptdisc}. 
Solutions of these equations correspond to configurations 
of the~$N$ eigenvalues~$u_i$ such that the value of the variation of the action with respect 
to~$u_i$ (the left-hand side of~\eqref{saddleptdisc}) is independent of~$i$, 
so that~$\lambda^a$ can be set to that value. 
So far we have repeatedly used the idea that the periodicity of the potential ensures that 
the configuration of eigenvalues wrapping any cycle of the torus~$E_\t = \IC/(\IZ\t+\IZ)$ 
in a string-like configuration is a solution. 
Now we show that more general solutions are possible. Going further, we classify all possible universal 
solutions, where \emph{universal} means that the values of the~$N$ eigenvalues at each node of the 
quiver are independent of the node, so that~$u^a_i=u_i$. Since the left-hand side of~\eqref{saddleptdisc} 
involves the differences~$u^a_{i}-u^b_{j}$, a non-generic solution would involve highly non-trivial cancellations, 
we will not explore this in detail here.

The universal solutions are classified by group homomorphisms of finite abelian groups of order~$N$ into 
the torus~$E_\t$ viewed as an abelian group. In other words, the set of eigenvalues~$\CU = \{u_j\}_{j=1,\dots, N}$ 
is the image of the finite abelian group in~$E_\t$. 
The basic element of the left-hand side of~\eqref{saddleptdisc} can be written as, for a given~$i$,
\be\label{groupForm}
\sum_{u \in \CU} V_{\chi}'(u - u_i) \,.
\ee
Since~$\CU$ is the image of a finite abelian group, we have that~$\{u - u_i\}_{u \in \CU} = \{u\}_{u \in \CU}$ 
as an equality of sets. (Since~$u_i$ is itself an element of~$\CU$, $u-u_i = u'-u_i \Rightarrow u=u'$ 
using the group operation. Therefore the set~$\{u - u_i\}_{u \in \CU}$ consists of~$N$ distinct elements of~$\CU$,
i.e., it is equal to~$\CU$.) 
In other words, the solution set~$\CU$ preserves the permutation symmetry of the action. 
Thus, we have that the expression~\eqref{groupForm} is really an average over the group,
\be\label{groupavg}
\sum_{u \in \CU} V_{\chi}'(u - u_i) \= \sum_{u \in \CU} V_{\chi}'(u) \,,
\ee
and, in particular, it is independent of~$i$. 
Since the left-hand side of~\eqref{saddleptdisc} is a linear combination of~\eqref{groupForm} 
for various multiplets, that is also independent of~$i$ and we have found a solution.

We thus see that a group homomorphism of the type described above leads to a solution.
In fact we argue that this structure is essentially 
necessary, namely that it follows upon making a certain assumption that we spell out below. 
Since the torus itself is an abelian group, the associativity and commutativity axioms hold automatically for~$u_i$.
Assume for the moment that one of the eigenvalues~$u_i$ is the identity element~$0$ on the torus. 
Then the~$SU(N)$ condition~$\sum_{i} u_i =0$ implies that any element~$u_i$ has 
an inverse equal to~$-\sum_{j\neq i} u_j$. 
The final closure axiom also follows if we make the assumption that the set~$\{u - u_i\}_{u \in \CU}$ 
itself is independent of~$i$, namely that there is no spontaneous breaking of permutation 
symmetry.\footnote{This assumption is motivated by the fact that it is not easy to have cancellation across multiplets 
in the saddle-point equations~\eqref{saddleptdisc}. Further, even within each multiplet,
a solution which spontaneously breaks the permutation symmetry would involve many 
non-trivial cancellations that depend on the details of the potential~$V_\chi$. 
It would be interesting to explore this direction further.}
If this is the case, 
upon choosing~$u_i$ to be the identity element in~$E_\t$, we find that~$\{u - u_i\}_{u \in \CU} = \CU$, so that,
in particular, for any pair of elements~$u_j$, $u_i$, $u_j-u_i \in \CU$. Since we have already shown the 
existence of an inverse, this proves the closure property. 
In the above arguments we have assumed that the set~$\CU$ contains the identity element. 
If this does not hold, we can translate~$\CU$ by the inverse of one of its elements to reach a set containing the 
identity. The solution in this case is labelled by the embedding of the finite group combined with the translation. 

\vspace{0.4cm}

Thus the problem reduces to finding all possible finite abelian groups~$G$ of order~$N$, and finding their 
embeddings into~$E_\t$. The solution to the first problem, as we review below, is well-known,
$G$ is a direct product of cyclic groups. If we replace the torus by a circle~$U(1)$,
the second problem reduces to one of finding characters of~$G$, whose solution is also well-known. 
This leads precisely to the uniform distribution of eigenvalues that we have been discussing throughout the paper.
In order to find a general solution, we map each cyclic factor of~$G$ into a different cycle of the torus 
representing a~$U(1)$. Now we spell this out in a little more detail. 

The structure theorem for finite abelian groups (see e.g. \cite{Conrad,GroupBerkeley}) 
says that any finite abelian group~$G$ is isomorphic to a product of cyclic groups.
If~$G$ has order~$N$, then we have
\be \label{Gprimefacs}
G \; \cong \; (\IZ/p_1^{r_1} \IZ) \times \dots \times (\IZ/p_m^{r_m} \IZ) \,, \qquad p_1^{r_1} \dots p_m^{r_m} \= N \,.
\ee
Here the prime power factors~$p_i^{r_i}$ of~$N$ are uniquely determined by the group (up to permutations), 
but the primes~$p_i$ need not be distinct. For example, when~$N=24$, the 
factorizations~$2 \times 2 \times 2 \times 3$, $2 \times 4 \times 3$, and $8 \times 3$ lead to non-isomorphic groups. 
The fact that~$\IZ/mn\IZ = \IZ/m\IZ \times \IZ/n\IZ$ when gcd$(m,n)=1$ 
implies that one can write the direct product in the following manner\footnote{Here the notation~$a|b$ for~$a,b \in \IZ$
means that~$a$ is an integer factor of~$b$.} 
\cite{GroupBerkeley},
\be  \label{GNfacs}
G \; \cong \;  \IZ/n_1 \IZ \times \dots \times  \IZ/n_k \IZ  \,, \qquad n_1 | n_2\,, \dots , n_{k-1}| n_k \,,
\qquad n_1 \dots n_k \= N \,.
\ee
The integers~$n_1, \dots , n_k$ (which need not be distinct) are uniquely determined by the group. 
In this notation the three different factorizations of 24 presented above are written 
as~$2 \times 2 \times 6$, $2 \times 12$, and~$24$.

Now we move to the second part of the problem, namely embedding the group into the torus. 
To start off with, any cyclic group should be embedded in some~$U(1)$ cycle on the torus. 
We also want to keep track of different ways of doing so, and we want to avoid redundancies,  
and here we see the utility of the above structure theorem. 
To illustrate the discussion take~$N=6$. In this case, one may imagine that we can distribute the~$6$ eigenvalues 
along some cycle, or we can have a two-dimensional~$3\times 2$ lattice structure with two periods. 
These two different embeddings correspond to the groups~$\IZ/6\IZ$ and~$\IZ/2\IZ \times \IZ/3\IZ$. 
But we know from the above discussion that these two groups are isomorphic, and, in particular from 
the cyclic~$\IZ_6$ description, that there should be one generator which produces the other eigenvalues 
in a string-like configuration. As an explicit example, the configuration~$\frac{m_1}{2}T_1+\frac{m_2}{3}T_2$, 
$m_1=0,1$, $m_2=0,1,2$, $T_1=(0,1)$, $T_2=(1,0)$, 
leads to a lattice structure on the torus, but in fact the configuration is the same
as the string~$\frac{m}{6} T$ with period~$T=(3,2)$.  

Thus, instead of considering arbitrary factorizations of~$N$, we can focus on either of the non-isomorphic 
factorizations given by~\eqref{Gprimefacs},~\eqref{GNfacs}. 
We choose~\eqref{GNfacs} to be concrete. For every factorization of that form, we 
associate~$k$ non-zero periods~$T_1, \dots T_k$. 
Then we distribute the~$n_1$ eigenvalues uniformly along the period~$T_1$, 
smear each of these eigenvalues into a uniform distribution with~$n_2$ along the 
period~$T_2$, and so on, iteratively, for all the~$k$ periods, so as to reach a saddle-point solution.
Note that here we can take some of the periods to be equal. 
This leads to a ``clumped" distribution of saddles. For example, 
if we have~$N=8 = 2 \times 4$, then choosing the periods for~2 and~4 to both be equal to~$T$ 
gives rise to the distribution~$\frac{m}{4} T$, $m=0,1,2,3$, with each point being populated twice. 
Note also that the same distribution can be obtained by choosing the period~$T$ for~4
and trivial period for~2. Since we have the condition that~$n_i | n_j$ for~$i<j$ in the factorization~\eqref{GNfacs}, 
any choice of trivial period is already counted by making certain periods equal. 
This is the reason why we did not allow for the trivial period.\footnote{The only exception to this is the 
distribution where all the eigenvalues are at the origin. This is a special configuration, and is 
discussed in detail in Appendix~D of~\cite{Cabo-Bizet:2019eaf}.} 

This finishes our discussion of the general saddles. It is instructive to compare other results in the literature in this light. 
The configurations introduced in~\cite{Hong:2018viz,Benini:2018ywd}, and also discussed in~\cite{ArabiArdehali:2019orz}, 
are classified by a set of three labels $(m,n,r)$ with $r=0,\ldots,n-1$ and $m n=N$.
It can be checked that every such configuration, taken as a set of points on the torus, has an abelian 
group structure of order~$mn$ under addition. Therefore these solutions 
are included in the above discussion.
On the other hand, the configurations discussed in this section contain new configurations which have 
clumped eigenvalues. For example,~$\mathbb{Z}_{2} \times \mathbb{Z}_{2}$ distributed along~$T_1=T_2$ 
(two clumps of two eigenvalues along the same cycle).

In the rest of this section we illustrate the calculation of the effective action for a simple solution
of this more general kind, namely a configuration corresponding to a factorization~$N=N_1 N_2$. 
As explained above, in order to be genuinely non-string-like, one needs~$N_1 | N_2$. However, 
if we are interested in calculating the large-$N$ effective action rather than enumerating distinct solutions, this 
divisibility property is less important and we can choose to impose it or not. What we want to study, rather,
is the nature of the action when one or both factors becomes very large, and the configuration
looks like a collection of strings, or like a surface, in these respective limits.

\subsection{Saddles with two factors}

Following the above discussion we assume $N=N_1 N_2\,$ with $N_{1}$ and $N_2$ positive and finite integers. 
For convenience we split the colour index $i=1,\ldots, N$ into the double index $(i_1,i_2)$ 
where $i_{1,2}=1,\ldots, N_{1,2}\,$. The new form of saddle point equations are equivalent to 
replacing, in the analysis presented in Section~\ref{sec:Saddles}, the colour indices and sums as follows, 
\be
i \,\mapsto\, (i_1,i_2)\,,\qquad \sum_{i=1}^{N}\,\mapsto \,\,\sum_{i_1=1}^{N_1}\,\sum_{i_2=1}^{N_2}\,.
\ee
In this notation, the saddle-point equations are, for fixed~$a$ and~$i_1i_2$,
\be \label{saddleEqnnew}
\begin{split}
\lambda^{a }&\=\sum_{j_1=1}^{N} \sum_{j_2=1}^{N} \, \Bigl( \partial V(u^a_{i_1 i_2}-u^a_{j_1j_2})
\,-\,\partial V(u^a_{j_1j_2}-u^a_{i_1i_2}) \Bigr) \\ 
&\qquad  \,+\,\sum_{j_1=1}^{N} \sum_{j_2=1}^{N} \, 
\Bigl( \, \sum_{\mathrm{fixed}\,a\to b} \! \partial V_{ab} (u^a_{i_1 i_2}-u^b_{j_1j_2}) \, -\, \sum_{\mathrm{fixed}\,a\leftarrow b}\!\partial V_{ba}(u^b_{j_1j_2}-u^a_{i_1i_2})\, \Bigr)\,, 
\end{split}
\ee
as well as its conjugate equation involving~$\wt \lambda^{a}$ and~$\overline{u^a_{i_1i_2}}$. 

Setting
\be \label{T12etc}
T_{1,2}\=m_{1,2}\, \t+n_{1,2} \,, \quad \text{gcd}(m_1,n_1)\=\text{gcd}(m_2,n_2)\=1\,,\quad (m_1,n_1) \neq (m_2,n_2)  \,,
\ee
we find that the following configurations solve the variational equations,
\be\label{saddlesNew}
u^{a}_{(i_1,i_2)}\=\Bigl(\frac{2i_1-1 - N_1}{2N_1}\Bigr)T_1\,+\,\Bigl(\frac{2i_2-1-N_2}{2N_2}\Bigr)T_2 \; 
\equiv \; u_{(i_1,i_2)}\,.  
\ee
To see this, we look at the right-hand side of~\eqref{saddleEqnnew} evaluated on the configuration~\eqref{saddlesNew}.
The first line on the right-hand side equals
\be 
\sum_{j_1=1}^{N} \sum_{j_2=1}^{N} \, \Bigl( \partial V \Bigl(\tfrac{i_1 - j_1 }{N_1}T_1\,+ \, \tfrac{i_2-j_2}{N_2}T_2\Bigr)
\,-\,\partial V \Bigl(\tfrac{j_1 - i_1 }{N_1}T_1\,+ \, \tfrac{j_2-i_2}{N_2}T_2\Bigr) \Bigr) \,,
\ee
which vanishes because of the periodicity of~$V$ exactly as in Section~\ref{sec:Saddles}. 
The second line on the right-hand side of~\eqref{saddleEqnnew} evaluated on the configuration~\eqref{saddlesNew}
equals
\be\label{SaddleLagrangeHom}
\begin{split}
\lambda^a &\=  \sum_{j_{1}=1}^{N_1}\, \sum_{j_{2}=1}^{N_2} \Bigl(  \,
 \sum_{\mathrm{fixed}\,a\to b} \!\partial V_{ab}\Bigl(\tfrac{i_1 - j_1 }{N_1}T_1\,+ \, \tfrac{i_2-j_2}{N_2}T_2\Bigr) \, - \sum_{\mathrm{fixed}\,a\leftarrow b}\!
\partial V_{ba}\Bigl(\tfrac{j_1 - i_1 }{N_1}T_1\,+ \, \tfrac{j_2-i_2}{N_2}T_2\Bigr)\, \Bigr) \\
& \=  \sum_{j_{1}=1}^{N_1}\, \sum_{j_{2}=1}^{N_2} \Bigl( \,\sum_{\mathrm{fixed}\,a\to b} \! 
\partial V_{ab}\Bigl(\tfrac{j_1 }{N_1}T_1\,+ \, \tfrac{j_2}{N_2}T_2\Bigr) \, - 
\sum_{\mathrm{fixed}\,a\leftarrow b}\!\partial V_{ba}\Bigl(\tfrac{j_1}{N_1}T_1\,+ \, \tfrac{j_2}{N_2}T_2\Bigr)\, \Bigr) \,,
\end{split}
\ee
where, again as in Section~\ref{sec:Saddles}, we have used periodicity of the potential to reach the 
second line. Thus the saddle-point equations simply determine the value of~$\lambda^a$ in a consistent manner. 

Thus we reach the conclusion that the configurations~\eqref{saddlesNew} are complex solutions of the 
variational problem at finite~$N$ arising from the generic family of $SU(N)$ $\mathcal{N}=1$ superconformal quiver 
theories. In the following subsections we move to taking large~$N$ limits.

\subsection{$N_2$ string-like saddles}
In this subsection we evaluate the large-$N$ action of a stack of string-like saddles. These configurations 
are obtained in a large $N=N_1 N_2$ limit such that $N_1$ is large and $N_2$ is finite. This limit is 
implemented by the following identifications
\be\label{limit2single}\begin{split}
\frac{i_{1}}{N_{1}}\mapsto x_{1},\quad \frac{1}{N_{1}}\mapsto dx_{1}, \quad u_{(i_1,i_2)} \mapsto 
u_{i_2}(x_1)\,, \\ \sum_{i_1=1}^{N_{1}}\sum_{i_2=1}^{N_2}\mapsto N_1 \int_0^1 dx_1 \sum_{i_2=1}^{N_2}\,.
\end{split}
\ee
In this continuum limit the configurations \eqref{saddlesNew}, \eqref{T12etc} become
\be
u_{i_2}(x_1)\= \Bigl(x_1-\frac{1}2 \Bigr) T_1\,+\,\Bigl(\frac{2i_2-1-N_2}{2N_2}\Bigr)T_2\,,
\ee
with the following effective action 
\be\label{PotentialOtherSaddle1}
\begin{split}
& S_{\text{eff}}(m_1,n_1;m_2,n_2) \\
& \qquad \=N_1^2\, \sum_{i, j=1}^{N_2}\, \int_0^1 dx\,  \int_0^1 dy\, 
\Bigl( \nu\, V\bigl((x-y)\,T_1 \,+\,\tfrac{(i\,-\,j)}{N_2} \,T_2 \bigr)  \,+\, \\
& \qquad \qquad \qquad \qquad \qquad \qquad \qquad \qquad 
\sllb  V_{ab}\bigl((x-y)\,T_1 \,+\,\tfrac{(i\,-\,j)}{N_2} \,T_2 \bigr)   \Bigr) \,,\\
& \qquad \=N_1^2\, N_2 \, \sum_{i=1}^{N_2}\, \int_0^1 dx\,   
\Bigl( \nu\, V\bigl(x\,T_1 \,+\,\tfrac{i}{N_2} \,T_2 \bigr)  \,+\, 
\sllb  V_{ab}\bigl(x\,T_1 \,+\,\tfrac{i}{N_2} \,T_2 \bigr)   \Bigr)\,.
\end{split}
\ee
In obtaining the second equality we have used the periodicity of the integrand in~$T_1$ to 
reduce the double integral to a single one, and the periodicity of the integrand in~$T_2$ 
to reduce the double sum to a single one.

Using the definition of the potentials $V$ and $V_{ab}$ in terms of the function $\log Q_{a,b}$
in~\eqref{PotentialOtherSaddle1},  we use the identities~\eqref{Id11},~\eqref{Id12} as before 
to calculate averages. The additional term~$\tfrac{i}{N_2} \,T_2 $ in the argument of~$V$ and~$V_{ab}$
and the related sum over~$i$ implies that the integrals over the~$P$ and~$Q$ functions are now sums 
of the following sort (for~$k=2,3$),
\be \label{intersum}
N_2 \sum_{i=1}^{N_2} \frac{ B_k \bigl(\{ m_1 d_{\llb}-n_1 c_{\llb} 
+ \frac{i}{N_2}(m_1 n_2\,-\,n_1 m_2)\} \bigr)}{m_1 (m_1 \tau+n_1)^{k-1}} \,.
\ee
Now, suppose we have~$\text{gcd}(N_2,m_1 n_2-n_1 m_2)=h$. 
Then the sum~\eqref{intersum} can be evaluated using the identity~\eqref{AverageBkgcd} 
with this value of~$h$ and $K=N_2$ and $\ell=m_1 n_2-n_1 m_2$. 
In this manner we obtain the effective action of these complex saddles,  
\be\label{SeffmnInt}
\begin{split}
& S_{\text{eff}}(m_1,n_1;m_2,n_2) \\
& \= \frac{8 \pi\i\t}{27}\,(\textbf{a}\,+\,3\, \textbf{c}) \,+\, \pi\i {\nu}N_1^2\,h^2 \, \frac{1}{6 m_1 (m_1\t+n_1)} \\ 
& \qquad +\pi \i  N_1^2   \, h^2 \, \sllb  \biggl(\frac{ B_3 \bigl(\{ (m_1 d_{\llb}-
n_1 c_{\llb}) \, \wt N_2 \} \bigr)}{3 \wt N_2 m_1 (m_1 \tau+n_1)^2} \,+\,\frac{c_{\llb} B_2 \bigl( \{\wt N_2  \,(m_1 d_{\llb}-
n_1 c_{\llb})\}  \bigr)}{m_1 (m_1 \tau +n_1)}\biggr)\\
&\qquad + \pi \ii \, \Phi'(m_1,n_1) \,,
\end{split}
\ee
where~$\wt N_2 = N_2/h$, and~$\Phi'$ is a real, $\tau$-independent function calculated in a similar manner 
as~$\Phi$ for the string-like ansatz. Note that the answer depends on $m_2$ and $n_2$ only in a discrete manner through~$\wt N_2$.

\subsection{Surface-like saddles} 

Now we analyse the situation~$N=N_1 N_2$ in the limit in which both~$N_1$ and~$N_2$ go to infinity. 
In this limit, these saddles fill the torus $\IC/(\IZ\t+\IZ)$. 
The answer is an integral of a doubly periodic function over the torus and hence it will be given 
by the constant coefficient of the double Fourier expansion of the integrand. 
Recall from~\eqref{DefVa} that the potentials~$V$, $V_{ab}$ are both a sum of three terms,
namely a constant term, a term proportional to~$\log P$, and a term proportional to~$\log Q$. 
As we now show, for these surface-like saddles, the terms proportional to $\log P$ and $\log Q$ have vanishing average 
over the torus and thus do not contribute to the effective action. The only contribution is the linear term in~$\t$.

In this case the continuum limit is
\be\label{limit2}\begin{split}
\frac{i_{1,2}}{N_{1,2}}\mapsto x_{1,2},\quad \frac{1}{N_{1,2}}\mapsto dx_{1,2}, \quad u_{(i_1,i_2)}
 \mapsto u(x_1,x_2)\,, \\ \sum_{i_1=1}^{N_{1}}\sum_{i_2=1}^{N_2}\mapsto N^2 \int_0^1 dx_1\int_{0}^1 dx_2 
\; \equiv \; N^2\int d^2x\,.
\end{split}
\ee
In this limit the configurations \eqref{saddlesNew} become
\be\label{saddles2}
u(x_1,x_2)\= \bigl(x_1-\frac{1}2 \bigr) \, T_1\,+\, \bigl(x_2-\frac{1}2 \bigr) \,T_2 \,.
\ee
The effective action equals the following integral 
\be\label{finalActionN1N2}
\begin{split}
S_\text{eff} & \= N^2 \int \frac{d^2 x}{T_1 T_2} \int \frac{d^2 y}{T_1 T_2} \, \biggl(\,  \nu  \, V(x-y) +
\sllb  \,  V_{ab}(x-y) \biggr) \,, \\
& \= N^2 \int \frac{d^2 x}{T_1 T_2}  \, \biggl( \, \nu \, V(x) 
+  \sllb  V_{ab}(x) \biggr) \,, \\
& \= 2\pi\i \t\Bigl(\frac{\text{Tr}R^3}{6}-\frac{\text{Tr}R}{12}\Bigr)\,=\,\frac{8 \pi\i\t}{27}\,(\textbf{a}\,+\,3\, \textbf{c})\,.
\end{split}
\ee
To obtain the second equality we use the periodicity of the potential as usual, and 
to obtain the third equality we use the definition of~$V$ and~$V_{ab}$ in terms of~$\log Q_{c,d}$ 
and finally use~\eqref{logpi},~\eqref{QKron}. 
Since the effective action in Equation~\eqref{finalActionN1N2} depends on~$\t$ only in a linear manner, 
its Legendre transform vanishes at large~$N$. Therefore the surface-like saddles do not carry large-$N$ entropy. 

The arguments of Section~\ref{sec:contour} regarding the equivalence of the meromorphic and the doubly periodic 
actions and the consequent contour deformation applies to the saddles discussed in this section as well. 
Further, it is clear that we can also calculate the index refined by the flavor chemical potentials 
for these more general class of saddles by combining the ideas of this section and those of 
Sections~\ref{sec:saddles_flavoured_index} and~\ref{sec:Action_Anomalies}. We do not spell out the details here.

\section{Outlook}\label{sec:outlook}

In this paper we explored the large-$N$ limit of the index \eqref{Indexseqt} using the elliptic extension method proposed in~\cite{Cabo-Bizet:2019eaf}. We obtained general results for $\mathcal{N}=1$ superconformal quiver theories with $SU(N)$ gauge group factors, which extend to supersymmetric theories with an R-symmetry. These include identifying infinite families of saddles and, for the $(m,n)$ saddles of Section~\ref{sec:Saddles}, evaluating their contribution to the large-$N$ index, both without and with chemical potentials for the flavor symmetries. We have shown that in specific regimes of the chemical potentials, such contribution is controlled by anomaly coefficients. We discussed how, when specialized to the classes of examples already considered in the literature, our findings agree with the different methods used there.

As remarked at various stages of the presentation, the method used in this paper leaves a constant phase, denoted by $\Phi$, undetermined in the exponential contribution of each saddle, and we have discussed in Section~\ref{sec:discussion_phase} to what extent this is a limitation. On the one hand, we have explained that this phase should be determined by matching it with the $\tau$-independent part of  the meromorphic action evaluated at the saddles. On the other hand, we have noticed that a comparison with gravitational results predicts a specific form of $\Phi$, at least for the cases where the contribution of one saddle can be mapped to a known gravitational solution. 
We have pointed out  that  the $\tau$-independent part of the meromorphic effective action computed in  the literature  \cite{Benini:2018ywd,Lezcano:2019pae,Lanir:2019abx}
  is in slight tension with the gravity predictions. 
However,  note that this comparison may not be entirely straightforward. Firstly, we already noticed that determining a constant phase in the large-$N$ 
saddle-point approximation \emph{a priori} requires control of $O(1)$ effects;  secondly, it may be necessary to better understand the relation between the index and different versions of the supersymmetric partition function, that differ by a prefactor   subtly dependent on the regularization prescription  \cite{Assel:2014paa,Assel:2014tba,Genolini:2016ecx,Papadimitriou:2017kzw,An:2017ihs,Cabo-Bizet:2018ehj,Closset:2019ucb}. In \cite{Cabo-Bizet:2018ehj} we showed that one such regularisation leads to a prefactor that coincides exactly with the entropy function.
It would also be very interesting to constrain~$\Phi$ by symmetry arguments, for instance the modular properties of the index recently discussed in \cite{Gadde:2020bov} may help.

We conclude mentioning a few ideas stemming from this work that we find interesting.
Firstly, there is a question of the existence of an exact finite-$N$ formula for the index~\eqref{Indexseqt}. 
One approach to obtain such an exact formula  would be to find a complete set of complex solutions 
to the saddle-point equations, and then use the theory of resurgence or the exact WKB method (see e.g.~\cite{Marino:2015yie,Aniceto:2018bis}).
As part of this program, one would need to 
find the complete effective action of the theory at each saddle, including quantum effects beyond 
the large-$N$ approximation. Relatedly, one could look for a localization-type argument which 
would reduce the calculation to one of a one-loop determinant. 

Secondly, it would be interesting to clarify the details of the relation of the elliptic extension approach followed here
to the Bethe ansatz approach. One concrete goal is the following. 
Both approaches involve a special set of eigenvalue configurations (saddles
and roots, respectively), and both involve the evaluation of a certain action at these configurations (effective
action and residue, respectively).  So far we understand that all known 
roots of the Bethe ansatz are saddles
of the elliptic approach (see the discussion in~\cite{Cabo-Bizet:2019eaf} and in Section~\ref{sec:familysaddles}
of the current paper). In this paper we found new solutions to the saddle-point equations. It would be interesting
to check whether they are also roots of the corresponding Bethe ansatz.  A related issue is which contour of 
integration is to be chosen and which saddles does it pick. While we have discussed a contour which can in 
principle pass through the saddles presented in this paper, a global analysis of the full contour deformation 
starting from the original integration along the real axis remains to be done.

Thirdly, and perhaps most relevant to the AdS/CFT duality, we have the question of the gravitational 
interpretation of our universal family of saddles. 
In addition to the saddles corresponding to  the known AdS$_5$ black hole solutions, 
we have found an infinite family of saddles with a $O(N^2)$ action that is entirely controlled by anomaly coefficients of the field theory, and such that the Legendre transform is also of order $O(N^2)$. Generally speaking these configurations should correspond to solutions to Euclidean supergravity. However, when there exists a limit in which 
the entropy and the charges are real,  it is natural to conjecture that the corresponding gravity solutions will admit a well-behaved Lorentzian continuation, and therefore
 they should correspond to configurations with a large horizon such as a black hole.
 It would be very interesting to find these solutions, if indeed they exist.
One is tempted to draw an analogy with  a similar phenomenon that is known to happen in AdS$_3$ gravity~\cite{Maldacena:1998bw}, 
although we expect the details in AdS$_5$ to be different.

\section*{Acknowledgements}

 The work of A.~C.~B.~and S.~M.~is supported by the ERC Consolidator Grant N.~681908,
``Quantum black holes:~A microscopic window into the microstructure of gravity''. 
The work of S.~M.~is also supported by the STFC grant ST/P000258/1. 
D.M.~would like to thank the Galileo Galilei Institute for hospitality while this work was being carried out.

\appendix

\section{Some useful special functions \label{App:defs_and_identities}}

Throughout this appendix we use this notation~$\e(x) = \rme^{2\pi \i x}$.
The Pochhammer symbol is defined for $w\in \mathbb{C}$, $q\in \IC$, $|q|<1$,  as 
\be \label{defPoch}
(w;q) \= \prod_{n=0}^\infty (1-wq^n)\,.
\ee
For~$q=\e(\t)$ and $\tau$ in the upper half-plane $\mathbb{H}$, we have 
\be
(q;q) \= \prod_{n=1}^\infty (1-q^n) \= q^{-\frac{1}{24}} \, \eta(\t) \,,
\ee
where $\eta(\tau)$ is the Dedekind eta function.

\vskip 0.4cm

The elliptic Gamma function~\cite{Felder,Spiridonov:2010em,Spiridonov:2012ww} is defined as, for~$z \in \IC$, $\s, \t \in \IH$,  
\be\label{GammaeDef}
\Ge(z;\s,\t) \=  \prod_{j,k=0}^{\infty}
\frac{1- \e \bigl(-z+\sigma (j+1)+ \tau(k+1) \bigr)}{1-\e( z +\sigma j + \tau k)} \,.
\ee
This is a meromorphic function in $z$, with simple poles at $z= -j\sigma -k\tau +\ell$ 
and simple zeros at $z=(j+1)\sigma+(k+1)\tau +\ell$, where $j,k\in \mathbb{Z}^{\geq 0}$ and $\ell \in \mathbb{Z}$. 

\vskip 0.4cm

The Bernoulli polynomials $B_{k}(z)$ for $z\in \mathbb{C}$ are defined through the following generating function,
\bea
\frac{t\, \rme^{z \,t}}{ (\rme^{t}-1)}\;=\;\sum_{k=0}^{\infty} B_{k}(z)\, \frac{t^k}{k!}\,.
\eea
They have the following Fourier series decomposition for~$k \ge 2$ and~$0 \le x<1$,  
\be
B_k(x) \= - \frac{k!}{(2\pi \i)^k} \, \sum_{j \neq 0} \frac{\e(j x)}{j^k} \,.
\ee
In particular we have, for~$z\in \IC$ and~$x \in \mathbb{R}$, 
\bea
& B_2(z) \= z^2-z+\frac{1}{6}\,, 
& \qquad B_2(\{x\})  \=  \frac{1}{2 \pi^2} \, \sum_{j \neq 0} \frac{\e(jx)}{j^2} \,, \label{B2Four} \\
& B_3(z) \= z^3-\frac{3 \,z^2}{2}+\frac{z}{2} \,,  
& \qquad B_3(\{x\})  \=  - \frac{3\,\i}{4\pi^3} \, \sum_{j \neq 0} \frac{\e(jx)}{j^3} \,. \label{B3Four} 
\eea

For $\ell \in \IZ$, $K \in \mathbb{N}$, gcd$(K,\ell)=1$, 
we have, for~$x \in \IR$ and $k\geq 2$, 
\be\label{AverageBk}
\begin{split}
K \sum_{i,\,j\,=\,1}^{K} \,B_k \Bigl(\bigr\{x\,+\,\ell\,\frac{i}{K}\bigl\} \Bigr)
&\=  - \frac{k!}{(2\pi \i)^k} \,  \sum_{p\neq 0}\frac{\e(p x)}{p^{k}} \,
	K \, \sum_{i\,=\,1}^{K} \e\Bigl(p\, \ell\,\frac{i}{K} \Bigr)  \\ 
&\=  - \frac{k!}{(2\pi \i)^k} \, \sum_{p\neq0}\frac{\e(p x)}{p^{k}} \,
	 K^2 \sum_{\tilde{p}_1=-\infty}^{\infty} \delta_{p, K \tilde{p}_1}  \\
&\=\frac{1}{K^{k-2}} \,B_{k}(\{K\, x\})\,.
\end{split}
\ee
When gcd$(K,\ell)=h$, so that~$(K,\ell) =h(\wt K, \wt \ell)$, we can use the following equality
\be
\sum_{i\,=\,1}^{K} \e\Bigl(p\, \ell\,\frac{i}{K} \Bigr)  \= h \sum_{i\,=\,1}^{\wt K} \e\Bigl(p\, \wt \ell\,\frac{i}{\wt K} \Bigr) 
\ee
on the right-hand side of the above result~\eqref{AverageBk} in order to obtain
\be\label{AverageBkgcd}
\begin{split}
K \sum_{i,\,j\,=\,1}^{N} \,B_k \Bigl(\bigr\{x\,+\,\ell\,\frac{i}{K}\bigl\} \Bigr)
&\=\frac{h^2}{K^{k-2}} \,B_{k}(\{\wt K\, x\})\,.
\end{split}
\ee

\vskip 0.4cm

The function~$P$ is defined, for~$z=z_1+\t z_2 \in \IC$, $\z=\e(z)$, $q=\e(\t)$,
\be \label{defP}
P(z;\t) \= q^{\half B_2(z_2)} \, (1-\z) \prod_{n=1}^\infty (1-q^n \z) \, (1-q^n \z^{-1})  \,.
\ee
The function~$|P|$ is invariant under the full Jacobi group~$SL_2(\IZ) \ltimes \IZ^2$, i.e., it is 
 invariant under the modular transformations~$\t \to \frac{a\t+b}{c\t+d}$, 
$z \to \frac{z}{c\t+d}$, as well as under translations by the lattice~$\IZ \t + \IZ$. 
These transformation properties are evident from the second Kronecker limit formula~\cite{Weil},
\be \label{KroneckerLimit}
- \log|P(z;\t)| \= \underset{s\to 1}{\text{lim}} \; 
\frac{\imt^s}{2 \pi} \!  {\underset{m,n \in \IZ \atop (m,n) \neq (0,0)}{\sum}} \; \frac{\e(n z_2 - m z_1)}{|m\t+n|^{2s}} \,.
\ee
Upon evaluating the two sides of this formula for~$z = (m\tau+n)x + c\tau + d$, 
we can calculate the following average values, for~$m \neq 0$,  
\be
\begin{split}
\int_0^1 dx \, \log |P\bigl(x(m\tau+n) +c \t + d \bigr)|  
& \= - \frac{\imt}{2\pi} \, \frac{1}{|m\t+n|^{2}} \, \sum_{p \in \IZ \atop p \neq 0} \, \frac{1}{p^2} \, \e\bigl( (n c - m d) p \bigr) \\
& \=  - \pi  \imt  \, \frac{1}{|m\t+n|^{2}} \, B_2 \bigl( \{m d-nc\} \bigr) \,.
\end{split}
\ee
Since~$P$ is a holomorphic function of~$\t$, we have 
\be\label{Id11}
\int^1_0 dx \log P\bigl((m\t+n)x+c\t+d\bigr) \= -\frac{\pi\i  B_2(\{m d - n c\})}{m (m\t+n)}+\pi\i\, \v_P (m,n) \,, 
\ee
where~$\v_P$ is a real,~$\t$-independent function. 
We present these results for coprime~$m,n$. 
More generally, $m$ and~$n$ on the right-hand side of~\eqref{Id11} are replaced by~$m/h$, $n/h$, where~$h=\text{gcd}(m,n)$.
For $m=0$, $n\neq0$, the evaluation of the integrals proceeds in a slightly different manner. We have
\be
\int_0^1 dx \, \log |P\bigl(nx +c \t + d \bigr)|  
\= - \frac{\imt}{2\pi} \,  \sum_{\ell \in \IZ \atop \ell \neq 0} \, \frac{1}{\ell^2} \, \e(\ell c) \=  - \pi \imt \, B_2 (\{c \}) \,,
\ee
which leads to 
\be\label{intlogPm0}
\int^1_0 dx \log P\left(nx+c\t+d\right) \= \pi\i \tau B_2(\{c\}) +\pi\i\, \varphi_{P}(0,n) \,.
\ee
We can summarize the integrals~\eqref{Id11}, \eqref{intlogPm0} for all~$m$,~$n$ by the following 
series\footnote{This series converges for~$z \notin \mathbb{Z} \, \t+\mathbb{Z}$. 
For such values of $z$ the highly oscillatory behaviour of high frequency Fourier modes suppresses 
their contribution, and consequently the double series converges. For~$z\in \mathbb{Z}\,\t+\mathbb{Z}$ 
there is no oscillatory suppression and the double series diverges. In fact the points~$z \in \mathbb{Z}\,\t+\mathbb{Z}$ 
are zeroes of~$P(z;\t)$, as can be seen directly from the definition~\eqref{defP}. This is consistent with 
the fact that for $\t_2>0$ the real part of the series in~\eqref{logpi} diverges 
to~$-\infty$ at these values.},
\be \label{logpi}
\log P(z;\t) 
 \=  -  \frac{\i}{2\pi} \! 
{\underset{m,n \in \IZ \atop m \neq 0}{\sum}} \; \frac{\e(n z_2 - m z_1)}{m(m\t+n)} + \pi\i \tau B_2(\{z_2\})
+ \pi\i \,\wt \Psi_P  (z) \, ,
\ee
where $\wt \Psi_P(z)$, is a real doubly periodic function of $z_1$ and $z_2$. 
The real part of \eqref{logpi} equals the expression of $\log |P(z;\t)|$ defined by the 
Kronecker limit formula \eqref{KroneckerLimit}. Thus, there exists a $\wt \Psi_P(z)$ 
for which the right-hand side of \eqref{logpi} equals the logarithm of \eqref{defP} in the domain $0\leq z_2<1$. 

\vskip 0.4cm

The doubly periodic function~$Q$ is developed in~\cite{Cabo-Bizet:2019eaf} in a similar manner as~$P$ 
above, i.e., start with an infinite product, then write a double Fourier expansion for~$\log |Q|$, then 
deduce the Fourier coefficients for~$\log Q$ using holomorphy in~$\t$. Here we present the result 
as a Fourier series for~$\log Q$, 
\be \label{QKron}
\log Q(z;\t) 
 \=  - \frac{1}{4\pi^2} \! 
{\underset{m,n \in \IZ \atop m \neq 0}{\sum}} \; \frac{\e(n z_2 - m z_1)}{m(m\t+n)^2}
\, +\,\frac{2 }{3}\pi \i \, \t  B_3\bigl( \{z_2\} \bigr) + \pi\i \,\wt \Psi_Q(z) \,,
\ee
where $\wt \Psi_Q(z)$ is a real doubly periodic function of $z_1$ and $z_2$.
Equivalently, we have, 
\bea\label{Id12}
\int^1_0 dx \log Q\bigl((m\t+n)x+c\t+d\bigr) \!&\=&\! \frac{\pi \i B_3(\{m d- n c\})}{3 m (m\t+n)^2}
+\pi\i \, \varphi_Q(m,n)\,, \quad m \neq 0 \,, \ \qquad\\
\int^1_0 dx \log Q\left(nx+c\t+d\right) \!&\=&\! \frac{2 }{3 }\pi \i\tau B_3(\{c\})+\pi\i\, \varphi_{Q}(0,n)\,. \label{intlogQm0}
\eea
where~$ \varphi_Q$ is a real,~$\t$-independent function that arises from integrating $\wt \Psi_Q(z)$.
Again without loss of generality we have presented the result for coprime $(m,n)$.

\vskip 0.4cm

Using the definition of the function~$Q_{c,d}$ in~\eqref{defQcd}, 
and putting together~\eqref{Id11}, \eqref{intlogPm0}, \eqref{Id12}, \eqref{intlogQm0} 
we can calculate the average value of~$\log Q_{c,d}$ to be
\be
\begin{split}
& \int_0^1 dx \, \log Q_{c,d}\bigl((m\tau+n)x \bigr)  \\
&\  \= \frac{\pi\i \t}{6} (2c^3- c) + \frac{\pi \i B_3(\{m d- n c\})}{3 m (m\t+n)^2}
+ c\,\frac{\pi\i  B_2(\{m d - n c\})}{m (m\t+n)} +\pi\i \bigl( \varphi_Q(m,n) -c\,  \v_P (m,n)\bigl)\,,
\end{split}
\ee
for~$m\neq 0$, and 
\be\label{intP_m=0}
\begin{split}
 \int_0^1\! dx \log \Qabz{c}{d}{nx} 
&=  \frac{\pi\i \t}{6} \Bigl( 2c^3- c  + 4 B_3(\{c\}) -  6c \,B_2(\{c\})  \Bigr) + \pi \i \bigl( \varphi_{Q}(0,n) - c\, \varphi_{P}(0,n)  \bigr)\,.
\end{split}
\ee
When $-1< c < 1$, plugging the definitions of the Bernoulli polynomials~\eqref{B2Four}, \eqref{B3Four} into  \eqref{intP_m=0}
shows that the expression in the bracket vanishes.
In Section~\ref{sec:largeN_action_unflavoured} we have~$c = r-1$, with $r$ the R-charge of the superfield under consideration. 
The condition $-1 < c< 1$ is therefore equivalent to requiring that the R-charges satisfy $0< r < 2$. In this case we have
\be\label{identities_m=0_final}
\int_0^1 dx \log \Qabz{c}{d}{nx} \, =\,  \pi \i\, \bigl( \varphi_{Q}(0,n) - c\, \varphi_{P}(0,n)  \bigr) \,.
\ee
In Section~\ref{sec:largeN_action_unflavoured} we use this identity to show that the action of the  $ (m=0,n\neq0)$ saddle
vanishes at order $O(N^2)$, up to a purely imaginary, $\tau$-independent term.

In the main text, to each supermultiplet in the quiver theory we associate a function $Q_{c,d}$, where for a chiral multiplet $c=r-1,\,d=-n_0\frac{r}{2}$, while for a vector multiplet $c=1,\,d=0$. When we compute the action of a given $(m,n)$ saddle, we denote by $\Phi$ the total contribution of the real, $\tau$-independent 
functions $\varphi_Q(m,n)$, $\varphi_P(m,n)$ appearing in the average value of $\log Q_{c,d}$, that is
\be \label{Phivarphirel}
\Phi = \nu\, \big(  \varphi_{Q}(m,n) - \varphi_{P}(m,n)\big)_{c=1,\,d=0} + \sllb  \big(  \varphi_{Q}(m,n) - c \,\varphi_{P}(m,n) \big)_{c=r_{ab}-1,\,d=-n_0\frac{r_{ab}}{2}}   \,.
\ee

\section{Anomaly coefficients and ABJ anomaly cancellation  \label{App:anomalies}}

The R-symmetry anomaly coefficients relevant to our discussion are  
\bea\label{tHooftA}\begin{split}
\text{Tr}R&\= \text{dim}\,\text{G} \,+ \sum_{\a \in \{\text{chirals}\}}\text{dim}\,\text{R}_\alpha\,(r_\alpha-1)\,\,, \\
\text{Tr}R^3 &\= \text{dim}\,\text{G}\,+\sum_{\a \in \{\text{chirals}\}}\text{dim}\,\text{R}_\alpha\,(r_\alpha-1)^3\,, 
\end{split}
\eea
where $\text{dim}\,\text{G}$ denotes the dimension of the gauge group G and is the gaugino contribution, while $\text{dim}\,\text{R}_\alpha$ denotes the dimension of the G-representation $\text{R}_\alpha$ under which the chiral superfields transform, and  $r_\alpha-1$ are the R-charges of the fermions in the chiral superfields.
The relation between these two quantities and the $\mathcal{N}=1$ Weyl anomaly coefficients $\bf{a}$ and $\bf{c}$ is  \cite{Anselmi:1997am} 
\bea\label{centralCharges}
\textbf{a}\= \frac{3}{32} \Bigl(3 \,\text{Tr}R^3\,-\,\text{Tr}R\Bigr)\,,\qquad \textbf{c}\= \frac{1}{32} \Bigl(9 \,\text{Tr}R^3\,-\,5\,\text{Tr}R\Bigr)\, .
\eea

For quiver gauge theories with gauge group $(SU(N))^\nu$ and matter in bifundamental or adjoint representations, we have
\be
\text{dim}\,\text{G} = \nu\,(N^2-1)\, , \qquad \quad \text{dim}\,\text{R}_\alpha = \left\{\begin{array}{ll} N^2 & \quad \mathrm{for~bifundamentals}\\ N^2-1 & \quad \mathrm{for~adjoints}\end{array} \right. \, .
\ee

Recall that for an $\mathcal{N}=1$ theory with gauge group $\text{G}$ and matter in representation $\text{R}_\alpha$,  the Gauge-Gauge-R-symmetry (ABJ)
 anomaly cancellation condition takes the form
\be
\mu (\mathrm{Adj}) \;+\!\!\sum_{\a \in \{\text{chirals}\}} \mu (\text{R}_\alpha) (r_\alpha -1) \,=\, 0\, ,
\label{ABJcond}
\ee
where $\mu (\mathrm{Adj})$ is the Dynkin index of the adjoint representation of $\text{G}$ and  $\mu (\text{R}_\alpha)$ are the Dynkin indices of the matter representations. 
  Note that this can be any R-symmetry, not necessarily the superconformal one. 
 We impose this condition as the R-symmetry that appears in the index should be non-anomalous. For quiver theories with gauge group $(SU(N))^\nu$ and matter in bifundamental or adjoint representations, focussing on a gauge group factor $SU(N)$, in our conventions we have 
 $\mu (\mathrm{Adj})=N$,  $\mu (\text{fund})=\tfrac{1}{2}$, hence for every node the condition (\ref{ABJcond}) reads
 \be
 N \,+\, \sum_{\a\, \in\, \mathrm{\{adjonts\}}}^\mathrm{fixed~node} N (r_{\a}-1) \,+ \!\!\sum^\mathrm{fixed~node}_{\a\, \in\, \mathrm{\{bifundamentals\}}}\!\!\!\tfrac{1}{2}N\,(r_{\a}-1) = 0 \, ,
 \ee
where the factor of $N$  in the last term comes from the fact that from the point of view of the node, each bifundamental contributes as $N$ fundamentals. 
Summing
 over all the  $\nu$ gauge group factors we obtain the \emph{exact} relation
 \be
  \nu \, + \sum^\mathrm{all~nodes}_{\a \,\in\, \mathrm{\{adjoints\}}}  (r_{\a}-1) \,+\!\! \sum^\mathrm{all~nodes}_{\a\, \in\, \mathrm{\{bifundamentals\}}}\!\!\!\!\!(r_{\a}-1) = 0 \, .
 \ee
 For the superconformal R-symmetry, this argument was given in \cite{Benvenuti:2004dw}, 
 based on vanishing of the $\beta$-function.
Comparing with the definition of $\text{Tr}R$ we obtain
\be
\begin{split}
\text{Tr}R  &\, =\, \nu \,(N^2-1)\; +\!\!\! \sum_{\a\,\in\,\{\text{adjoints}\}}\!\!(N^2-1) (r_{\a}-1) \; +\!\!\! \sum_{\a\,\in\,\{\text{bifundamentals\}}}\!\!\! N^2 (r_{\a}-1) \\
& \,=\,  - \,\nu\; -\!\!\! \sum_{\a\,\in\,\{\text{adjoints}\}}\!\! (r_{\a}-1)\, .
\end{split}
\label{precisetrR}
\ee
We therefore conclude that for the $(SU(N))^\nu$ quiver gauge theories of interest to us, the ABJ anomaly cancellation condition implies that for any R-symmetry $\text{Tr}R=O(1)$ in the large $N$ limit.

\section{Equivalent rewritings of the index}\label{app:rewriting_index}

In this appendix we make some simple manipulations on the index, aimed at clarifying the relation of our formulation in the main text with previous related work, such as \cite{Benini:2018ywd,Amariti:2019mgp,Lezcano:2019pae,Lanir:2019abx}.

We start from our expression \eqref{our_flavored_index}, that we repeat here for convenience: 
\be\label{our_flavored_index_app}
\mathcal{I}\=   {\rm Tr}\,  \rme^{\pi \i (n_0+1) F} \, \rme^{-\beta \{\mathcal{Q},\overline{\mathcal{Q}}\}+
2\pi \i \,\sigma  J_1 + 2\pi \i\, \tau  J_2 + 2\pi\i\,(\varphi^{\fli} \widetilde{Q}_{\fli} + \varphi^d  \widetilde{Q}_d)} \,.
\ee
We recall that $\widetilde{Q}_d$ is the R-charge, that $\widetilde{Q}_i$, $\fli=1,\ldots,d-1$, are non-R-charges, and that the R-symmetry chemical potential is fixed to $\varphi^d = \frac{\sigma+ \tau -n_0}{2}$. In  this expression, the integer $n_0$ appearing in $\varphi^d$ also controls the boundary conditions for the fermion fields around the Euclidean time circle via the insertion of $\rme^{\pi \i (n_0+1) F}$; these are 
 periodic when $n_0$ is even and anti-periodic when it is odd.

Let us assume that there exists a charge assignement such that the new global $U(1)$ charge
\be\label{relation_two_Rcharges}
Q_d = \frac{1}{2}\,\widetilde{Q}_d - \frac{1}{d} \sum_{\fli=1}^{d-1} \widetilde{Q}_{\fli}\,
\ee
takes {\it integer} values on {\it all} bosonic fields in the quiver. Since $\widetilde{Q}_d$ is an R-charge and the $\widetilde{Q}_{\fli}$ commute with the supercharges, we have that $2Q_d$ is still an R-charge.  Using these two properties, we can model the fermion number operator as
\be\label{2Qd_is_F_app}
2Q_d= F \quad {\rm (mod\ 2)}
\ee 
on all fields. In terms of the new charge $Q_d$, the index \eqref{our_flavored_index_app} takes the form
\be\label{formal_index_Bethe}
\mathcal{I}\=   {\rm Tr}\,  (-1)^F   \rme^{-\beta \{\mathcal{Q},\overline{\mathcal{Q}}\}+
2\pi \i\, \sigma  (J_1+Q_d) + 2\pi \i\, \tau  (J_2+Q_d) }\rme^{ 2\pi \i\, \Delta^{\fli} \widetilde{Q}_{\fli} }\,,
\ee
where we introduced the shifted non-R-symmetry chemical potentials
\be\label{dictionaryDeltavarphi_us}
\Delta^{\fli} \,=\, \varphi^{\fli} + \frac{\sigma+ \tau -n_0}{d}\,,\qquad \fli=1,\ldots, d-1\,.
\ee
This is manifestly a supersymmetric index, and the chemical potentials $\Delta^{\fli}$  match the variables used in~\cite{Benini:2018ywd,Lanir:2019abx}.\footnote{The variables $\Delta^{\fli}$ appearing in \cite{Benini:2018ywd,Lanir:2019abx} are expressed as $\Delta^{\fli} = \xi^{\fli} + \frac{\sigma+\tau}{d}$, where $\xi^{\fli}$ are the original non-R-symmetry chemical potentials. These agree with \eqref{dictionaryDeltavarphi_us} upon identifying $\xi^{\fli} = \varphi^{\fli} - \frac{n_0}{d}$. The charges used in \cite{Benini:2018ywd} for $\mathcal{N}=4$ SYM are identified with those used here as: $r=\widetilde{Q}_d$, $q_{\fli}=\widetilde{Q}_{\fli}$, with $\fli=1,2$. 
 The rational values of the R-charges given in \cite{Lanir:2019abx}  for  many examples of quiver gauge theories are the values of $\widetilde{Q}_d$, not of $Q_d$. The latter takes instead integer values on all bosonic fields.} Note that the integer $n_0$ appearing in \eqref{our_flavored_index_app} has been reabsorbed in the definition of the $\Delta^{\fli}$. That $n_0$ can be reabsorbed in a redefinition of the flavor chemical potentials can also be seen from expression \eqref{our_index_alternative}, upon using \eqref{relation_two_Rcharges} and noting that $\rme^{2\pi\i n_0(J_2+Q_d)} = \rme^{2\pi\i n_0 F}=1$; again the crucial point is that here the R-symmetry $2Q_d$ is chosen in a particular way, such that \eqref{2Qd_is_F_app} is satisfied.

In Section \ref{sec:democratic_basis} we have introduced a new basis $Q_I$, $I=1,\ldots,d$, for the charges, satisfying the commutation relation \eqref{commrelQ}, and such that 
\be
\varphi^{\fli} \widetilde{Q}_{\fli} + \varphi^d  \widetilde{Q}_d \,=\, \Delta^IQ_I\,,
\ee
with the new chemical potentials $\Delta^I$ satisfying the constraint
\eqref{eq:constraintDelta}, namely
\be\label{varphi_Delta_app}
\sum_{I=1}^d \Delta^I = \tau+\sigma -n_0\,,
\ee
which follows from requiring that the two sides of \eqref{varphi_Delta_app} have the same commutation relation with the supercharge.
In terms of the new charges, the index \eqref{our_flavored_index_app} reads 
\be
\mathcal{I} = {\rm Tr}\,  \rme^{\pi \i (n_0+1) F}\,\rme^{-\beta \{\mathcal{Q},\overline{\mathcal{Q}}\}} \rme^{2\pi\i\, \left(\sigma J_1 + \tau J_2 +\Delta^I Q_I\right)}\,.
\ee
Again demanding that $Q_d$ satisfies \eqref{2Qd_is_F_app}, one can reabsorb $\rme^{\pi \i (n_0+1)F}= \rme^{2\pi \i (n_0+1)Q_d}$ into the redefinition of $\Delta^d$, taking $\Delta^d_{\rm new}= \Delta^d_{\rm old} + n_0+1$, so that
\be\label{index_AmaritiEtAl}
\mathcal{I} = {\rm Tr}\,\rme^{-\beta \{\mathcal{Q},\overline{\mathcal{Q}}\}} \rme^{2\pi\i\,\left(\sigma J_1 + \tau J_2 +\Delta^I Q_I\right)}\,,
\ee
and the new chemical potentials $\Delta^I$ satisfy the constraint
\be
\sum_{I=1}^d \Delta^I = \sigma+\tau +1\,.
\ee
Precisely this expression for the index and the constraint between the chemical potentials was used in  \cite{Amariti:2019mgp} (see also \cite{Lezcano:2019pae}). In \cite{Amariti:2019mgp}, the change of basis implemented on the charges was the one given by Eq. \eqref{explicit_basis_charges}, which implies \eqref{relation_two_Rcharges}.  

In summary, we have shown that our expression for the index  can be recast both in the form \eqref{formal_index_Bethe} and \eqref{index_AmaritiEtAl}, upon making a choice for the R-charge $2Q_d$ so that \eqref{2Qd_is_F_app} is satisfied.

\bibliography{Saddles}
\bibliographystyle{JHEP}

\end{document}